\documentclass[aps,prd,twocolumn,superscriptaddress,showpacs,preprintnumbers,amsmath,amssymb]{revtex4}

\usepackage{graphicx} 
\usepackage{dcolumn}  
\usepackage{colordvi}

\graphicspath{{ps}}

\renewcommand{\arraystretch}{1.1}

\newcommand {\zac}{ \begin{eqnarray}}
\newcommand {\kon}{ \end{eqnarray}}
\def \dm{$\Delta M$~}
\def\simle{\mathrel{\rlap{\raise 0.511ex \hbox{$<$}}{\lower 0.511ex \hbox{$\sim$}}}}

\def \ssld{0.23\textwidth}
\def \ssls{0.48\textwidth}
\def \mix{$D^0$-$\overline{D}{}^0$ mixing~}
\def \d{$D^0$~}

\def \dstp{$D^{*+}$~}
\def \t{$t_{xy}$}

\begin{document}

\title{\large
Improved search for $D^0$-$\overline{D}{}^0$ mixing using semileptonic
decays at Belle}

\affiliation{Budker Institute of Nuclear Physics, Novosibirsk}
\affiliation{University of Cincinnati, Cincinnati, Ohio 45221}
\affiliation{Department of Physics, Fu Jen Catholic University, Taipei}
\affiliation{The Graduate University for Advanced Studies, Hayama}
\affiliation{Hanyang University, Seoul}
\affiliation{University of Hawaii, Honolulu, Hawaii 96822}
\affiliation{High Energy Accelerator Research Organization (KEK), Tsukuba}
\affiliation{Institute of High Energy Physics, Chinese Academy of Sciences, Beijing}
\affiliation{Institute of High Energy Physics, Vienna}
\affiliation{Institute of High Energy Physics, Protvino}
\affiliation{Institute for Theoretical and Experimental Physics, Moscow}
\affiliation{J. Stefan Institute, Ljubljana}
\affiliation{Kanagawa University, Yokohama}
\affiliation{Korea University, Seoul}
\affiliation{Kyungpook National University, Taegu}
\affiliation{\'Ecole Polytechnique F\'ed\'erale de Lausanne (EPFL), Lausanne}
\affiliation{Faculty of Mathematics and Physics, University of Ljubljana, Ljubljana}
\affiliation{University of Maribor, Maribor}
\affiliation{University of Melbourne, School of Physics, Victoria 3010}
\affiliation{Nagoya University, Nagoya}
\affiliation{Nara Women's University, Nara}
\affiliation{National Central University, Chung-li}
\affiliation{National United University, Miao Li}
\affiliation{Department of Physics, National Taiwan University, Taipei}
\affiliation{H. Niewodniczanski Institute of Nuclear Physics, Krakow}
\affiliation{Nippon Dental University, Niigata}
\affiliation{Niigata University, Niigata}
\affiliation{University of Nova Gorica, Nova Gorica}
\affiliation{Osaka City University, Osaka}
\affiliation{Osaka University, Osaka}
\affiliation{Panjab University, Chandigarh}
\affiliation{Saga University, Saga}
\affiliation{University of Science and Technology of China, Hefei}
\affiliation{Seoul National University, Seoul}
\affiliation{Sungkyunkwan University, Suwon}
\affiliation{University of Sydney, Sydney, New South Wales}
\affiliation{Toho University, Funabashi}
\affiliation{Tohoku Gakuin University, Tagajo}
\affiliation{Tohoku University, Sendai}
\affiliation{Department of Physics, University of Tokyo, Tokyo}
\affiliation{Tokyo Institute of Technology, Tokyo}
\affiliation{Tokyo Metropolitan University, Tokyo}
\affiliation{Tokyo University of Agriculture and Technology, Tokyo}
\affiliation{Virginia Polytechnic Institute and State University, Blacksburg, Virginia 24061}
\affiliation{Yonsei University, Seoul}

\author{U.~Bitenc}\affiliation{J. Stefan Institute, Ljubljana} 
\author{I.~Adachi}\affiliation{High Energy Accelerator Research Organization (KEK), Tsukuba} 
\author{H.~Aihara}\affiliation{Department of Physics, University of Tokyo, Tokyo}
\author{K.~Arinstein}\affiliation{Budker Institute of Nuclear Physics, Novosibirsk} 
\author{T.~Aushev}\affiliation{\'Ecole Polytechnique F\'ed\'erale de Lausanne (EPFL), Lausanne}\affiliation{Institute for Theoretical and Experimental Physics, Moscow} 
\author{A.~M.~Bakich}\affiliation{University of Sydney, Sydney, New South Wales} 
\author{V.~Balagura}\affiliation{Institute for Theoretical and Experimental Physics, Moscow} 
\author{I.~Bedny}\affiliation{Budker Institute of Nuclear Physics, Novosibirsk} 
\author{K.~Belous}\affiliation{Institute of High Energy Physics, Protvino} 
\author{A.~Bondar}\affiliation{Budker Institute of Nuclear Physics, Novosibirsk} 
% ...
\collaboration{The Belle Collaboration}

\date{\today}

\begin{abstract}

A search for mixing in the neutral $D$ meson system has been performed using 
semileptonic $D^0\to K^{(*)-}e^+\nu$ and $D^0\to K^{(*)-}\mu^+\nu$ decays. Neutral $D$ mesons from
$D^{\ast+}\to D^0\pi_s^+$ decays are used and  the flavor at production is
tagged by the charge of the slow pion. The measurement is performed using 
492\,fb$^{-1}$ of data recorded by the Belle detector. From the yield
of right-sign and wrong-sign decays arising from non-mixed and mixed
events, respectively, we measure the ratio of the time-integrated
mixing rate to the unmixed rate to be $R_M = (1.3 \pm 2.2 \pm 2.0)\times 10^{-4}$.
This corresponds to an upper limit of $R_M < 6.1\times 10^{-4}$ at the $90\%$ C.L.

\end{abstract}


\maketitle
\tighten

{\renewcommand{\thefootnote}{\fnsymbol{footnote}}}
\setcounter{footnote}{0}


\section{INTRODUCTION}

The phenomenon of mixing has been well established in the
$K^0$-$\overline{K}{}^0$, 
$B^0$-$\overline{B}{}^0$ and 
$B_s^0$-$\overline{B}_s{}^0$
systems. Recently, evidence for mixing in the
$D^0$-$\overline{D}{}^0$
system has been obtained with a statistical significance of more than
three standard deviations for the first time \cite{belleKK,babarKpi}.
In addition, several new measurements help constrain the relevant
mixing parameters \cite{LP07, hfag}.
The parameters used to characterize 
$D^0$-$\overline{D}{}^0$ mixing are $x = \Delta m /
\overline{\Gamma}$ and $y = \Delta \Gamma / 2\overline{\Gamma}$, where
$\Delta m$ and $\Delta \Gamma$ are the differences in mass and decay
width between the two neutral charmed meson mass eigenstates, and
$\overline \Gamma$ is the mean decay width. The mixing rate 
within the Standard Model is expected to be small \cite{small_mix}:
the largest predicted values, which include the impact of long
distance dynamics, are of the order 
$|x|, |y| \simle 10^{-2}$. 

For $x, y \ll 1$ and negligible $CP$ violation, the time-dependent mixing
probability for semileptonic $D^0$ decays has the following form \cite{Xing}:
\begin{equation}
{\cal{P}}(D^0\to\overline{D}{}^0\to X^+\ell^-\overline{\nu}_\ell)\propto R_M
~t^2~e^{-\Gamma t},
\label{eq_time_dep}
\end{equation}
where $R_M$ is the ratio of the time-integrated mixing probability
to the time-integrated non-mixing probability:
\begin{equation}
R_M
= {\int_0^\infty
  dt~{\cal{P}}(D^0\to\overline{D}{}^0\to X^+\ell^-\overline{\nu}_\ell)\over 
\int_0^\infty dt~{\cal{P}}(D^0\to X^-\ell^+\nu_\ell)}
\approx{x^2+y^2\over 2}.
\label{eq3}
\end{equation}

The mixing rate $R_M$ can be measured directly by using semileptonic
decays of $D^0$ mesons. The most stringent constraint from
semileptonic decays, $R_M <
1.0\times 10^{-3}$ at the 90\% confidence level, comes from our previous
measurement \cite{Urban}. 
Other measurements of $R_M$ using semileptonic
decays are less sensitive \cite{Babar, Cleo, E791}, whereas
results from hadronic decays are more precise \cite{liming-kpi,
babar-k2pi, babar-k3pi}.
In this paper we present an improved search for 
$D^0$-$\overline{D}{}^0$
mixing using semileptonic decays of charmed mesons, which 
supersedes our previous  measurement \cite{Urban}. 
We measure $R_M$ in a 492\,$\rm{fb}^{-1}$ data sample
recorded by the Belle detector at
the KEKB asymmetric-energy $e^+e^-$ collider \cite{pospesevalnik}, at a
center-of-mass (cms) energy of 10.58\,GeV.  
The Belle detector \cite{detektor} is a large-solid-angle magnetic spectrometer that
consists of a silicon vertex detector (SVD), a 50-layer central drift
chamber (CDC), an array of aerogel threshold Cherenkov counters
(ACC), a barrel-like arrangement of time-of-flight scintillation
counters (TOF), and an electromagnetic calorimeter (ECL) comprised of
CsI(Tl) crystals located inside a superconducting solenoid coil that
provides a $1.5$\,T magnetic field. An iron flux-return located outside
of the coil is instrumented to detect $K^0_L$ mesons and to identify
muons (KLM). 
 Two different inner detector configurations were
used. 
The first 140\,$\rm{fb}^{-1}$ of data were taken using a 2.0\,cm
radius beam-pipe and a 3-layer silicon vertex detector (SVD-1), and the
subsequent  
352\,$\rm{fb}^{-1}$ were taken using a $1.5$\,cm radius beam-pipe, a 
4-layer silicon detector (SVD-2) and a small-cell inner drift chamber
\cite{SVD-2}.

To study signal and background distributions we use Monte Carlo (MC)
simulated samples \cite{MC} in which the number of selected
events is about 2.7 
times larger than in the data sample.

\section{RECONSTRUCTION OF $D^0$ DECAYS}

We select $D^0$ mesons arising from $D^{*+} \to D^0 \pi^+_s$ decays
and reconstruct them as $D^0 \to K^{-} \ell^+ \nu_\ell$, where $\ell^+$
can be either an electron or muon  \cite{naboj}.
 The notation $\pi^+_s$
denotes a $slow$ pion, 
i.e.,
the pion that originates from the $D^{*+}$.
The average momentum of this pion is only about 0.23\,GeV$/c$, whereas
the average momentum of the lepton and kaon from the signal
decay are 
0.96\,GeV$/c$ and 1.52\,GeV$/c$, respectively.
The momenta given in this paper are measured in the laboratory
frame, unless otherwise stated; momenta measured in the cms frame
are denoted with an asterisk, e.g., $p^*$.
The reconstruction of $D^0$ mesons in this specific decay chain  
enables 
tagging of 
the $D^0/\overline{D}{}^0$ meson flavor at production
using the charge of the slow pion $\pi_s^\pm$.

There are
three detected particles in the final state: $\pi_s^+$, $K^-$ and
$\ell^+$, where $\ell^+$ can be either a muon or an electron. The
non-mixed decay results in a charge combination
 $\pi_s^+$ $K^-$ $\ell^+$, which 
we refer to 
as the Right-Sign (RS) charge combination.
The mixing process results in $\pi_s^+$ $K^+$ $\ell^-$, which 
we refer to 
as the Wrong-Sign (WS) charge combination, as summarized
in Table~\ref{tab-defrsws}.

\begin{table}[htbp]
\begin{center}
\caption{ The definition of the Right-Sign (RS) and Wrong-Sign (WS)
charge combinations. }
\begin{tabular}{c c c}
\hline
\hline
charge combination & process	& name	\\
\hline
$\pi_s^+,~K^-,~\ell^+$	& non-mixed	& Right-Sign, RS \\
$\pi_s^+,~K^+,~\ell^-$	&  mixed	& Wrong-Sign, WS \\
\hline
\hline
\end{tabular}
\label{tab-defrsws}
\end{center}
\end{table}

Because the neutrino is not directly reconstructed, the masses
of the \d and \dstp candidates are smeared.
However, by calculating the difference between the two masses, the
uncertainty due to the neutrino four momentum cancels to a large
extent. Thus
\zac
\Delta M \equiv M(\pi_s K \ell \nu) - M(K \ell \nu),
\kon
the reconstructed invariant mass difference between the $D^{*+}$ and
the $D^0$ meson, is the most appropriate observable 
to extract the number of signal events. For signal events, the
distribution of $\Delta M$ peaks at 0.145\,GeV$/c^2$, the mass
difference between the \dstp and $D^0$ meson (see
Fig.~\ref{fig-dmpKl}).

\subsection{Selection criteria} 

Among all the different processes occurring in $e^+ e^-$ collisions,
hadronic final states are selected  with an 
efficiency above 99\%. The 
selection is based on the energy of the 
charged tracks and
neutral clusters, total
visible energy in the  cms system, the 
$z$ component (opposite to the positron beam direction) of the total
cms momentum, and the position of the 
reconstructed event vertex \cite{hadron_B}.

Using MC simulation, the criteria to select
the signal decays are optimized to 
give the best significance for the extracted number of mixed (WS) events,
$N_{\rm WS}^{\rm sig}/\sigma_{N_{\rm WS}^{\rm bkg}}$.
The uncertainty
$\sigma_{N_{\rm WS}^{\rm bkg}}$ is due to the fluctuation of the background in
the region $\Delta M < 0.16$\,GeV/$c^2$; the fluctuation of the
signal events at the rate of our previously measured upper limit
\cite{Urban} is negligible. Hence we maximize
\zac
\frac{N_{\rm WS}^{\rm sig}}{\sqrt{N_{\rm WS}^{{\rm bkg}, \Delta M <
0.16}}}. 
\label{eq-sens}
\kon
Since the kinematic properties of mixed and non-mixed events are the
same, in the optimization the RS signal is used instead of the WS
signal. The optimal values for the selection criteria in some
observables are 
correlated and hence the final criteria are obtained by iterative
optimization. 
In the optimization,
selection criteria based on the $D^0$ proper decay time are also
included, as described in Sec.~\ref{lifetime}.

We suppress
$D^0$ mesons arising from $\Upsilon(4S)\to B\overline{B}$
events in order to avoid the situation in which the selected sample would
be composed of two subsamples with different kinematic properties.
These $B\overline{B}$ events  have different kinematic properties from the 
decays of $D^0$ mesons produced in $e^+ e^- \to c \bar c$ (continuum
events), and a different
apparent decay length between the interaction point and the $D^0$
decay vertex, because 
of the finite $B$ lifetime.
As a result, the $D^0$ mesons from this source have slightly different
resolutions in kinematic variables, and their proper decay time 
cannot be
measured in the same way as for $D^0$ mesons from the continuum.
Since the fraction of $D^0$ mesons from $B$ decays is smaller than
that from the continuum  production, and the
background contribution from $B$ decays is large,
the sensitivity to mixed events is not reduced by rejecting
candidates from $B$ decays.

The quantity used to discriminate between $B\overline{B}$ events
(spherical) and 
continuum events (jet-like)  is 
 the ratio of
the second to zeroth Fox-Wolfram moment, $R_2$ \cite{fox-wolfram}.
To suppress candidates from $B$ decays we demand $R_2>0.2$. A further
effective rejection of $B\overline{B}$ events is described below.

Tracks with an
impact parameter with respect
to the interaction point in the radial
direction, $dr<1$~cm, and in the beam direction, $|dz|<2$~cm,
are considered as $\pi_s^\pm$ candidates. These criteria remove
badly reconstructed tracks and tracks not 
arising from the interaction point.
A slow pion candidate is required to have a momentum smaller than
600\,MeV/$c$. To reduce the background from electrons, we
require the electron identification likelihood 
(based on
the information from the CDC, ACC and ECL \cite{e_id})
 of a $\pi_s$ candidate to be $\mathcal L_e < 0.1$, which selects
slow pions with an efficiency of 96\% and rejects 72\% of electrons.
The total efficiency of the slow pion selection criteria and
tracking is 51\%.

Electron candidates are required to have momenta
greater than 250\,MeV/$c$ and an electron identification likelihood
 $\mathcal L_e > 0.95$.
The efficiency of the identification criterion is 76\%; 
in total
around 46\% of all generated signal electrons are retained.
Muon candidates are required to have  momentum
greater than 650\,MeV/$c$ and the muon identification likelihood
(based on information from the KLM and properties of the track \cite{muid})
 $\mathcal L_\mu > 0.97$; the latter criterion selects muons in
the chosen momentum range with an efficiency of 67\%.
These two requirements are highly correlated
since the identification efficiency of muons with momenta lower
than 600\,MeV/$c$ is very poor.
The efficiency of the above selection criteria and tracking is
30\%.

Kaon candidates are required not to satisfy the lepton selection
criteria.
Kaons from $D^0 \to K e \nu$ decays should have $p > 850\,{\rm MeV/c}$
and kaons from $D^0 \to K \mu \nu$ decays  $p > 600\,{\rm MeV/c}$.
The difference in this requirement is due to 
the correlation between lepton momenta and
 kaon momenta that enters through other kinematic variables, and
due to different background contributions in both decay modes.
A combined likelihood for a given track to be a
$K^\pm$, $\pi^\pm$ or $p^\pm$ is obtained based on the
information from the TOF, CDC and ACC \cite{hadron_B}.
Kaon candidates are selected using 
$\frac{\mathcal L(K)}{\mathcal L(K)+\mathcal L(\pi)} > 0.51$ 
(efficiency of 87\% for signal kaons in the selected momentum
range) and 
 $\frac{\mathcal L(K)}{\mathcal L(K)+\mathcal L(p)} > 0.01$
(efficiency of 99\% for signal kaons).
Around 42\% of all  generated kaons in the electron
decay mode, and around 48\% in the muon decay mode pass the selection
criteria.

At this stage, about
17.0\% of all generated $D^{*+}\to\pi_s^+ D^0, D^0 \to
K^-e^+\nu_e$ decays, 
and about
12.5\% of all generated $D^{*+}\to\pi_s^+ D^0, D^0 \to
K^-\mu^+\nu_\mu$ decays, 
are reconstructed. 
Further criteria are applied to improve the sensitivity to mixed
events. In the following, these criteria are described and for each of
them the signal loss and the background rejection factors are given.

\begin{figure}[htbp]
\begin{center}
\includegraphics[width=\ssld]{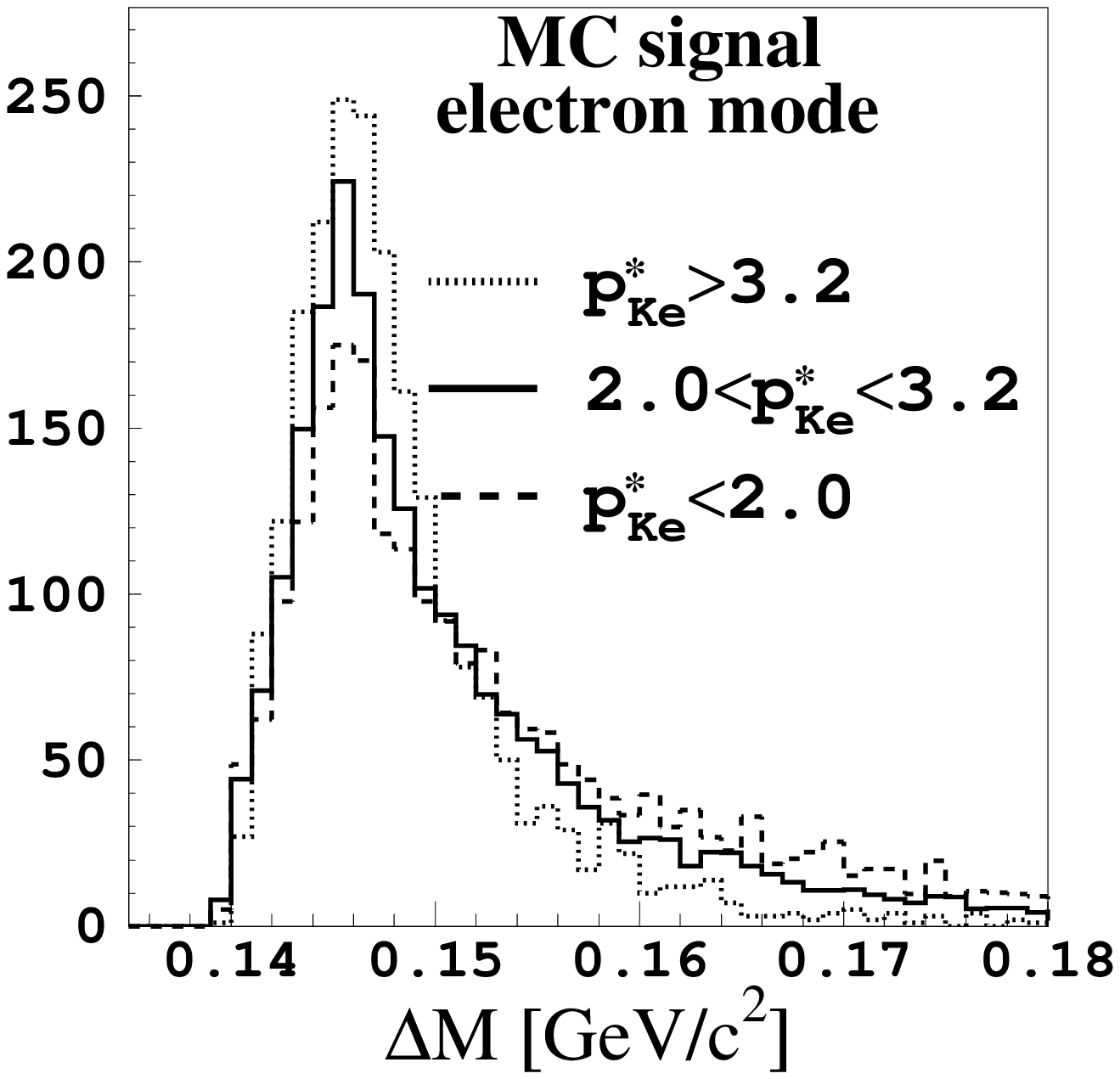}
\includegraphics[width=\ssld]{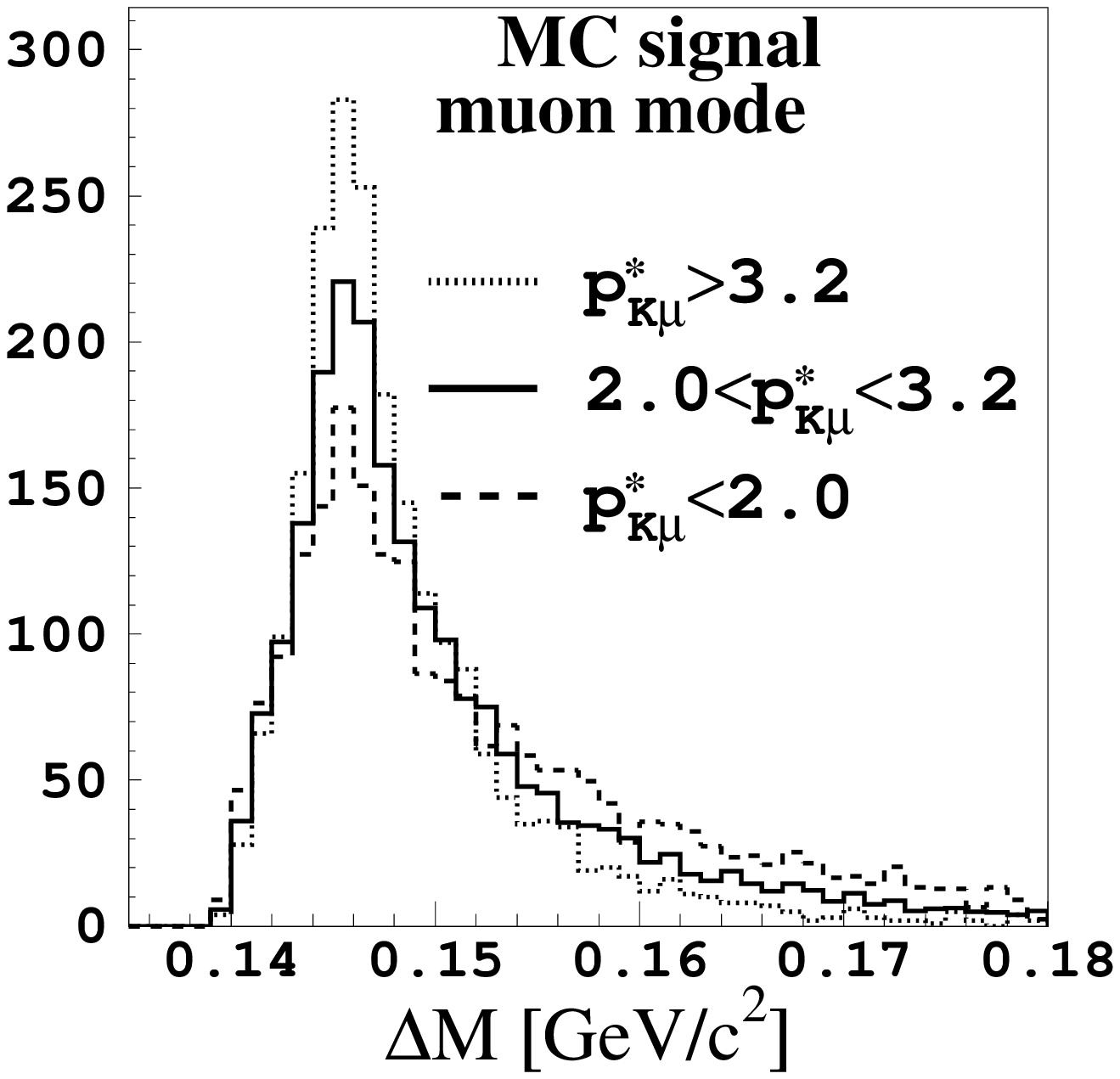}
\end{center}
\caption{ $\Delta M$ distribution for MC-simulated signal
events in different $p^*_{K\ell}$ bins: 
$p^*_{K\ell} ~< ~2.0$\,GeV/$c$ (dashed line), 
2.0\,GeV/$c~ <p^*_{K\ell} < 3.2$\,GeV/$c$ (solid line) and
 $p^*_{K\ell}>3.2$\,GeV/$c$
(dotted line). The
histograms are normalized to the same area. The
resolution is improved at higher values of momentum. The left plot is
for the electron decay mode and the right one for the muon decay mode.}
\label{fig-dmpKl}
\end{figure}

\begin{figure}[htbp]
\begin{center}
\includegraphics[width=\ssld]{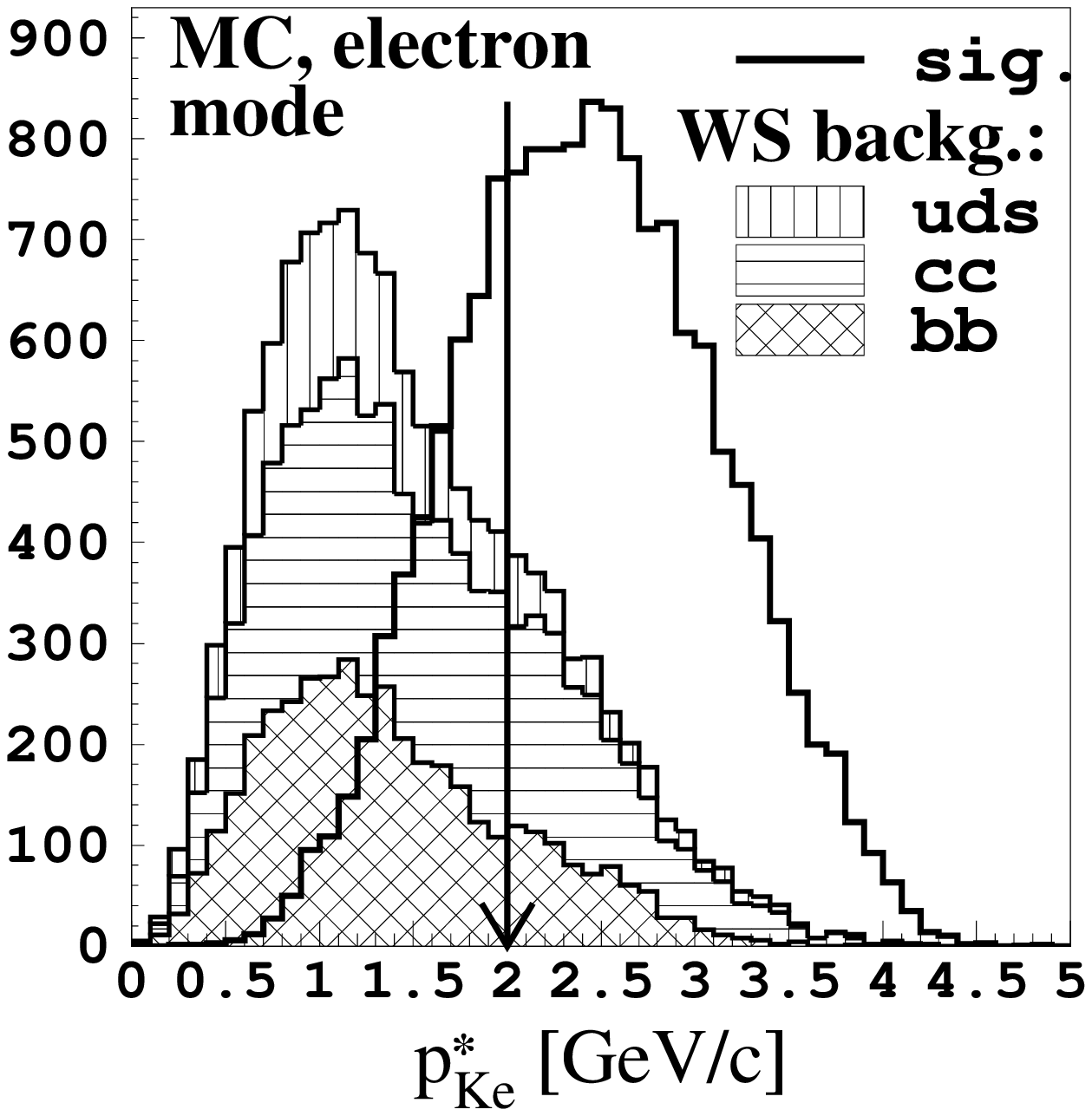}
\includegraphics[width=\ssld]{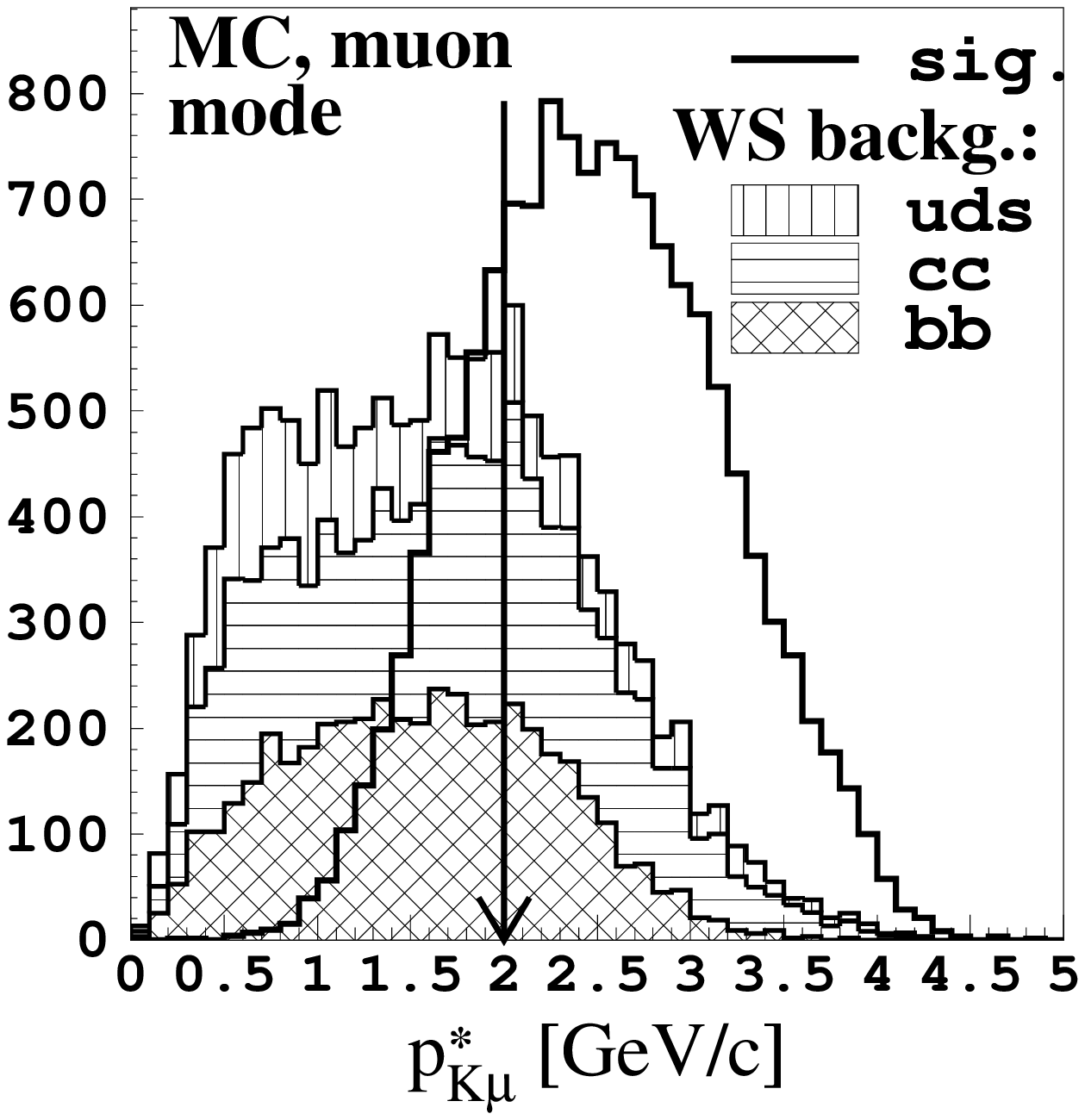}
\end{center}
\caption{MC simulated distribution of $p^*_{K\ell}$ for signal (solid
line) and
background  events ($b \bar b$, $c \bar c$ and $uds$
components of background are shown). The arrow shows the value of
the $p^*_{K\ell}$ requirement.}
\label{fig-pKl}
\end{figure}

The most effective requirement is the one on the sum of the kaon and lepton
momenta, calculated in the cms system, $p^*_{K\ell}$, see Fig.~\ref{fig-pKl}.
Its optimized value  is between 1.7 and 1.9\,GeV$/c$. 
However, simulated signal
events show a clear improvement in the $\Delta M$ resolution at higher
$p^\ast_{K\ell}$ values, see Fig.~\ref{fig-dmpKl}.
The full width at half maximum (FWHM) of the simulated \dm
distribution reduces from 8.6\,MeV/$c$ for $p^*_{K\ell} < 2.0$\,GeV/$c$ to
6.9\,MeV/$c$ for $p^*_{K\ell} > 3.2$\,GeV/$c$ in the electron decay mode,
and from 7.7\,MeV/$c$ to 5.2\,MeV/$c$ 
for the same $p^*_{K\ell}$ intervals
in the muon decay mode.  Hence $p^*_{K\ell}$
is required to be at least $2.0$\,GeV/$c$, a value that also
eliminates a large fraction of
$D^0$ meson decays arising from $\Upsilon(4S)\to B\overline{B}$
events. This requirement results in a signal loss of 28\% in the
electron decay mode and 23\% in the muon 
decay mode, 
while rejecting 76\% of the total background in the electron decay mode and
67\% of the total background in the muon decay mode. 

\begin{figure}[htbp]
\begin{center}
\includegraphics[width=\ssld]{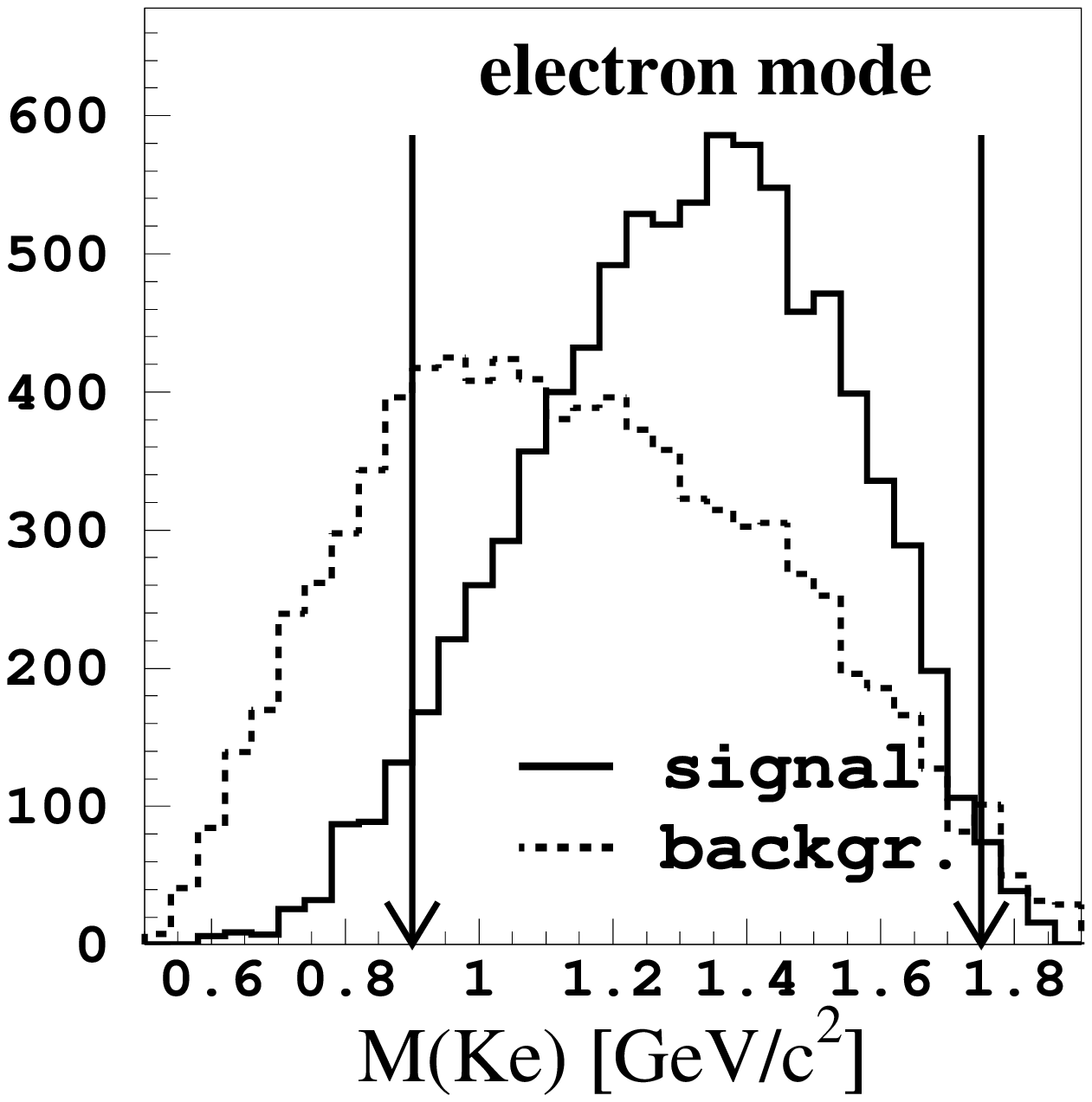}
\includegraphics[width=\ssld]{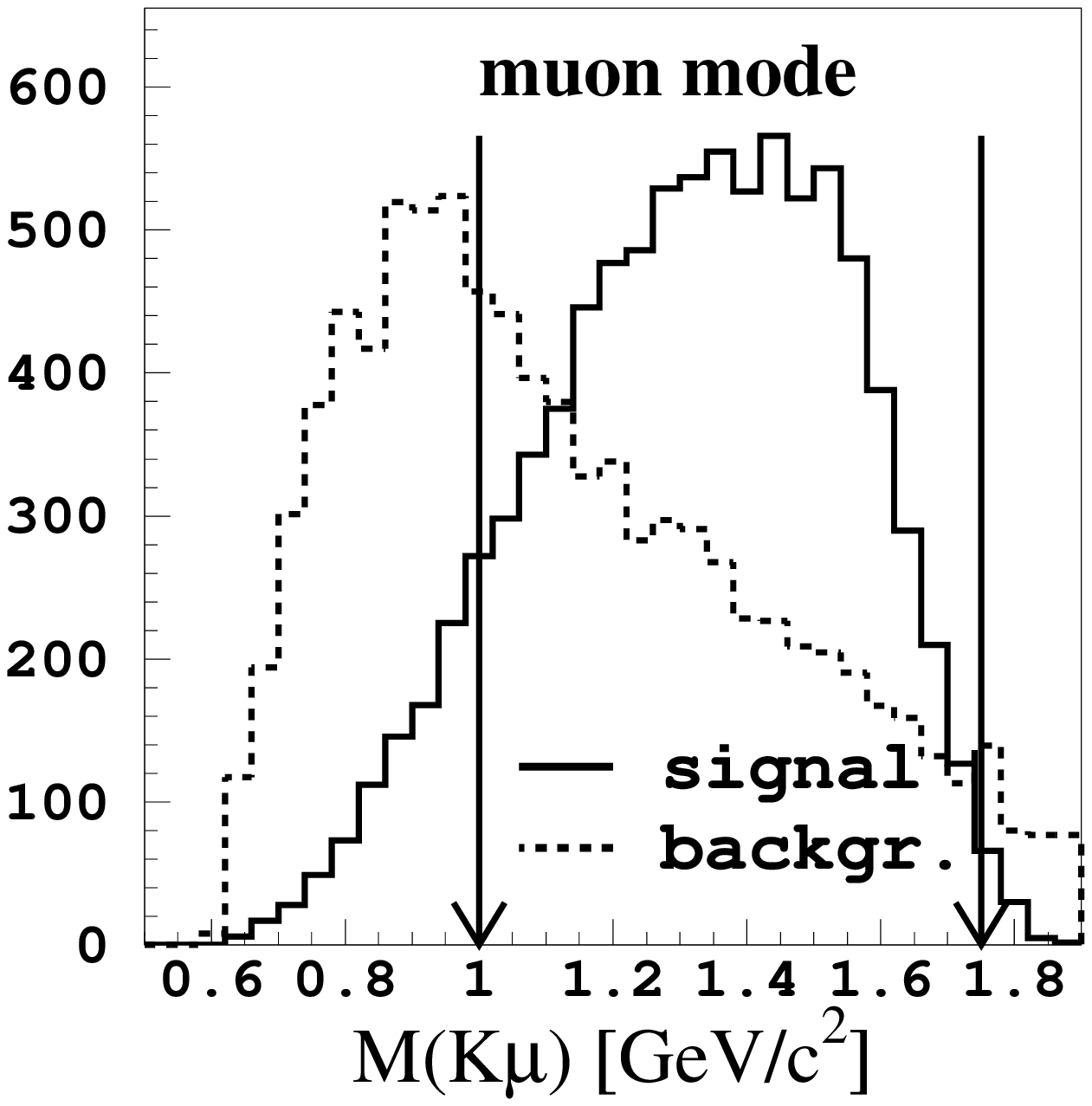}
\end{center}
\caption{ The distribution of $M(K\ell)$ for signal (solid line) and
for background events (dashed), normalized to 
the same number of entries. The arrows show the values of
the selection criteria.
The left plot is for the electron decay mode
and the right one for the muon decay mode.}
\label{fig-mKl}
\end{figure}

We apply a selection on the invariant mass of the 
kaon-lepton system.
For the electron decay mode the optimal range is 
$0.9{\rm \,GeV}/c^2<M(Ke)<1.75{\rm \,GeV}/c^2$ and for the muon decay mode
$1.0{\rm \,GeV}/c^2<M(K\mu)<1.75{\rm \,GeV}/c^2$, see Fig~\ref{fig-mKl}.
In the electron decay mode this requirement rejects 25\% of the total
background at a cost
of losing 5.5\% of signal events. In the muon decay mode the signal loss
is higher, 12\%, but so is the background rejection, 44\%.

$D^0$ decays to two mesons in the
final state are an important source of background.
 There are four such decays (see Table
\ref{tab-2bd}). Their branching fractions are at least 
one order-of-magnitude larger than the effective branching fraction for 
$D^0 \to K^+ \ell^- \overline{\nu}_\ell$. Although  particle 
identification reduces their presence in the 
final sample, it is still important to suppress this
background, because
its \dm
distribution has a shape similar to  that of the
signal. Background exhibiting a peak around
0.145\,GeV$/c^2$ is called {\it peaking background} and reduces the
sensitivity to mixed events much more than the non-peaking
background.

\begin{table}[htbp]
\begin{center}
\caption{ Two-body decays, representing a source of a peaking
background. The symbol ``$\Rightarrow$'' represents a misidentification.}
\begin{tabular}{c c c}
\hline
\hline

decay mode	& Br [$10^{-3}$] \cite{PDG} & contribution to WS	\\
\hline

$D^0 \to K^-\pi^+$ & $38.0 \pm 0.7$	 & $K^-\Rightarrow\ell^-$,
$\pi^+\Rightarrow K^+$ \\
$D^0 \to K^+\pi^-$ & $0.143 \pm 0.004$	 & $\pi^-\Rightarrow\ell^-$ \\
$D^0 \to K^-K^+$   &   $ 3.84 \pm 0.10 $ & $K^-\Rightarrow\ell^-$ \\
$D^0 \to \pi^-\pi^+$ & $ 1.36 \pm 0.03 $ & $\pi^-\Rightarrow\ell^-$,
$\pi^+\Rightarrow K^+$ \\

\hline
\hline
\end{tabular}
\label{tab-2bd}
\end{center}
\end{table}

In the electron decay mode, the requirement on $M(K\ell)$ rejects 90\% 
of the background arising from the doubly Cabibbo suppressed decay $D^0\to
K^+\pi^-$,  through misidentification of the pion as an electron; in
the muon decay mode the suppression rate for this background is 98\%. 
In both decay modes it completely eliminates the background 
due to misidentification of both pions from  $D^0\to \pi^-\pi^+$.

The Cabibbo-favoured decays $D^0 \to K^- \pi^+$ contribute to the WS
background if the kaon is misidentified as a lepton and the pion is
misidentified as a kaon. To suppress this type of background, the
invariant mass of the 
kaon-lepton system, $M_{\pi K}(K\ell)$, is calculated with the pion mass
assigned to the 
kaon candidate and the kaon mass assigned to the lepton candidate. If
$|M_{\pi K}(K e) - m_{D^0}| < 10\,{\rm MeV}/c^2 $ 
in the electron decay mode, and
$|M_{\pi K}(K\mu) - m_{D^0}| < 15\,{\rm MeV}/c^2 $ in the muon decay mode, the
$K-\ell$ candidate is rejected. Here $m_{D^0}$ is the mass of
the $D^0$ meson, 1.8645\,GeV/$c^2$ \cite{PDG}.
In the electron decay mode, this
requirement rejects $(64 \pm 2)\%$ of the WS background from $D^0 \to
K^- \pi^+$ decays, and $(85 \pm 3)\%$ in the muon decay mode.

To suppress the contribution from $D^0 \to K^+ K^-$ decays ($K^-$
being misidentified as a lepton),
the invariant mass of the 
kaon-lepton system, $M_{K K}(K\ell)$, is calculated with the kaon mass
assigned to both candidates. 
If
$|M_{K K}(K e) - m_{D^0}| < 10\,{\rm MeV}/c^2 $ 
in the electron decay mode, and
$|M_{K K}(K\mu) - m_{D^0}| < 15\,{\rm MeV}/c^2 $ in the muon decay mode, the
$K-\ell$ candidate is rejected. In the electron decay mode, this
requirement rejects $(70 \pm 5)\%$ of the WS background from $D^0 \to
K^- K^+$ decays, and in the muon decay mode $(89 \pm 1)\%$.

The requirements on $M_{\pi K}(K \ell)$ and $M_{K K}(K \ell)$  result in a 
signal loss of 3\% in the electron decay mode and 2\% in the muon
decay mode. The rejection of the total background in both decay modes is
similar to the signal loss. 

\subsection{Rejection of $\gamma \to e^+ e^-$}

An important source of background is due to electrons from photon
conversions:
either the electron candidate, the slow pion candidate, or both, 
may be due to $\gamma \to e^+ e^-$ tracks.

\begin{figure}[htbp]
\begin{center}
\includegraphics[width=\ssld]{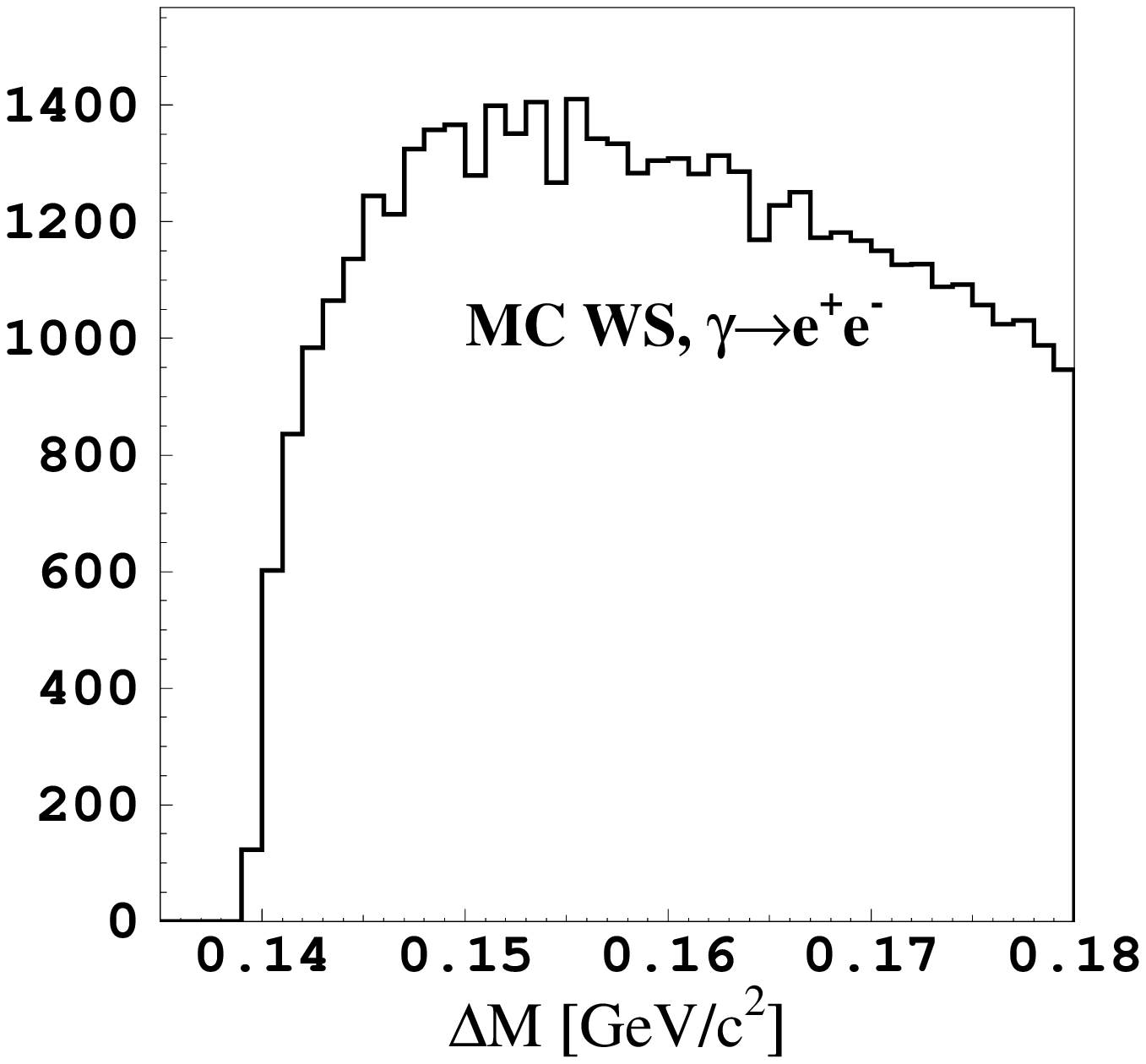}
\includegraphics[width=\ssld]{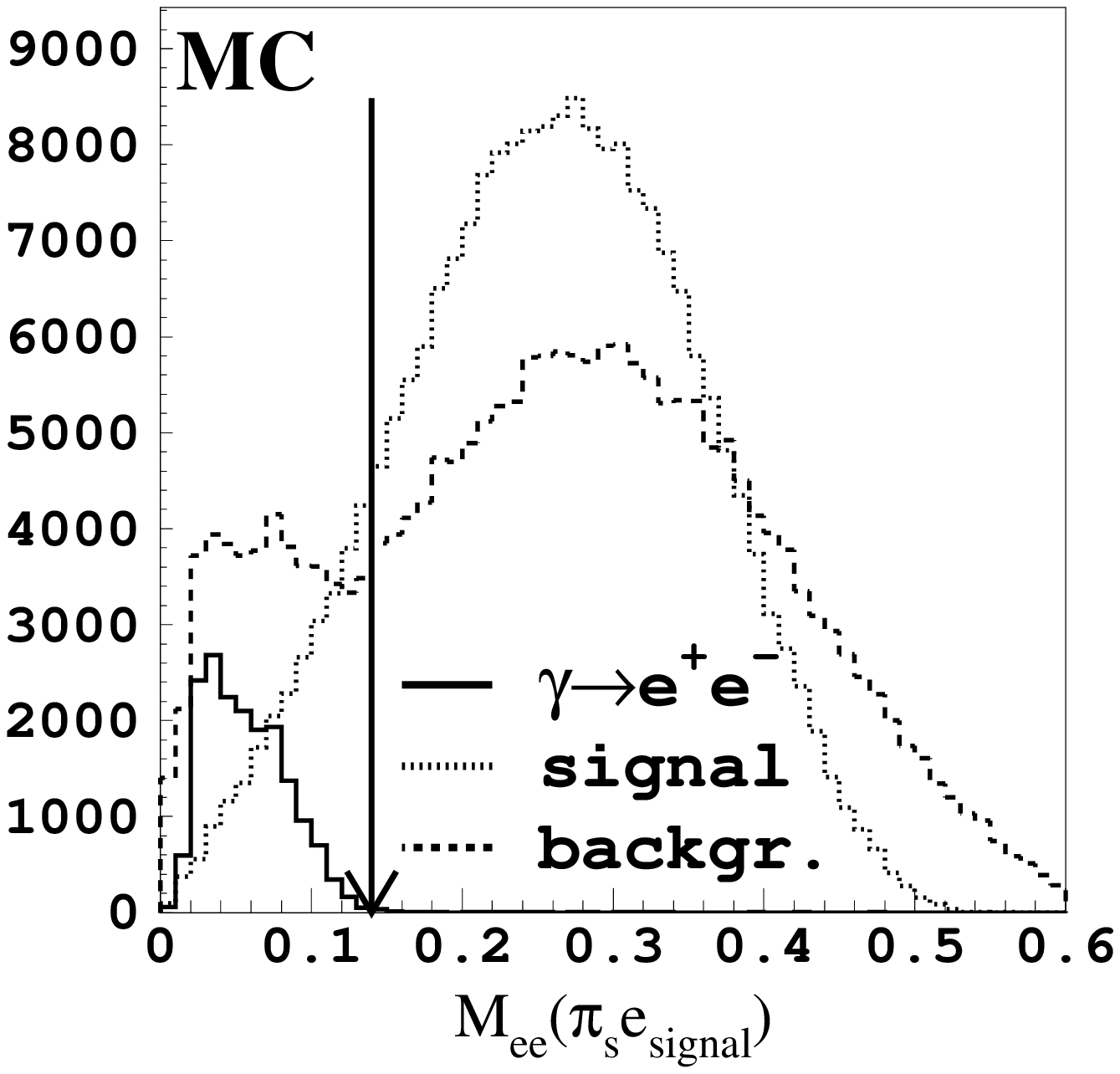}
\end{center}
\caption{
Left: the $\Delta M$ distribution of $\gamma \to e^+e^-$ events, where
one of the electrons is selected as the slow pion candidate and the
other as the electron candidate,
before applying the selection on $M_{ee}(\pi_s e_{\rm signal})$.
Right: the $M_{ee}(\pi_s e_{\rm signal})$ distribution for these events
(solid line),  for signal events (dotted line) and for the total WS
background (dashed line). The arrow shows the value of the selection
criterion.}
\label{fig-gammaee}
\end{figure}

In the electron WS sample, both the slow pion and the signal electron candidates
can come  from $\gamma \to e^+e^-$, and such events tend to have low
$\Delta M$ values (Fig.~\ref{fig-gammaee}, left). 
To suppress this background we calculate $M_{ee}(\pi_s e_{\rm signal})$, the
invariant mass of the $\pi_s-e_{\rm signal}$ system
with the electron mass assigned to both tracks.
We reject candidates with $M_{ee}(\pi_s 
e_{\rm signal}) < 0.14$\,GeV$/c^2$. The requirement rejects more than 99\% of
this background, see Fig.~\ref{fig-gammaee}, right. To retain equal
reconstruction efficiencies for the mixed and non-mixed events, this
requirement is implemented in both RS and WS samples.

Background from $\eta/\pi^0 \to \gamma \gamma \to e^+
e^- ~e^+e^-$, where one electron from the first photon and one
electron from
the other photon are taken as the electron and slow pion candidate,
exhibits similar behavior to the case where the electrons are both
from the same photon. This background is also successfully eliminated
by the above selection.

Assuming that the signal electron candidate comes from $\gamma \to e^+e^-$,
in the electron decay mode we perform a search for the other 
electron $e_2$ among all the
other tracks in the event with the opposite charge to the signal electron
candidate.
 If $M(e_{\rm signal} e_2)$, the mass of the
$e_{\rm signal}-e_2$ system, is below 80\,MeV$/c^2$, the electron
candidate is rejected. 

Assuming that the slow pion candidate 
is a misidentified electron from
$\gamma \to e^+e^-$,
we perform a search for the other electron ($e_2$) among all the
other tracks in the event.
The other electron should have the opposite charge to the slow pion
candidate, and an electron likelihood $\mathcal L_e > 0.8$ to reduce
rejection of true signal slow pions. If $M_{ee}(\pi_s
e_2)$, the mass of the
$\pi_s-e_2$ system with the electron mass assigned to both tracks,
is below 80\,MeV$/c^2$, the slow pion candidate is rejected. This photon
conversion rejection is performed for slow pion candidates in both
the electron and muon decay 
modes and results in around 0.4\% signal loss and rejects around
2.4\% of the total background.

In total, the rejection of photon conversion  in the electron decay mode
results in a 14\% signal loss and 32\% rejection of the total WS background.


\subsection{Neutrino reconstruction}
\label{ch-neutrino}

Four-momentum conservation in $e^+e^-$ collision implies 
\begin{equation}
P_\nu=P_{\rm cms}-P_{K \ell} -P_{\rm rest}
\label{eq4}
\end{equation}
for the signal decay, where $P_{\rm cms}$ stands for the cms
four-momentum of the $e^+e^-$ system and $P_{\rm rest}$ indicates the
four-momentum of all detected particles except the  charged
kaon and the lepton candidates \cite{momenta}. Eq.~(\ref{eq4}) is true if all the particles produced in
the $e^+e^-$ 
collision are detected. As the Belle detector covers nearly the entire
solid angle around the interaction point,  neutrino reconstruction can be
successfully performed.

The variable
$P_{\rm rest}$ is calculated using all the 
remaining charged
tracks (except the kaon and lepton candidates) with $dr<2$~cm
and $|dz|<5$~cm,
and photons with an energy above $70$\,MeV. Mass is assigned to a
track according to the following criteria:
\begin{itemize}

\item A track is assigned the electron mass if its electron likelihood is
$\mathcal L_e > 0.9$.

\item A track is assigned the muon mass if $\mathcal L_e < 0.9$ and
its muon likelihood is $\mathcal L_\mu > 0.9$.

\item A track is assigned the kaon mass if $\mathcal L_e < 0.9$,
$\mathcal L_\mu < 0.9$  and $\frac{\mathcal L(K)}{\mathcal
L(K)+\mathcal L(\pi)} > 0.5$. 

\item A track is assigned the proton mass if 
$\mathcal L_e < 0.9$, $\mathcal L_\mu < 0.9$, $\frac{\mathcal L(K)}{\mathcal L(K)+\mathcal
L(\pi)} < 0.5$ and $\frac{\mathcal L(p)}{\mathcal L(p)+\mathcal
L(\pi)} > 0.5$.

\item In all other cases the track is assigned the charged pion 
mass \cite{nu_reco_mass_assignment}.

\end{itemize}

\begin{figure}[h]
\begin{center}
\includegraphics[width=\ssld]{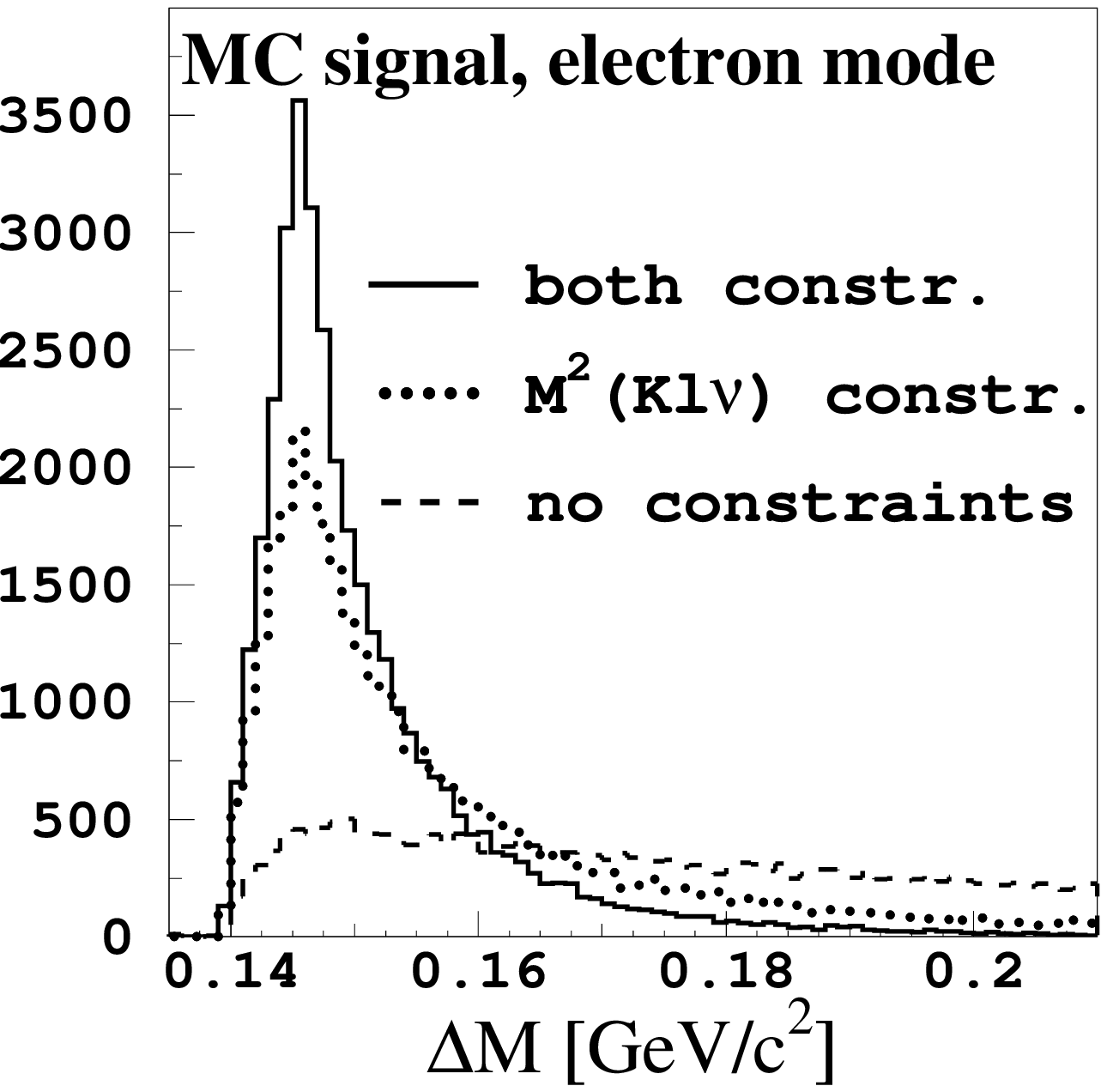} 
\includegraphics[width=\ssld]{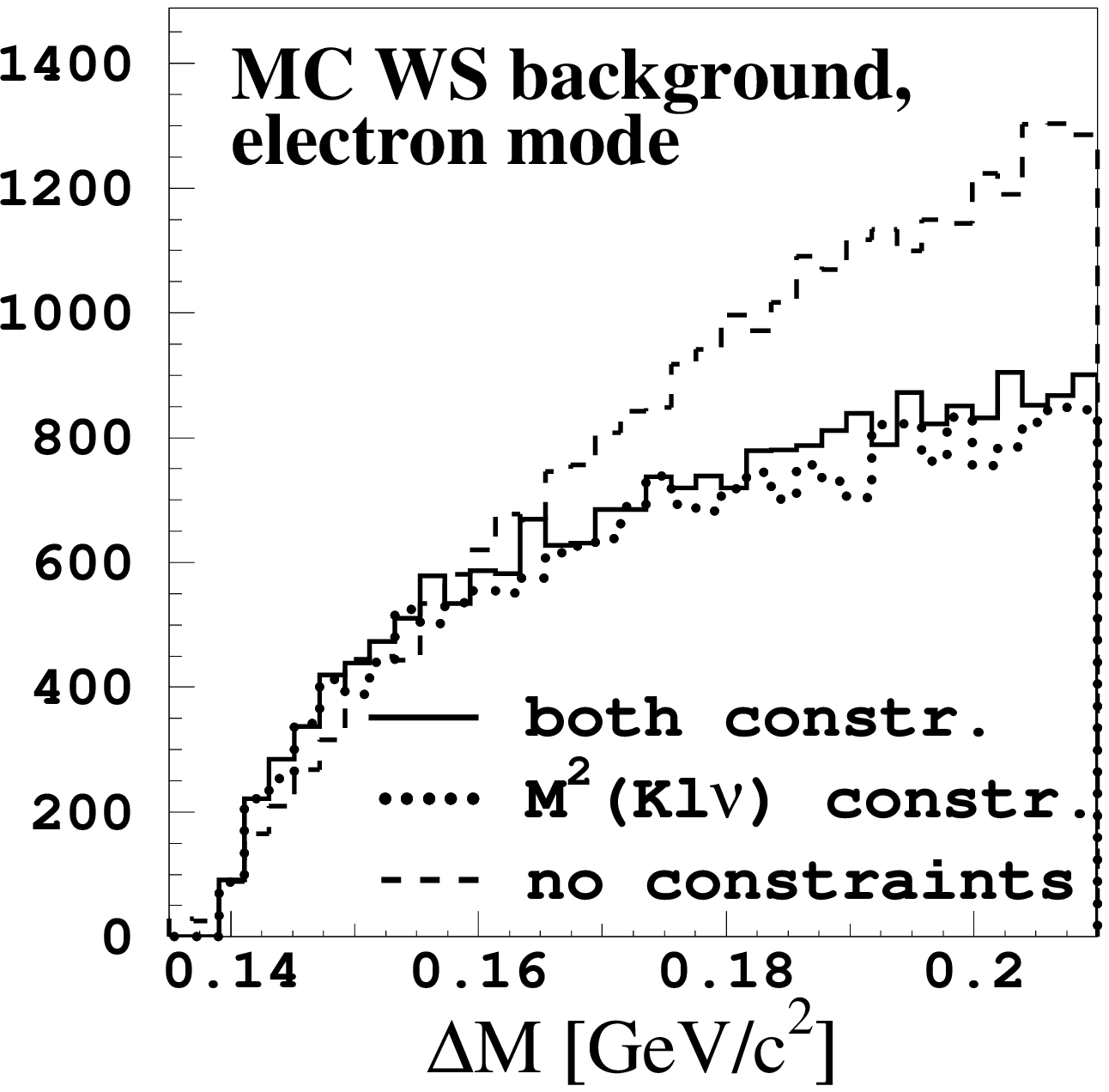}
\end{center}
\caption{ Distribution of $\Delta M$ for signal (left) and background
events (right): 
with the first approximation for the
neutrino four-momentum (dashed), after applying the
constraint on the $D^0$ 
mass (dotted) and with the final neutrino momentum, obtained
as described in the text (solid line).
Selection criteria on $M^2(K\ell\nu)$ and $M_\nu^2$ have been omitted.
The plot is for the electron decay mode; the distributions in
the muon decay mode are similar.} 
\label{fig-dmnu-s}
\end{figure}

A first
approximation for the neutrino four-momentum $P_\nu$ is obtained
using Eq.~(\ref{eq4})
and the resulting $\Delta M$ distribution for signal
events is shown in
Fig.~\ref{fig-dmnu-s} (left) with the dashed line. It peaks at around
$0.148$\,GeV/$c^2$,  a value close to the $D^{\ast +}-D^0$ mass
difference, $0.145$\,GeV/$c^2$, and has a  
FWHM of 58\,MeV/$c^2$. 

\begin{figure}[t]
\begin{center}
\includegraphics[width=\ssld]{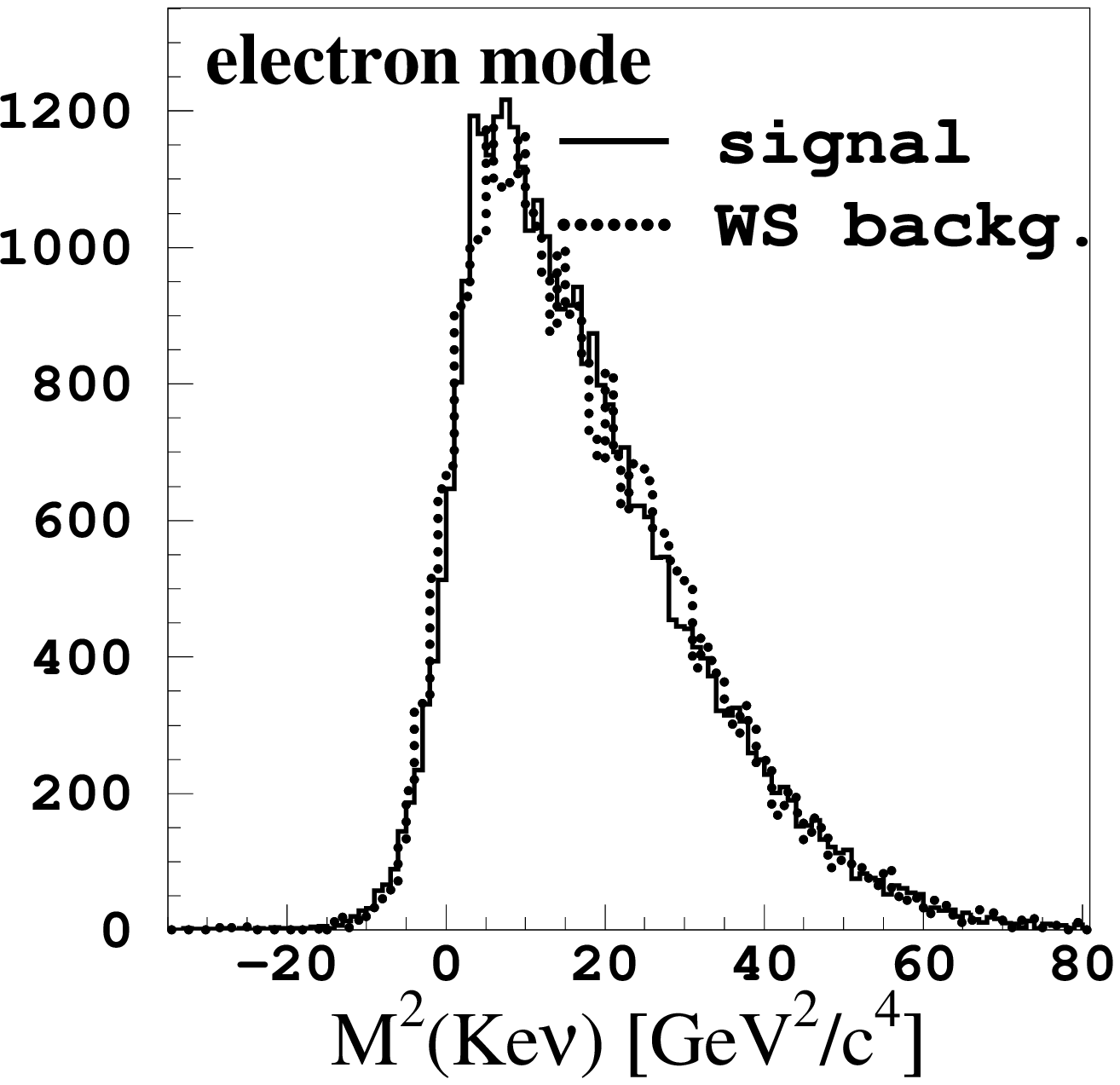}
\includegraphics[width=\ssld]{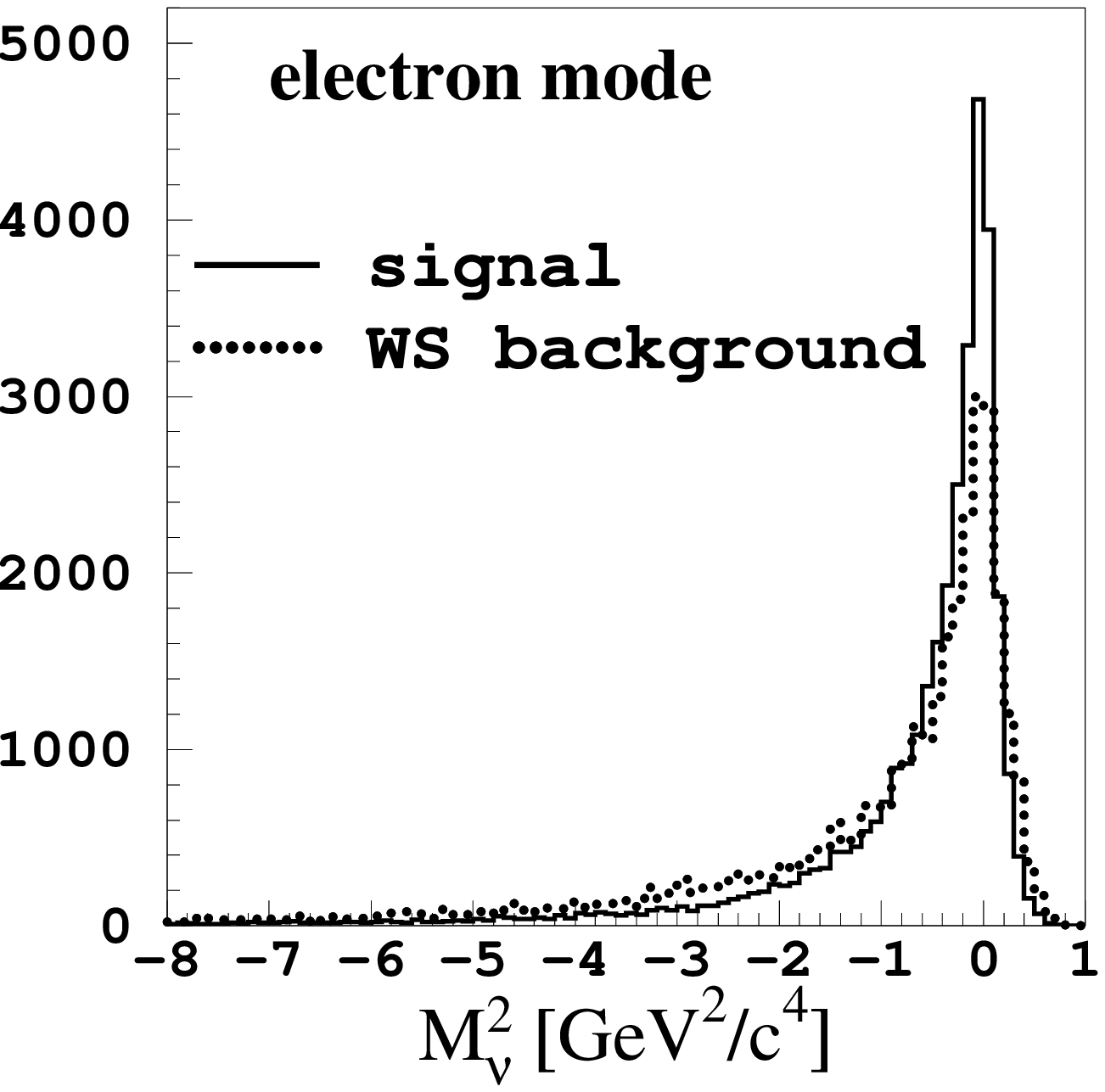}
\end{center}
\caption{ Left: distribution of $M^2(K\ell\nu)$, calculated using the first
approximation neutrino momentum. Right: distribution of $M^2_\nu$,
calculated with the neutrino 
momentum obtained after applying the constraint on $M^2(K\ell
\nu)$. The solid line is for signal and the  
dotted one for the background.
The distributions are shown for the electron decay mode; in the muon
decay mode they are similar with a slightly smaller root mean square values.}
\label{fig-md02}
\end{figure}

\begin{figure}[t]
\begin{center}
\includegraphics[width=\ssld]{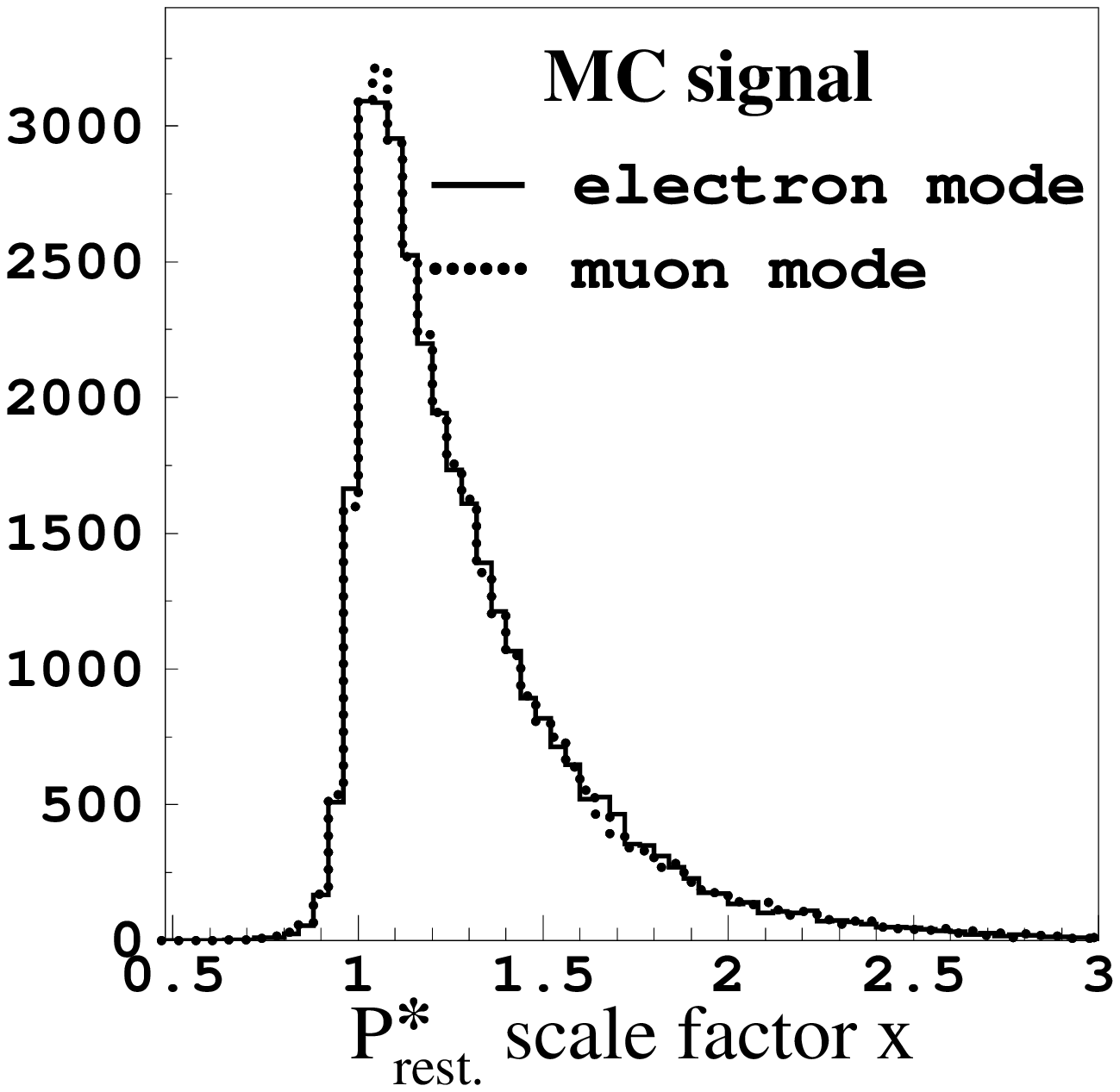} 
\includegraphics[width=\ssld]{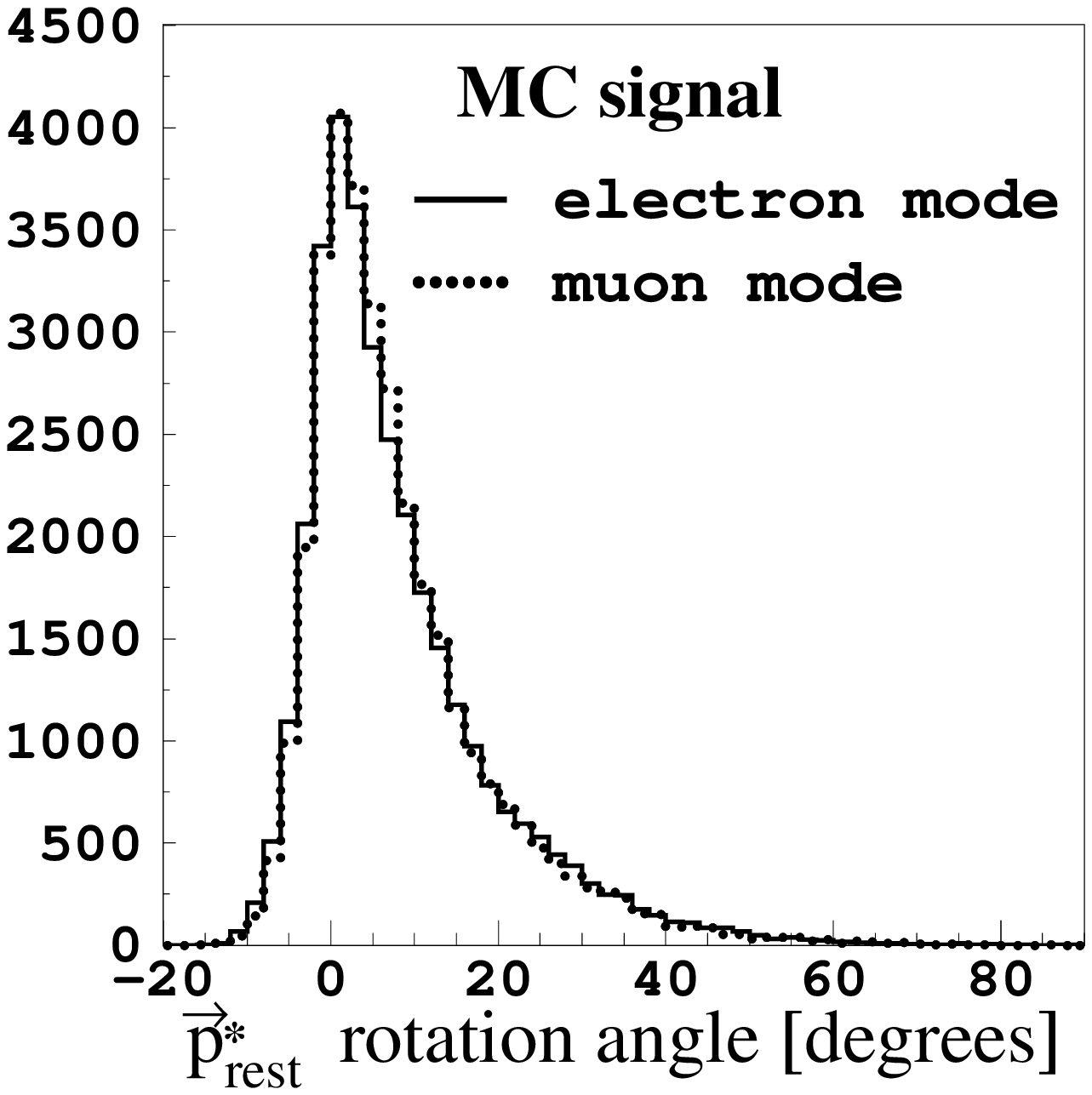}
\end{center}
\caption{Left: distribution of the scale factor $\xi$ obtained by
the constraint on $M^2(K \ell \nu)$. Right: distribution
of the rotation angle for $\vec p^*_{\rm rest}$, obtained by
the constraint on $M_\nu^2$. Both are for signal events;
the solid line shows the distributions for electron
decays and the dotted one for muon decays.} 
\label{fig-x}
\end{figure}

Two kinematic constraints are used to improve the resolution on the
neutrino momentum. 
To simplify the expressions, we performed the calculation in the
cms system, since $\vec p^*_{\rm cms} \equiv 0$.
First, the squared invariant
mass of the selected 
particles is calculated using $M^2(K \ell \nu)=(P^*_\nu + P^*_{ K
\ell})^2/c^2$. 
 The distribution of $M^2(K \ell \nu)$ is shown in Fig.~\ref{fig-md02}, left.
For
signal events, the invariant mass should equal $m_{D^0}$. To reject poorly reconstructed events, exhibiting a
large FWHM of the final \dm distribution, only
candidates with $-25$\,GeV$^2/c^4<M^2(K \ell \nu) < 64$\,GeV$^2/c^4$ are
retained. 
For the selected events, $P^*_{\rm rest}$ is
rescaled by a factor $\xi$ requiring
\begin{equation}
M^2(K \ell \nu)=(P^*_{\rm cms} - \xi P^*_{\rm rest})^2/c^2\equiv m^2_{D^0}~.
\label{eq5}
\end{equation}
The neutrino four-momentum is then recalculated as $P^*_\nu=P^*_{\rm
cms}-P^*_{K \ell} - \xi  P^*_{\rm rest}$, and a corrected $M(\pi_s K
\ell \nu)$ is obtained, where $M(K \ell \nu)$ has been forced to equal
$m_{D^0}$. With this correction, the \dm distribution has a
FWHM of 11\,MeV/$c^2$ in the electron decay mode and 10\,MeV/$c^2$ in the
muon decay mode; the improvement is shown in
Fig.~\ref{fig-dmnu-s} (left). 
The distribution of the scale factor $\xi$ for events in the finally
selected sample is shown in the left plot of Fig.~\ref{fig-x}. It
peaks at around 1.04 
for the electron 
decay mode and 1.06 for the muon decay mode. The average $\xi$ in both
decay modes is around  1.3.  

As a second kinematic constraint, the square of the missing
mass, $M_\nu^2$, is used. 
 The distribution of $M^2_\nu$ is shown in Fig.~\ref{fig-md02}, right.
For events satisfying
$-5$\,GeV${}^2/c^4<M_\nu^2<0.5$\,GeV${}^2/c^4$,
the
angle $\alpha$ between the direction of $\vec p^*_{\rm rest}$ and the
direction of $\vec p^*_{K \ell}$ is corrected in order to yield 
\zac
(P^*_\nu)^2 = (P^*_{\rm cms}-P^*_{K\ell}-\xi P^*_{\rm rest})^2 \equiv 0;
\kon
expressed in terms of energies and magnitudes of three-momenta this yields
\begin{widetext}
\begin{equation}
M_\nu^2c^4=(E^*_{\rm cms}-E^*_{K \ell}-\xi E^*_{\rm rest})^2-p^{*2}_{ K
\ell}c^2-\xi^2 p^{*2}_{\rm rest}c^2-2p^*_{K \ell} \xi p^*_{\rm rest}c^2\cos{\alpha}\equiv 0.
\label{eq6}
\end{equation}
\end{widetext}
The angle $\alpha$ is corrected by rotating $\vec p^*_{\rm rest}$
in the plane determined by the vectors $\vec p^*_{\rm rest}$ and
$\vec p^*_{ K \ell}$. The distribution of the correction angle
$\alpha_{\rm NEW} - 
\alpha_{\rm OLD}$ for the finally selected signal events is shown 
in the right plot of
Fig.~\ref{fig-x}. It has a peak at around $1^\circ$ and an average
value of $8^\circ$. 
The final neutrino four-momentum is calculated 
with the rescaled and rotated $P^*_{\rm rest}$, using Eq.~(\ref{eq4}).

The requirements on $M^2(K\ell\nu)$ and $M^2_\nu$ result in a signal
loss of 4.5\% in the electron decay mode and 4.1\% in the muon
decay mode while rejecting 9.7\% in of background in the electron decay mode
and 8.9\% of the background in the muon decay mode. 

The $\Delta M$ distribution obtained using the neutrino four-momentum
after the use of kinematic constraints is shown in
Fig.~\ref{fig-dmnu-s} (left) as the solid line. The resolution is
significantly improved,
with the FWHM being about 6.6\,MeV/$c^2$ for the electron decay mode and
6.2\,MeV/$c^2$ for the muon decay mode.
Using the MC-simulated background events,
it has been verified that such a neutrino reconstruction does
not induce any peaking in the background $\Delta M$ distribution. From
the right plot in 
Fig.~\ref{fig-dmnu-s} it can be seen that the number of background
events in the signal region ($\Delta M < 0.16\,{\rm GeV/}c^2$) after
applying the constraints only slightly exceeds the number of events without the
constraints (by 8.1\% in the electron decay mode and 16.6\% in the muon
decay mode).

For \dstp candidates we require $\Delta M < 0.18\,{\rm GeV}/c^2$,
which retains  97.4\% signal events in the electron decay mode
and 97.8\% in the muon decay mode.

\subsection{$D^0$ combined with both $\pi_s^+$ and $\pi_s ^-$}
\label{sec-Dppipi}

After all the above requirements are applied,
a small fraction of events contain a $D^0$ candidate that has been
combined with slow pions of opposite charges to form both RS and WS 
$D^*$ candidates.
Events in which such candidates are found are rejected. This veto
results in a signal loss of about 
2.6\%  in both electron and muon decay mode. For the former it
rejects  18\% of background events and for the latter
 about 13\%.

At this stage the efficiency for
reconstructing the signal with $\Delta M<0.18$\,GeV/$c^2$ is found
to be $(8.0\pm 0.3)\%$ for the electron decay mode and $(7.2 \pm 0.3)\%$ for
the muon decay mode. The quoted errors include a small variation of
efficiency depending on the detector conditions.\\ 

A summary of the applied selection criteria is presented in
Table \ref{tab-selcr}.
Since the non-mixed and mixed processes have the
same kinematics, the WS and RS 
efficiencies for all the described selection criteria are the same.

\vskip 1cm

\begin{table*}[htbp]
\begin{center}
\caption{ Summary of all the applied selection criteria.}
\begin{tabular}{c | c | c}
\hline
\hline
	& electron decay mode	& muon decay mode	\\
\hline
\hline
event	& \multicolumn{2}{c}{$R_2 > 0.2$}	\\
\hline
$\pi_s$	& \multicolumn{2}{c}{impact parameter: $|\delta r| < 1$~cm,
$|\delta z| < 2$~cm} \\
	& \multicolumn{2}{c}{$p_{\pi_s} < 0.6\,{\rm GeV}/c$} \\
	& \multicolumn{2}{c}{electron likelihood $<$ 0.1} \\
	& \multicolumn{2}{c}{$M_{ee}(\pi_s e_2) > 80\,{\rm MeV}/c^2$} \\
\hline
$\ell$	& $p_{e} > 0.25\,{\rm GeV}/c$ & $p_{\mu} > 0.65\,{\rm GeV}/c$
\\
	& electron likelihood $>$ 0.95	& muon likelihood $>$ 0.97	\\
	& $M(e_{\rm signal} e_2) > 80\,{\rm MeV}/c^2$	&  	\\
\hline
$K$	& $p_{K} > 0.85\,{\rm GeV}/c$ & $p_{K} > 0.6\,{\rm GeV}/c$ \\
	& \multicolumn{2}{c}{$\mathcal L(K)/\left[\mathcal L(K)+\mathcal L(\pi)\right] > 0.51$ }\\
	& \multicolumn{2}{c}{$\mathcal L(K)/\left[\mathcal L(K)+\mathcal L(p)\right] > 0.01$ }\\

\hline

$K-\ell$ & \multicolumn{2}{c}{$p^*_{K\ell} > 2.0\, {\rm GeV}/c$}	\\

	& $0.9{\rm \,GeV}/c^2<M(Ke)<1.75{\rm \,GeV}/c^2$ &
	$1.0{\rm \,GeV}/c^2<M(K\mu)<1.75{\rm \,GeV}/c^2$ \\

	& $|M_{\pi K}(K e) - m_{D^0}| < 10\, {\rm MeV}/c^2 $ &
	$|M_{\pi K}(K \mu) - m_{D^0}| < 15\, {\rm MeV}/c^2 $ 	\\

	& $|M_{K K}(K e) - m_{D^0}| < 10\, {\rm MeV}/c^2 $ &
	$|M_{K K}(K \mu) - m_{D^0}| < 15\, {\rm MeV}/c^2 $ 	\\

\hline

$\pi_s-e$ & $M_{ee}(\pi_s e_{\rm signal}) > 0.14{\rm \,GeV}/c^2 $	& / 	\\

\hline

$\nu$ recon. & \multicolumn{2}{c}{$-25$\,GeV$^2/c^4~<M(K \ell \nu)^2 < 64$\,GeV$^2/c^4$}	\\

	&
\multicolumn{2}{c}{$-5$\,GeV${}^2/c^4~<P_\nu^2<0.5$\,GeV${}^2/c^4$} \\

\hline

\dstp	& \multicolumn{2}{c}{$\Delta M < 0.18\,{\rm GeV}/c^2$} \\
	& \multicolumn{2}{c}{reject $D^0$ combined with both $\pi_s^+$ and $\pi_s^-$} \\
	& \multicolumn{2}{c}{$1.6 < t_{xy} < 9.0$} \\

\hline
\hline

\end{tabular}
\label{tab-selcr}
\end{center}
\end{table*}

\section{PROPER DECAY TIME}
\label{lifetime}

As the proper decay time distribution of WS
background events tends to have lower values than that of WS signal events,
the proper decay time of a $D^0$ meson can be used to
select possible mixed events with a higher purity. 
Since the information on the proper decay time is used only to
increase the sensitivity to WS events, modeling of the proper decay
time distribution is not detailed. 
We do not account for the fact
that the associated signal decays (defined in Sec. \ref{sec_sig}) have a
slightly different proper decay time
resolution function than the signal decays, or for the fact that the resolution
depends slightly on the true value of the proper decay time.
The effects of these two assumptions were studied carefully;
the differences between data and
the modeling functions are taken into account in the
systematic uncertainties and lead to a negligible change of the
final result.

The dimensionless proper decay time (proper decay time in units of
$\tau_{D^0} = (410.1 \pm 1.5)$ps~\cite{PDG}) is calculated from the \d
flight 
distance $l$ and its momentum $\vec p_{D^0}$:
\zac
t_{D^0} = \frac{m_{D^0} l}{\tau_{D^0}  p_{D^0}}.
\kon 
The $D^0$ momentum is calculated by summing the momenta of the
daughter particles. 
The $D^0$ flight distance is the distance between the $D^0$ production
vertex, $\vec r_{\rm prod}$, and its decay vertex, $\vec r_{\rm dec}$. The
decay vertex is obtained by fitting 
the kaon and lepton tracks to a common
vertex. The production vertex is 
obtained by extrapolating the 
$D^0$ momentum vector to the $e^+e^-$
interaction region.
The position and width of this region are determined over a 
large number of $e^+e^-$ interactions for which the KEKB beam conditions
do not change significantly.

According to MC simulation, the resolution on the proper decay time
is improved if the \d flight distance is calculated as the
projection  of the $\vec r_{\rm dec} - \vec r_{\rm prod}$  vector 
 on the normalized \d momentum vector. Since the interaction region is much
narrower in the radial direction, we use only the radial components
($x$ and $y$) to measure the proper decay time.
 The radial flight distance $l_{xy}$ is calculated as
\zac
l_{xy} = \frac{( r^x_{\rm dec}-r^x_{\rm prod},r^y_{\rm dec}-r^y_{\rm prod} ) \cdot (p_{D^0}^x,p_{D^0}^y)}{\sqrt{(p_{D^0}^x)^2+(p_{D^0}^y)^2}}.
\kon
The proper decay time is then evaluated  as
\zac
t_{xy} = \frac{m_{D^0} l_{xy}}{\tau_{D^0} \sqrt{(p_{D^0}^x)^2+(p_{D^0}^y)^2 }}.
\kon
The observed $t_{xy}$ distribution is smeared due to
the experimental resolution. As the data recorded with the SVD-2
configuration has a slightly narrower resolution function 
than the data taken with the SVD-1 configuration,
we perform the measurements of $R_M$
separately for both subsamples. Thus we have four subsamples: the
electron subsamples are denoted by e-1 and e-2 for the SVD-1 and
SVD-2 configurations, respectively. Similarly the muon subsamples will be
denoted as $\mu$-1 and $\mu$-2.

\subsection{Distribution of signal events}

We obtain the resolution function for signal events from the data,
using the RS decays. To be able to do so, we first determine
the shape of the $t_{xy}$ distribution for RS background events, 
which is shown in 
Fig.~\ref{fig-effRSe}, right, as the dashed line. This
distribution  is also obtained from the data, as described in the 
following.

The normalized function that describes the $t_{xy}$
distribution of RS background events is
\begin{widetext}
\zac
\mathcal F^{\rm RS}_{\rm bkg} = \left(1-f^b_w\right) \cdot  \left\{\left[f^b_e E(t;\tau^b) + (1-f^b_e)
\delta(t)\right]\otimes\left[f^b_1 L(t;b^b) + (1-f^b_1)
L_a(t;b^b_l,b^b_r)\right]\right\}+f^b_w \cdot L(t;b^b_w).
\label{eq-bkg}
\kon
\end{widetext}
It is composed as a sum of an
exponential function $E(t;\tau^b)$ with the decay time $\tau^b$ and a
delta function (their fractions are determined by the parameter
$f_e^b$),
convolved with a detector
resolution. The latter is phenomenologically
described by a sum of the Lorentz function $L(t;b^b)$ and an 
asymmetric Lorentz
function $L_a(t;b^b_l,b^b_r)$  (both 
are explicitly given in the Appendix, Eq.~(\ref{eq-L}) 
and (\ref{eq-La}));
$b^b$, $b^b_l$ and $b^b_r$ are their width parameters and $f_1^b$
determines their fractions in the resolution function.
A wide Lorentz function is added to describe the decay
times measured from badly reconstructed tracks (outliers); its width
parameter is $b^b_w$ and its fraction in the total sample $f^b_w$. 
The convolutions are
performed numerically by substituting the integral with a sum. 
It has been
verified that the numerical accuracy is satisfactory, {\it i.e.} not
affecting the result.

To determine the eight free parameters of the function from the data, 
we divide the $t_{xy}$ range into 15 intervals as shown in 
Fig.~\ref{fig-effRSe}, left. In each of these 15 intervals we extract
the number of RS background events, $N^{i}_{\rm RS,bkg}$,
 by performing fits to the \dm distribution of RS events. The fit
to the \dm distribution is described in detail in Sec.~\ref{sec-dm-fit}.
The errors include the systematic error due to the finite 
number of MC simulated events
and the uncertainty of the correlated background fraction in the total RS
background.

If we divide $N^{i}_{\rm RS,bkg}$ in one of the 15
$t_{xy}$ intervals by the total number of background events 
$N^{\rm tot}_{\rm RS,bkg}$ (the sum
over the 15 intervals), the obtained fraction is expected to agree
with  the integral of 
$\mathcal F^{\rm RS}_{\rm bkg}$ (Eq.~\ref{eq-bkg}) over that $t_{xy}$ interval.

Hence we calculate
$N^{i}_{\rm RS,bkg} / N^{\rm tot}_{\rm RS,bkg}$
in all 15 $t_{xy}$ intervals
and determine  the eight free parameters of 
$\mathcal F^{\rm RS}_{\rm bkg}$  by a $\chi^2$ fit
 to these fractions. The value of the fitting function in each 
$t_{xy}$ interval is calculated as
the integral of $\mathcal F^{\rm RS}_{\rm bkg}$ over that $t_{xy}$
interval. The fractions and the result of the fit for the e-2 subsample
are shown in Fig.~\ref{fig-effRSe}, left plot; the obtained 
background distribution is
shown with the dashed line in the right plot.
The reduced $\chi^2$ values of the fits are
reasonable, ranging from 0.4 to 4 for 7 degrees of freedom.
 The uncertainties on the fitted parameters of the RS
background proper decay time distribution are taken into account when
calculating the uncertainty of the result, $R_M$.

\subsubsection{RS signal distribution}
\label{sec-RSsig-txy}

The proper decay time distribution for RS events is described by 
\zac
\mathcal F^{\rm RS} = \mathcal N_{\rm tot} (f_s \cdot \mathcal F_{\rm sig}^{\rm RS} +
(1-f_s) \cdot\mathcal F_{\rm bkg}^{\rm RS}),
\label{eq-a}
\kon
where $\mathcal N_{\rm tot}$ is the total number of RS events and $f_s$ is
the signal fraction, obtained from a fit to $\Delta M$ in the entire 
proper decay time region. The $f_s$ values are 
$(69.9 \pm 0.2)\%$,
$(70.8 \pm 0.1)\%$,
$(62.7 \pm 0.2)\%$ and
$(62.6 \pm 0.1)\%$ for e-1, e-2, $\mu$-1 and $\mu$-2 subsamples,
respectively. 
The function $\mathcal F_{\rm sig}^{\rm RS}$  
describes the shape of the RS signal events,
which is an exponential convolved with the resolution function,

\zac
\mathcal F^{\rm RS}_{\rm sig} = E(t;\tau^s ) \otimes \mathcal R_{\rm sig}.
\label{eq-rssigtxy}
\kon
The resolution function is parameterized as $ \mathcal R_{\rm sig} =  $
\zac
 f^s_1 L_a(t;b^s_l,b^s_r) + f^s_2
G(t;\sigma) + (1-f^s_1-f^s_2) L(t;b^s_w).
\kon

\begin{figure}[htbp]
\begin{center}
\includegraphics[width=\ssld]{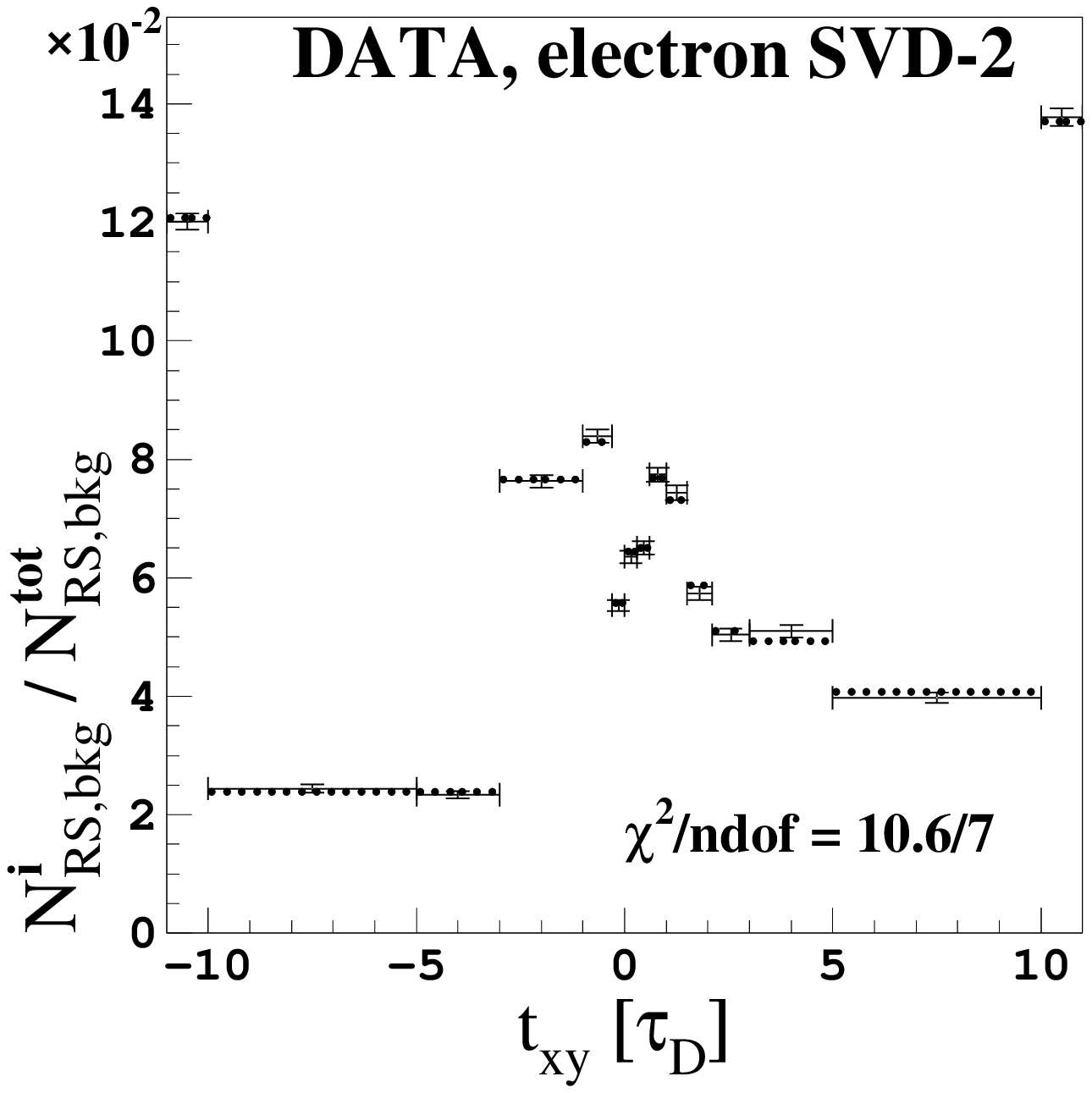}
\includegraphics[width=\ssld]{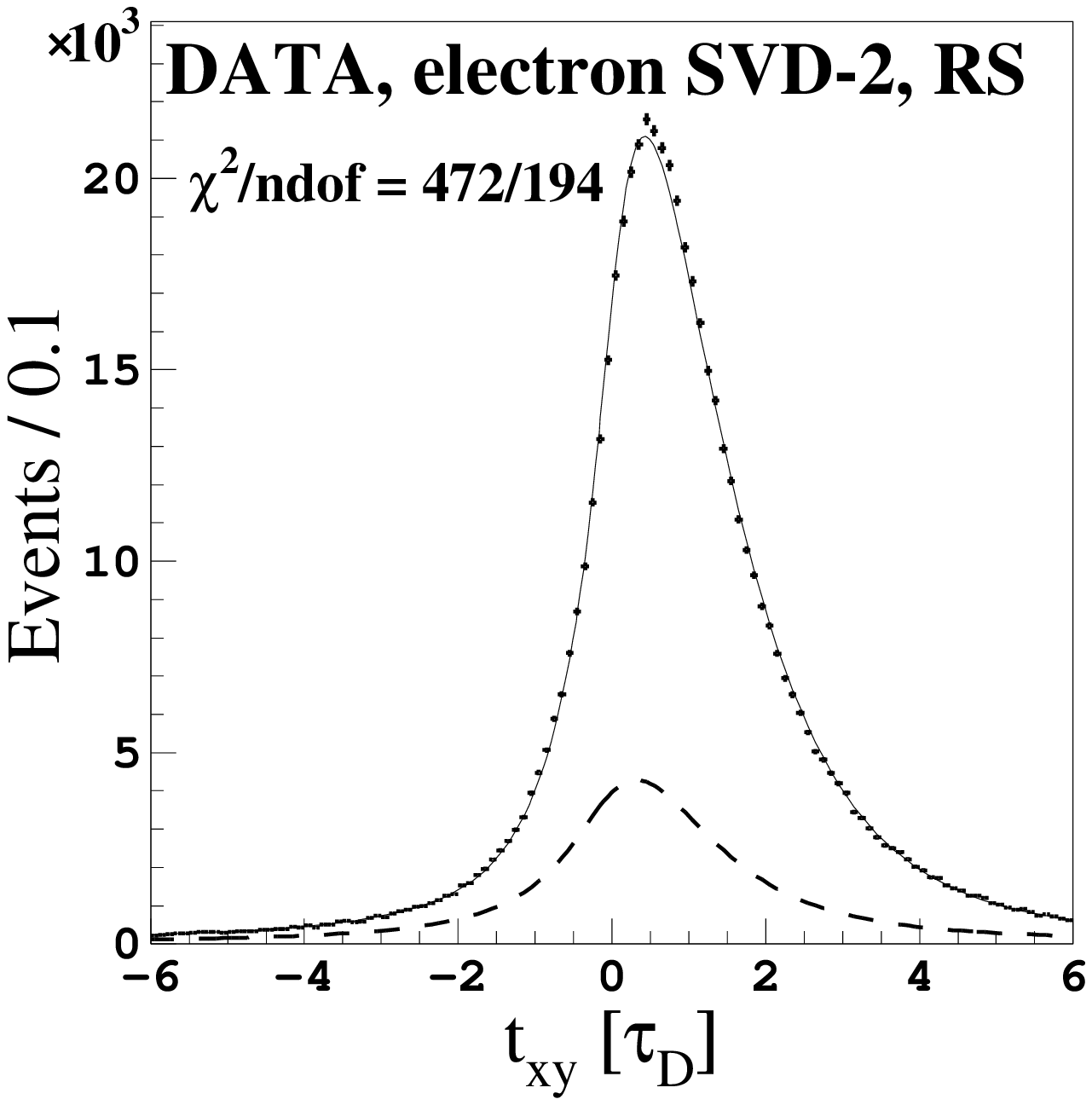}
\end{center}
\caption{Left: the fraction of RS background events in 15 proper decay
time intervals (error bars) and the result of the fit,
described in the text (dotted lines). Note that the $t_{xy}$ intervals are
not equidistant, which causes the apparently non-smooth shape of the function.
The first and the last interval include
events with $t_{xy} < -10$ and $t_{xy} > 10$, respectively.
Right: the $t_{xy}$ distribution of RS events (error
bars) with the result of the fit (solid line) for the e-2 subsample. The dashed
line shows the RS background distribution, $\mathcal F^{\rm RS}_{\rm bkg}$, obtained
from the data as explained in the text.}
\label{fig-effRSe}
\end{figure}

Here $G(t;\sigma)$ is the Gaussian function.
The six free parameters of the resolution function for the RS signal
events are $f_1^s$, $f_2^s$, $b^s_l$, $b^s_r$, $\sigma$
and $b^s_w$. They 
are obtained by a $\chi^2$ fit of $\mathcal F^{\rm RS}$ to 
the proper decay time distribution of RS events. In this fit, the 
parameters of $\mathcal F_{\rm bkg}^{\rm RS}$ are fixed to the values previously
obtained and $\tau^s$, the dimensionless $D^0$ decay time, is fixed to
1.0. An example of a fit result is shown in the right plot of
Fig.~\ref{fig-effRSe}. 
The reduced $\chi^2$ values range between 1.2 and 2.4 for 194 degrees
of freedom and exhibit a slight
disagreement between the fitting model and the data. The
disagreement is accounted for in the systematic error evaluation,
resulting in a negligible change of the final result.

\subsubsection{WS signal distribution}
\label{sec-WSsig-txy}

Since decays of the mixed and of the unmixed \d mesons have the same
kinematic properties, the proper decay time resolution function for
both is assumed to be the same. 
Hence from the RS signal resolution function, $\mathcal R_{\rm sig}$,
 the proper decay time
distribution for WS signal events is calculated: 
\zac
\mathcal F^{\rm WS}_{\rm sig} = \mathcal A t^2 e^{-t/\tau} \otimes \mathcal
R_{\rm sig},
\label{eq-ws-sig}
\kon
where $\mathcal A$ is the corresponding normalization constant and
$\tau$ is fixed to 1.

\begin{figure}[htbp]
\begin{center}
\includegraphics[width=\ssld]{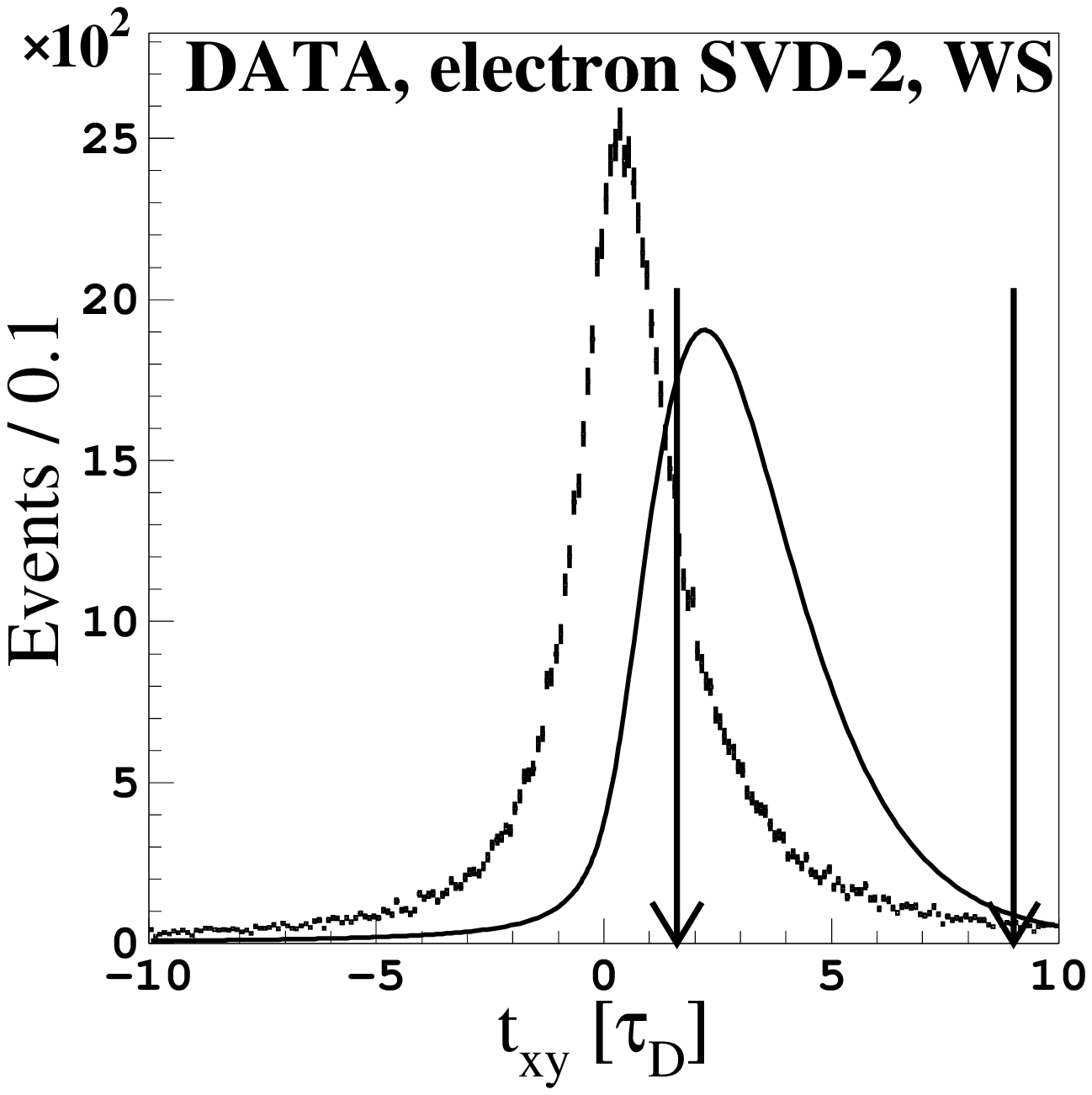}
\includegraphics[width=\ssld]{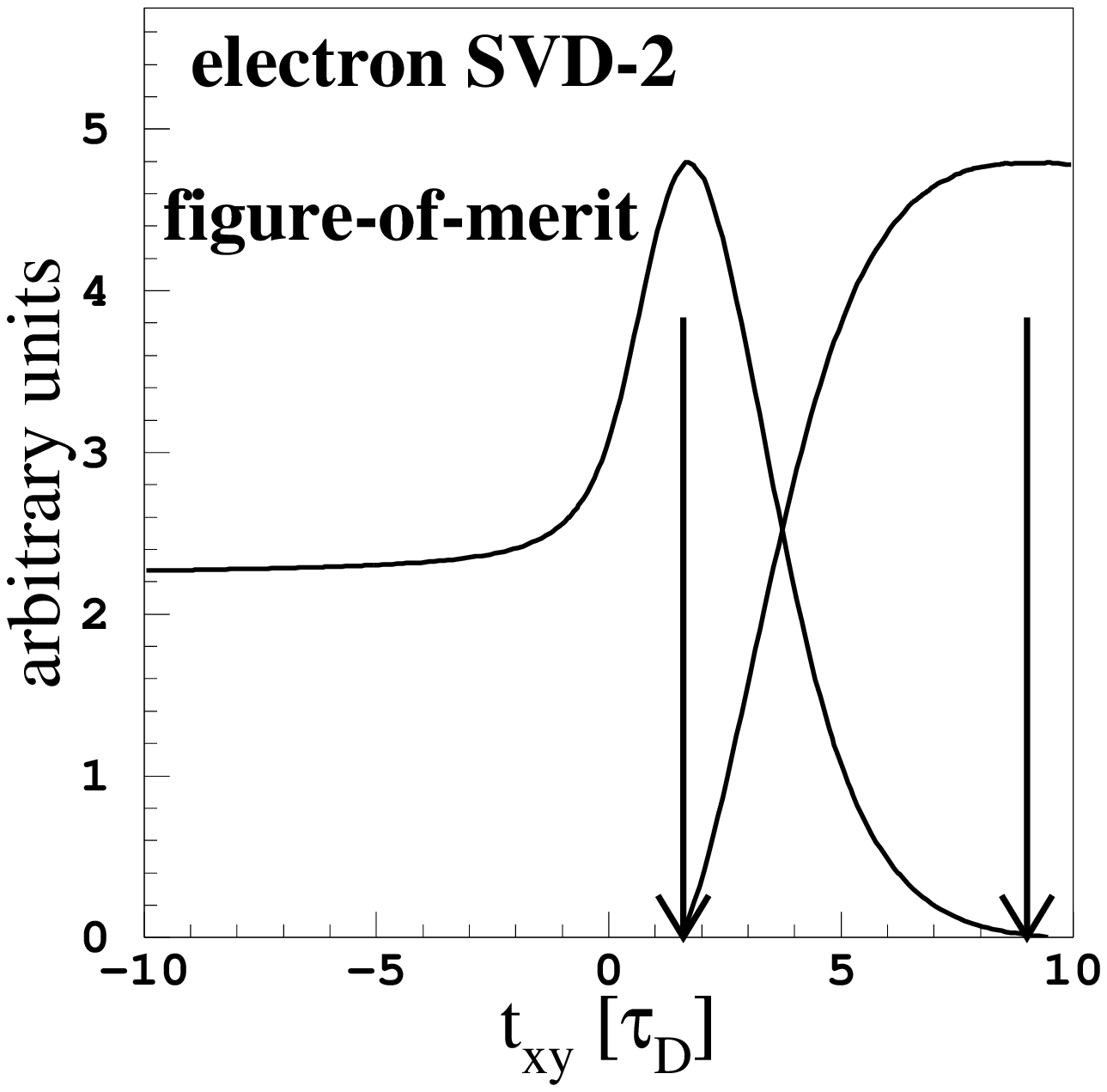}
\end{center}
\caption{Left: the proper decay time distribution of
the WS data events (error bars) and the 
distribution for
the WS signal events (solid line), obtained from the data according to
Eq.~(\ref{eq-ws-sig}).
The distributions are for the e-2 subsample. Right: the figure-of-merit 
dependence on the lower and upper limit of the selected $t_{xy}$
interval. When plotting the dependence on the lower limit, the upper
limit was set at its optimal value and vice versa.
The arrows show the value of
the selection criterion, Eq.~(\ref{t-int}).
Distributions for other three subsamples are similar.}
\label{fig-txyopt-e}
\end{figure}

To select the $t_{xy}$ interval with the highest sensitivity to mixed events,
the ratio given in Eq.~(\ref{eq-sens}) is maximized. In this optimization we
use the calculated distribution for WS signal events, $\mathcal F^{\rm WS}_{\rm sig}$
(Eq.~(\ref{eq-ws-sig})); for the background we use
the 
distribution of the WS events from the data. Even if the latter contains some
mixed events, the effect on the result of the optimization is
negligible, since $D^0$ mixing is small.
The distributions of background events, WS signal events and the
figure-of-merit can be seen in Fig.~(\ref{fig-txyopt-e}). 
The optimal proper decay time intervals for the four subsamples range
between 1.6--1.7  and 8.9--9.5. 
In order to keep the measurement method uniform, we 
select a common interval for all four subsamples,
\zac
1.6 < t_{xy} < 9.0.
\label{t-int}
\kon
In this interval, about 70\% of WS signal events are
selected, while rejecting about 80\% of background events (the values
are similar in each of the four subsamples).

\subsection{Extraction of $R_M$ and further improvement}
\label{sec-txy-c}

Because the proper decay time distribution of the RS signal events
($e^{-t/\tau_D}$) is different from that of the WS signal events 
($t^2 e^{-t/\tau_D}$),
after the application of a selection based on proper decay time, the
ratio $R_M$ is obtained as
\zac
R_M = 
\frac{N_{\rm WS}}{N_{\rm RS}} =
\frac{N^i_{\rm WS}}{N^i_{\rm RS}}\frac{\epsilon^i_{\rm
RS}}{\epsilon^i_{\rm WS}},
\label{eq-rm}
\kon
where $N_{\rm WS,RS}$ are the numbers of extracted signal events without
the $t_{xy}$ selection, and $N^i_{\rm WS,RS}$ are the
numbers of extracted signal events in the selected $t_{xy}$ interval.
The superscript $i$ labels different $t_{xy}$ intervals.
The efficiencies $\epsilon^i_{\rm RS}$ are obtained by integrating
the proper decay time distribution of the RS signal events, $\mathcal
F^{\rm RS}_{\rm sig}$, over the selected $t_{xy}$ interval.
Similarly, the efficiencies $\epsilon^i_{\rm WS}$ are obtained from
the calculated proper decay time distribution of the WS signal events,
$\mathcal F^{\rm WS}_{\rm sig}$. 
The ratios $\epsilon^i_{\rm RS}/\epsilon^i_{\rm WS}$ are listed in
Table \ref{tab-mu2}.
The errors on the efficiencies quoted there include the uncertainty on
the fraction of the RS correlated background (as defined in
Sec.~\ref{sec-bkg}), the statistical and systematic
uncertainty on the signal 
fraction in the RS sample, the statistical and  systematic
uncertainties in the parameters of the RS background and signal $t_{xy}$
distributions and
the uncertainties in the world averages of $\tau_{D^0}$
and $m_{D^0}$ \cite{PDG}.
The resulting errors on the measured parameter $R_M$ 
are included in the systematic uncertainty and are
  negligible. Imperfections in modelling the decay time
distributions are included as a separate source of a systematic
uncertainty as described below.

To further exploit the proper decay time information,
we divide the chosen $t_{xy}$
range (Eq.~(\ref{t-int})) into six intervals,
with boundaries at 1.6, 2.0, 2.5, 3.1, 4.0, 5.6, and 9.0.
The binning is chosen so as to have approximately the same number
of events in each interval. The mixing rate is measured in
each of the six intervals and the measurements are expected to be
consistent. Due to the additional proper
decay time information
the sensitivity of the final result is expected to be
improved in comparison with the sensitivity of a single measurement in
the total $1.6 < t_{xy} < 9.0$ range.

\section{SIGNAL YIELD EXTRACTION}
\label{sec_sig}

According to the MC simulation, the selected sample of RS events
includes 
many candidates from semileptonic decays other than 
$D^0 \to K^- \ell^+ \nu$, combined with the correctly reconstructed
slow pion. The most important of these decays are:
\begin{itemize}
\item $D^0 \to K^- \pi^0 \ell^+ \nu$,
\item $D^0 \to K^{*-} \ell^+
\nu_\ell$, followed by $K^{*-} \to K^- \pi^0$,
\item $D^0 \to \pi^- \ell^+ \nu_\ell$,
\item $D^0 \to \rho^- \ell^+
\nu_\ell$, followed by $\rho^- \to \pi^- \pi^0$,
\item $D^0 \to K^{*-}
\ell^+ \nu_\ell$, followed by $K^{*-} \to \overline{K}{}^0 \pi^-$.
\end{itemize}
The final state lepton in these decays is of the same charge as in
the decay $D^0 \to K^- \ell^+ \nu$, so its charge can be 
used to tag mixing in the same way. In the last three decays with a
$\pi^-$ instead of the $K^-$ in the 
final state, the pion is misidentified as a kaon.
The selected sample also includes candidates where the \d semileptonic
decay is correctly reconstructed, but the slow pion
decays in flight to a muon, $\pi_s^+ \to \mu^+ \nu_\mu$, and then the muon is
misidentified as the slow pion. The muon has the same charge as the
slow pion and can be used to tag the \d flavor at production in the
same way as the slow pion. Hence all these processes are treated as 
part of the signal and will be referred to as {\it associated signal} decays.

\begin{figure}[htbp]
\begin{center}
\includegraphics[width=\ssld]{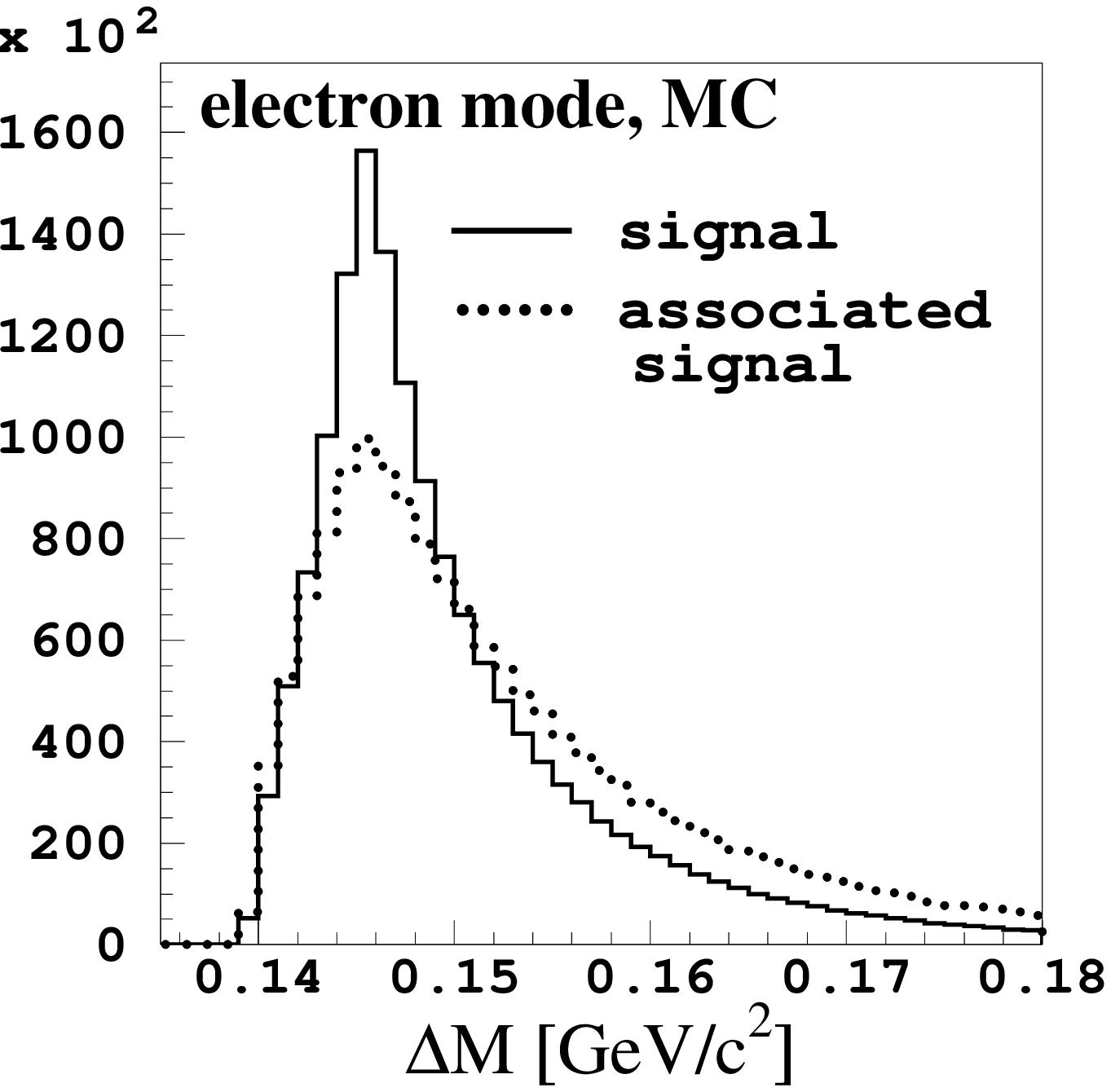}
\includegraphics[width=\ssld]{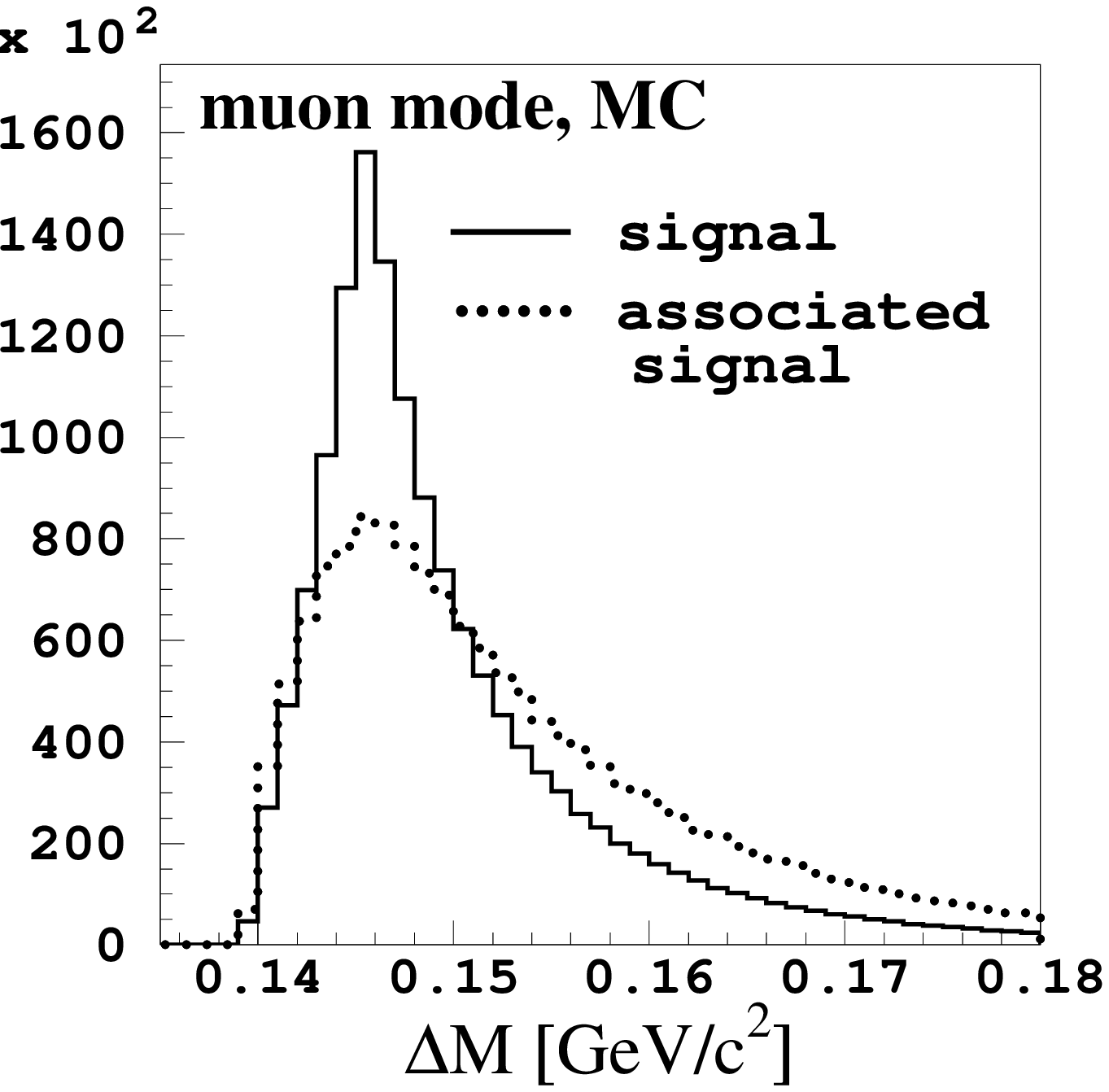}
\end{center}
\caption{The simulated $\Delta M$ distribution for the signal decays
(solid line) and for the 
associated signal decays (dotted line), for both the (left) electron and
(right) muon decay modes.}
\label{fig_asssig}
\end{figure}

According to MC simulation,
the associated signal decays have a similar \dm distribution to the
$D^0 \to K^- \ell^+ \nu$ decays.
Due to unreconstructed or misidentified particles in these final
states, the FWHM of the distribution is larger, 12.3~MeV$/c^2$ in the
electron decay mode and 
14.3~MeV$/c^2$ in the muon decay mode (see Fig.~\ref{fig_asssig}).
For the same reasons, the proper decay time distribution
is slightly different from that for the signal.
The fraction of the associated signal decays in
the sample of all reconstructed signal decays can be found in
Table~\ref{tab-asssig}.
 There is also a
small fraction of signal events (around 1\%)  from
$B\overline{B}$ events. Due to the lower average momentum of $D$ mesons
from $B$ decays, the $\Delta M$ distribution for signal
events from $B$ decays is slightly wider than that for signal events
from $c \bar c$ events; it is similar to the $\Delta M$ distribution
of the associated signal decays.

\begin{table}[htbp]
\begin{center}
\caption{The associated signal fraction and the
fraction of signal from $B$ decays in the total signal, in [\%],
as obtained from MC simulation. The
quoted uncertainties are statistical only. The fractions are shown for the
entire proper decay time interval and for $1.6 < t_{xy} < 9.0$.}
\begin{tabular}{ c | r@{$\pm$}l r@{$\pm$}l | r@{$\pm$}l r@{$\pm$}l}
\hline
\hline
	& \multicolumn{4}{c|}{assoc. sig. [\%]}		&
	\multicolumn{4}{c}{sig. from $B$ decays [\%]}		\\
	& \multicolumn{2}{c}{all $t_{xy}$}	& \multicolumn{2}{c|}{$1.6-9.0 $}	&
	  \multicolumn{2}{c}{all $t_{xy}$}	& \multicolumn{2}{c }{$1.6-9.0 $}	\\
\hline
$e$	& 16.58 & 0.05	& 17.7 & 0.1	& 1.20 & 0.01 & 1.18 & 0.03	\\
$\mu$	& 11.54 & 0.04	& 12.3 & 0.1	& 1.07 & 0.01 & 1.04 & 0.03	\\

\hline
\hline
\end{tabular}
\label{tab-asssig}
\end{center}
\end{table}

\subsection{Background}
\label{sec-bkg}

\begin{figure}[htbp]
\begin{center}
\includegraphics[width=\ssld]{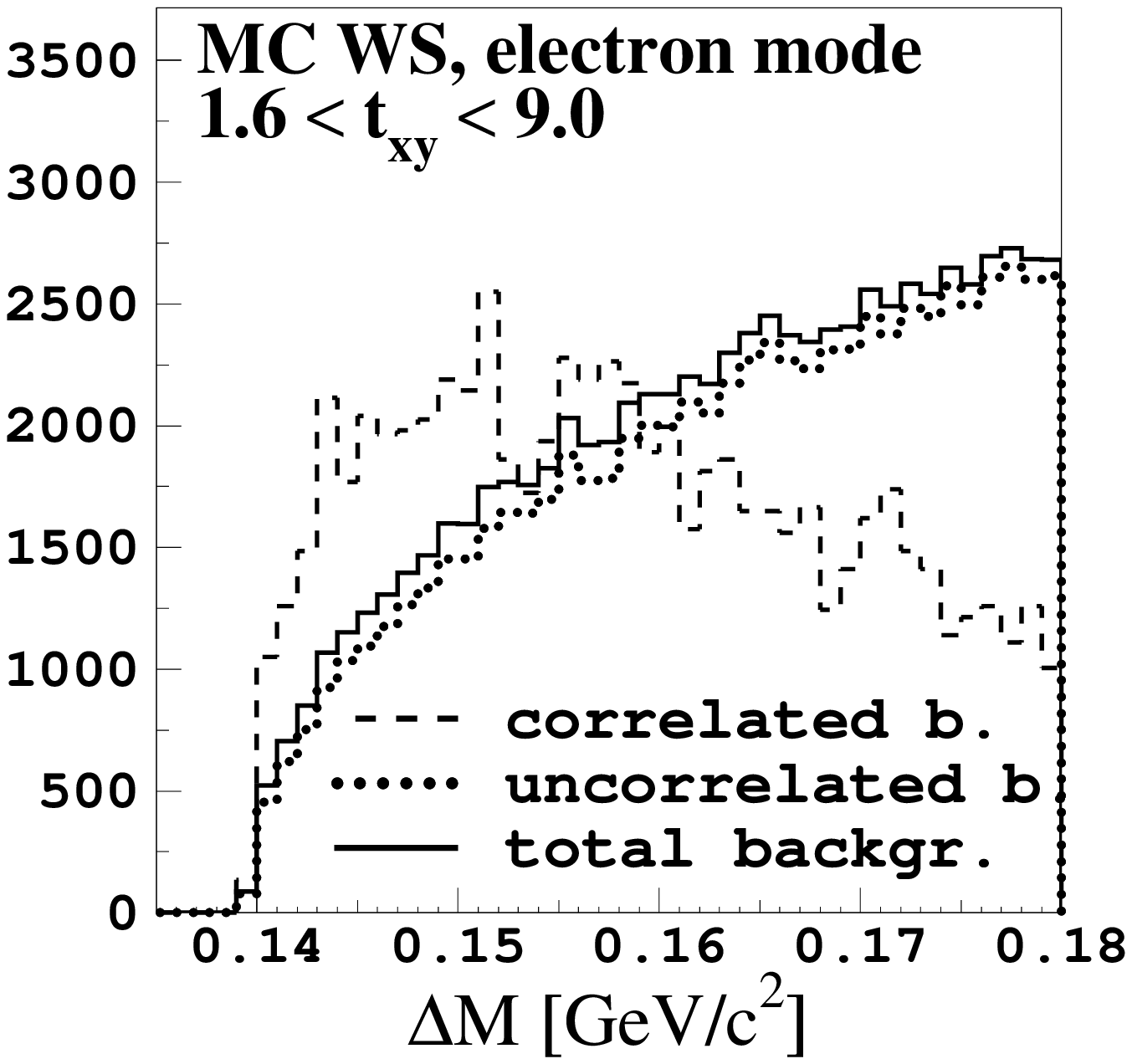}
\includegraphics[width=\ssld]{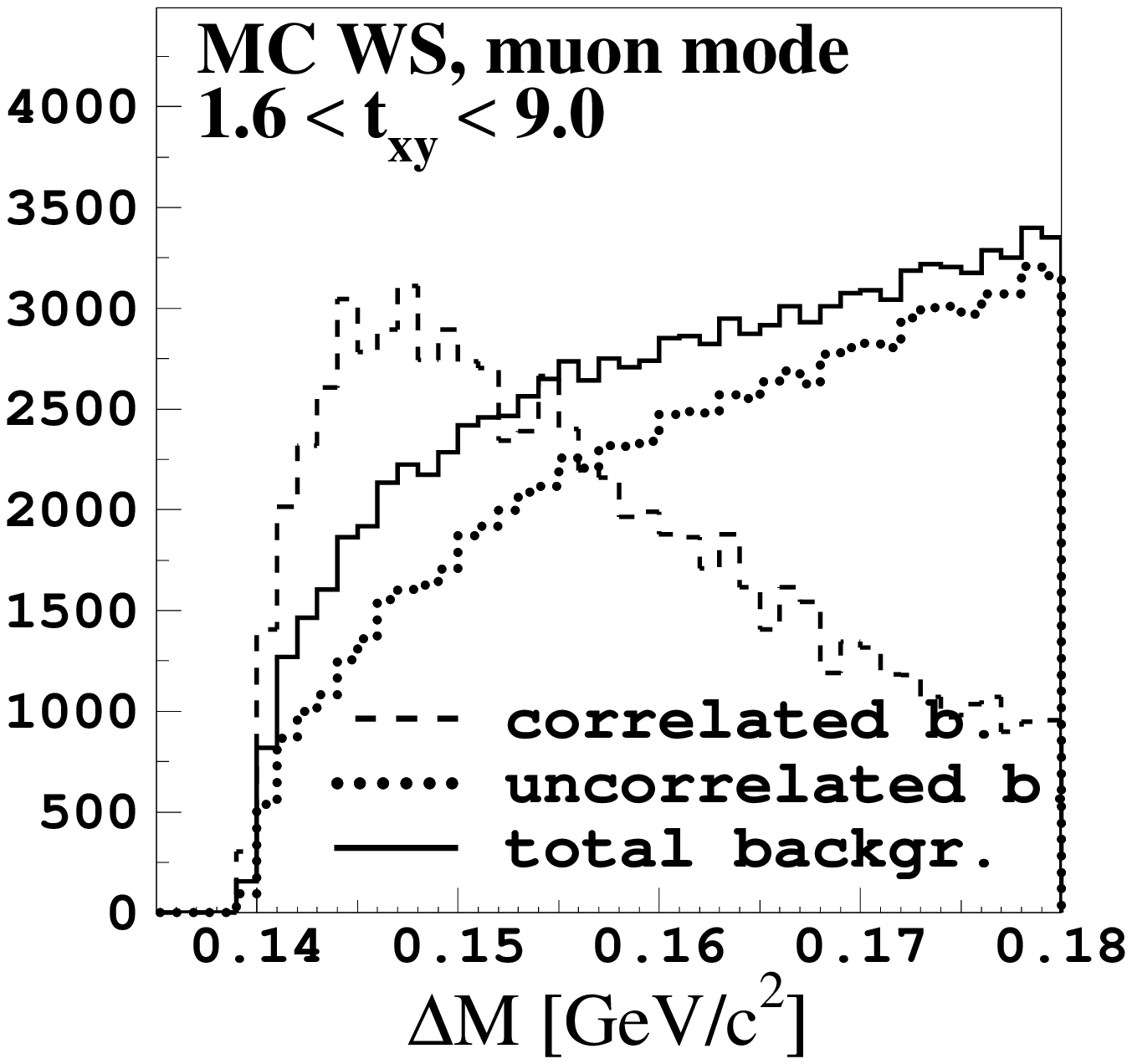}
\end{center}
\caption{The simulated $\Delta M$ distribution for WS background
events. The solid line shows the total background, the dotted line
shows the uncorrelated background and the dashed 
one shows the correlated background, multiplied by a factor of 15 and 5
in the electron and muon samples, respectively. The plots are for the selected
proper decay time interval. The left plot is for the electron decay mode
and the right one for the muon decay mode.
}
\label{fig-WSbkg}
\end{figure}

The background is divided into two categories: the correlated
background and the uncorrelated background. The \dm shapes of both
background components can be seen in Fig.~\ref{fig-WSbkg}.

Correlated background is the background 
where the lepton candidate or the kaon candidate,
or both, originate from the same decay
chain as the slow pion candidate. 
The angular correlation
between the slow pion and \d candidates leads to a
concentration of events at low values of $\Delta M$.

The remaining, uncorrelated background has a $\Delta M$
distribution that rises steadily from threshold, as the
available phase space increases. This component is
dominant, especially in the WS sample (see Table
\ref{tab-corb}). The fraction in
the muon decay mode is larger than 
in the electron decay mode, because the probability for a kaon or a pion to
be misidentified as a muon is larger than the probability to be
misidentified as an electron.
The fraction is larger in the RS sample than in the WS
sample due to the larger branching fractions of the Cabibbo favored
decays. 
Selecting the proper decay time interval $1.6 < t_{xy} <
9.0$, decreases the fraction of the correlated background in the total
background.
According to MC simulation,
the correlated background 
has three components: background from 
$D^{*+} \to \pi^+_s D^0$ decays, which has the largest fraction and is
the most strongly peaked of the three,
background from 
$D^{*0} \to \gamma D^0$, $\gamma \to e^+ e^-$ decays in which one
of the two electrons from $\gamma$ conversion is
taken as a slow pion candidate, and background from 
$K^0_S \to \pi^+ \pi^-$ decays, where one of the pions is 
taken as the slow pion candidate and the other 
is assigned as a kaon or lepton candidate.

\begin{table}[htbp]
\begin{center}
\caption{The fraction of the correlated background in the total
background in [\%], as obtained from MC simulation, for the
electron and muon decay mode, in both RS and WS samples.
 Fractions for the entire proper decay time interval and
for the selected interval are shown. The quoted uncertainties are statistical
only.}
\begin{tabular}{ c | r@{$\pm$}l r@{$\pm$}l | r@{$\pm$}l r@{$\pm$}l}
\hline
\hline
	& \multicolumn{4}{c|}{Right Sign sample}		&
	\multicolumn{4}{c}{Wrong Sign sample}		\\
	& \multicolumn{2}{c}{all $t_{xy}$}	& \multicolumn{2}{c|}{$1.6-9.0 $}	&
	  \multicolumn{2}{c}{all $t_{xy}$}	& \multicolumn{2}{c }{$1.6-9.0 $}	\\
\hline
$e$	& 34.49 & 0.07	& 25.33 & 0.15	&  7.22 & 0.05 &  5.64 & 0.10	\\
$\mu$	& 40.07 & 0.06	& 39.62 & 0.14	& 14.87 & 0.06 & 14.76 & 0.14	\\

\hline
\hline
\end{tabular}
\label{tab-corb}
\end{center}
\end{table}

The \dm distribution for the total RS background can be seen in
Fig.~\ref{fig-dmRS_1-9} as the dashed line.
To fit the \dm distribution of the data as explained in Sec.
\ref{sec-dm-fit}, the $\Delta M$ distribution 
of the RS background events is obtained from MC simulation.

The \dm distribution for the total WS background is shown in
Fig.~\ref{fig-WSbkg} as the solid histogram. One can see
that the difference in shape between the uncorrelated background
and the total background is smaller in the electron decay mode than
in the muon decay mode. This is both due to the larger 
correlated background component in the muon  decay mode (see
Table~\ref{tab-corb}) and due to its shape, shown in the same figure.

To check the MC simulation and to avoid systematic errors arising from
any discrepancy with the data, the uncorrelated background
in the WS sample was described using the data. 
This background
is modeled by combining slow pion candidates and \d candidates
from different events.
Technically this is done
by {\it embedding} slow pion candidates into other events
according to the following procedure:

\begin{itemize}
\item All slow pion candidates
from an event in which a $D^0$ candidate was found, are taken to be
embedded into other events.
\item We denote by $\mathcal N$ the number of slow pion candidates
which form a WS combination in their original event, and
their charge by $\mathcal Q$. Slow pions are embedded 
only into events with the same value of $\mathcal N$
and only the embedded 
slow pions of charge $\mathcal Q$ are used to 
form the WS combinations.
\item For each value of $\mathcal N$, slow pion candidates from
$\mathcal N_A$ different events are stored to be embedded into other
events.
\item Each $D^0$ candidate 
is combined with slow pions from several other events; slow pions
from a maximum of $\mathcal N_A$ events are used. Once a
combination satisfying all the \dstp requirements is obtained
(including $\Delta M < 0.18~{\rm GeV}/c^2$), further combinations are
not formed.
\item Once an embedded slow pion is used to form a combination
satisfying all the \dstp requirements, none of the remaining slow
pions from the same event is embedded into other events.
\end{itemize}

With these requirements, the \dm distribution of the sample of embedded
slow pions slightly depends on $\mathcal N_A$, the maximum number of
events from which the embedded slow pions are taken and tested with a
single \d candidate.
The dependence is due to 
the \dm dependence on the slow pion momentum.  Slow pions with
higher momenta tend to form \dstp candidates with slightly higher \dm
values, hence their probability to form a \dstp candidate with $\Delta
M < 0.18~{\rm GeV}/c^2$ is smaller. Increasing $\mathcal N_A$
enables these slow pions to be tested with a larger number of \d
candidates and enhances the
probability to form a combination with $\Delta
M < 0.18~{\rm GeV}/c^2$. Thus increasing
$\mathcal N_A$ slightly enhances the contribution at higher \dm values.

The most appropriate $\mathcal N_A$ value is determined from the data by
observing samples of $t_{xy} < 0.0$. In this proper decay time region,
the expected fraction of the mixed
signal events in the WS background is 
much smaller 
(around 18 times in the electron and around 27 times in
the muon decay mode) 
than in the region $1.6 < t_{xy} < 9.0$. Hence it is safe to assume that the
WS data sample of $t_{xy} < 0.0$ contains no mixed signal events. For
$t_{xy} < 0.0$ we
combine the SVD-1 and SVD-2 subsamples and compare the \dm distribution
of the WS data with the \dm distribution of the subsample used to describe
the WS background, $i.e.$ embedded slow pions with the addition of the MC
correlated background events (as explained at the end of this section). The
\dm distributions are compared by observing the value of 
\zac
r_{155} = N_{\Delta M < 0.155} / N_{\Delta M < 0.18},
\label{eq-r155}
\kon
the ratio of the number of events with $\Delta M < 0.155~{\rm
GeV}/c^2$ and the number of events with $\Delta M < 0.18~{\rm
GeV}/c^2$. This is a representative observable that is used to
characterize the \dm distribution by a single number.
The values for the data and for the
background with three different values of $\mathcal N_A$ are shown in
Table~\ref{tab-NA}.

\begin{table}[htbp]
\begin{center}
\caption{Comparison of the $r_{155}$ values for the WS data and the
modeled WS background with $t_{xy} < 0.0$, using different values of
$\mathcal N_A$.}
\begin{tabular}{c | c c | c c}
\hline
\hline
$t_{xy} < 0.0$	& \multicolumn{2}{c|}{electron mode}	&
\multicolumn{2}{c}{muon mode} \\
		& $\mathcal N_A$ & $r_{155}[\%]$		& $\mathcal
N_A$ & $r_{155}[\%]$	\\
\hline
data		&		& $25.66 \pm 0.18$	& 	& $29.00 \pm 0.16$	\\

background 	&	40	& $25.80 \pm 0.05$	& 20	& $29.11 \pm 0.05$	\\
	 	&	45	& $25.70 \pm 0.05$	& 25	& $29.00 \pm 0.05$	\\
	 	&	50	& $25.61 \pm 0.05$	& 30	& $28.90 \pm 0.05$	\\
\hline
\hline
\end{tabular}
\label{tab-NA}
\end{center}
\end{table}

From Table~\ref{tab-NA} one can see that for the electron decay mode the
best agreement between the data and the background ($t_{xy} < 0.0$)
is for $\mathcal
N_A = 45$ and in the muon decay mode for $\mathcal N_A = 25$. 
The dependence of the final result on the $\mathcal N_A$ value is
taken into account
when evaluating the systematic uncertainty.

As a cross-check, we combine SVD-1 and SVD-2 subsamples and compare the 
$\Delta M$ distribution of the MC uncorrelated WS background
with that of the embedded slow pions. Their agreement is good.

To obtain the final \dm distribution of WS background events,
the \dm shape of the WS-correlated background is taken from MC
simulation and added to the sample
of the embedded slow pions in the same fraction as found by MC
simulation. The
uncertainty on this fraction is taken into account
 when evaluating the
systematic uncertainty.

\subsection{Fit to $\Delta M$ distribution}
\label{sec-dm-fit}

To extract the signal yield,
we perform a binned maximum likelihood fit to 
 the $\Delta M$ distribution, assuming a Poisson distribution of
events in $\Delta M$ bins and thus maximizing
\begin{equation}
{\cal L} = \prod_{j=1}^{N_{\rm bin}}\frac{e^{-{\mu}(\Delta M_j)}\cdot
  {(\mu}(\Delta M_j))^{N_j}}{N_j !}.
\label{eq_fit}
\end{equation}
Here $N_j$ is the number of entries in the $j$-th bin and $\mu(\Delta
M_j)$ is the expected number  
of events in this bin, given by
\begin{equation}
{\mu}(\Delta M_j)={\cal{N_R}}\left [ f_s P_s(\Delta M_j)+(1-f_s) P_b
(\Delta M_j)\right ].
\label{eq_predicted}
\end{equation}
$P_s$ is the signal $\Delta M$ distribution
obtained from MC simulation. $P_b$ is the background $\Delta M$ distribution
composed as described above. The signal fraction  $f_s$ is the only
free parameter in the fit. $\mathcal N_R$ is the number of entries 
 in the fitted histogram. $N_{\rm bin}=45$ is
the number of intervals in the \dm distribution.
The quoted $\chi^2$ values are obtained using
$
\chi^2 = \sum_{j=1}^{N_{\rm bin}} \frac{(N_j - \mu(\Delta M_j))^2}{\sigma_j^2},
$
where $\sigma_j$ includes the statistical uncertainties
of the fitting histograms, 
$\sigma_j^2 = N_j + \sigma^2_{P_s,j} + \sigma^2_{P_b,j}$.

\subsection{The RS signal yield}
\label{sec-RSyield}

\begin{figure}[htbp]
\begin{center}
\includegraphics[width=\ssld]{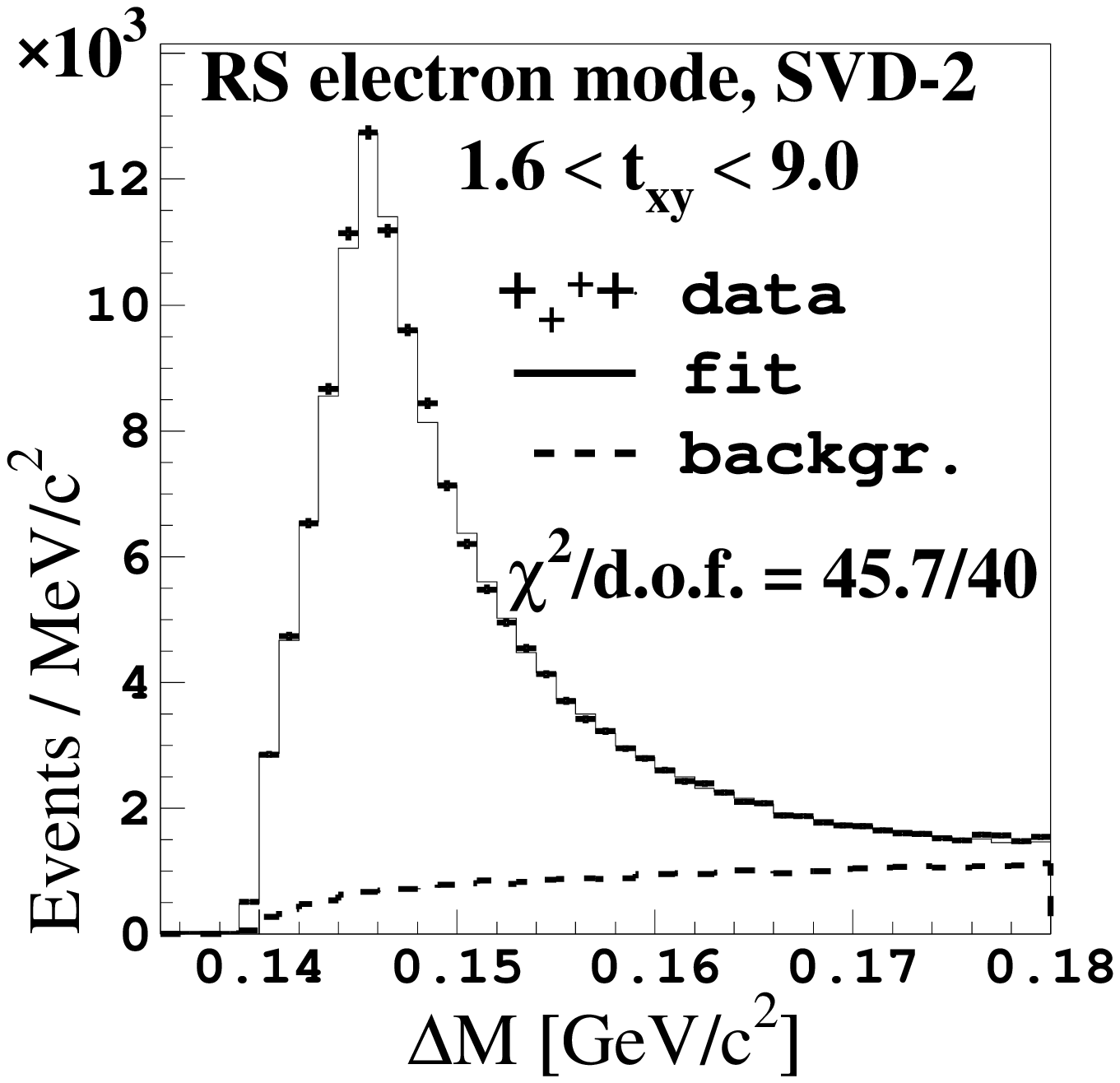}
\includegraphics[width=\ssld]{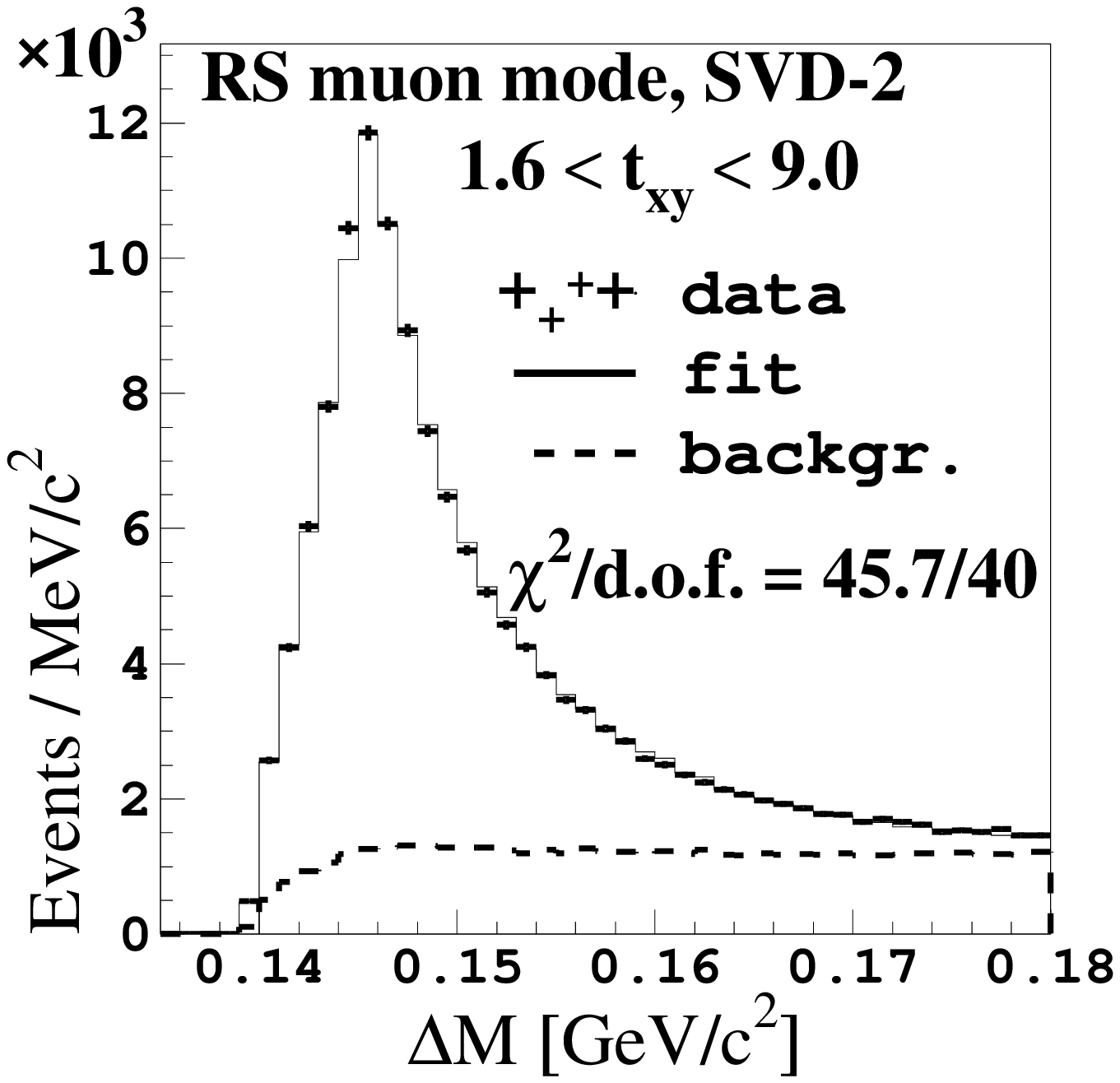}
\end{center}
\caption{The \dm distribution of the RS events for $1.6 < t_{xy} <
9.0$, SVD-2. The dashed line represents the background, the 
solid line is the result of the fit, described in the text,
and the points with error bars are the data. The left plot is for the
e-2 subsample, the right one for the $\mu$-2 subsample.}
\label{fig-dmRS_1-9}
\end{figure}

The fit to the \dm distribution in the RS sample is performed as
described above; examples of the fit result are shown in
Fig.~\ref{fig-dmRS_1-9}.
In the total $t_{xy}$ range, the signal fraction $f_s$ is
about 70\% in the electron decay mode and about 63\% in the muon decay mode.
The fraction is largest for 
$1.6 < t_{xy} < 2.0$ (82\% in the electron decay mode and 74\% in the muon
decay mode) and decreases at larger $t_{xy}$ values: for
$5.6 < t_{xy} < 9.0$ it is 62\% in the electron decay mode and 54\% in
the muon decay mode.
The $\chi^2$ values of the fits in the individual $t_{xy}$ intervals are
in good agreement with the expectation for 40 degrees of freedom.
In the total \t~range, the reduced $\chi^2$ values are 
larger than expected
(values of 1.5--2.6 for 40 degrees of freedom).
This is explained by a difference in the
amount of  
associated signal between the data and the MC simulation. Repeating
the fits with a fraction of the associated signal as the second free
parameter yields reduced $\chi^2$ values around 1.0 also for the
total \t~region. 
This effect is considered in the estimate of
the systematic uncertainty.
The numbers of RS signal events are given in 
Table \ref{tab-mu2}.

\section{RESULT}
\label{ch-result}

\begin{figure*}[htbp]
\begin{center}
\includegraphics[width=\ssls]{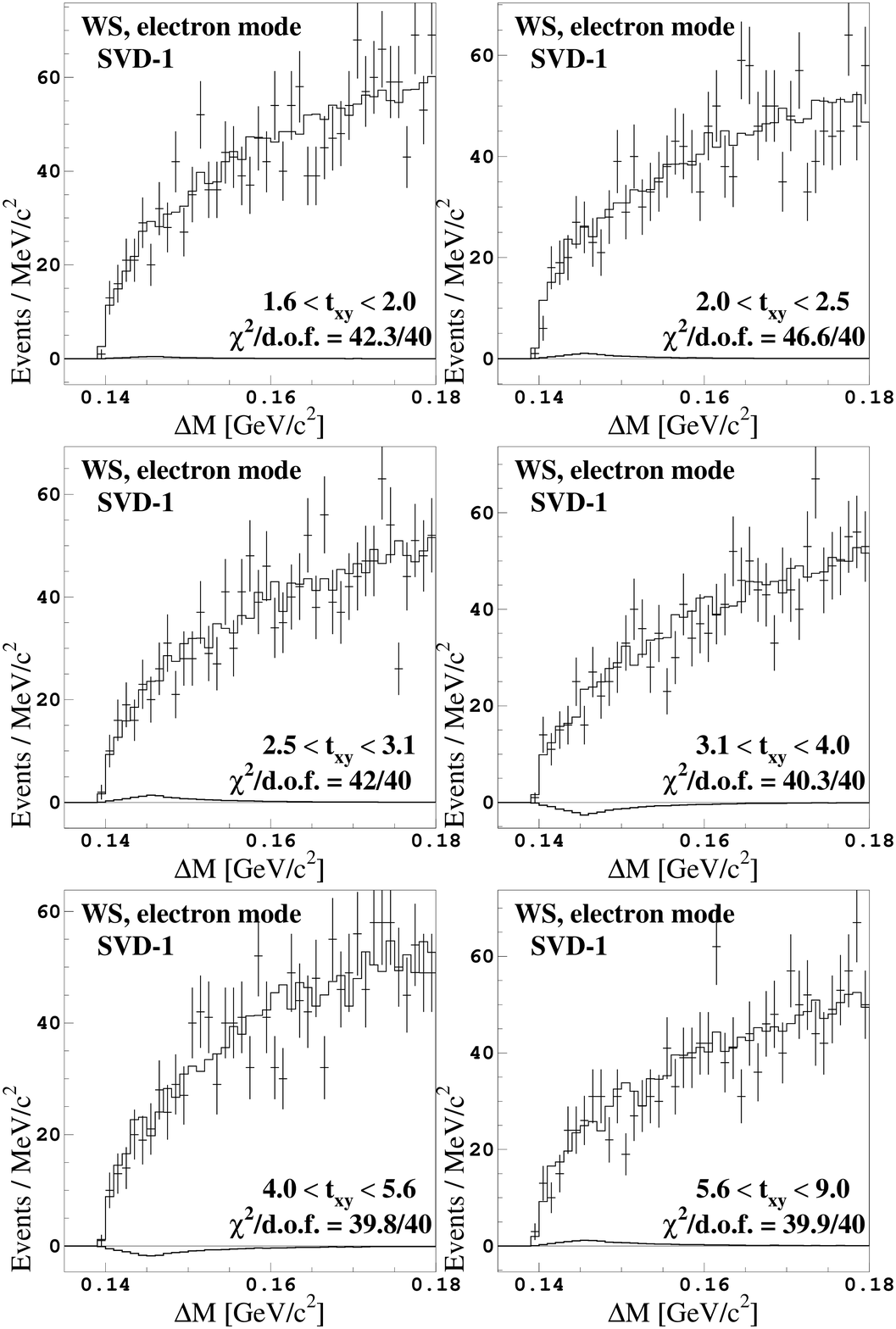}
\qquad
\includegraphics[width=\ssls]{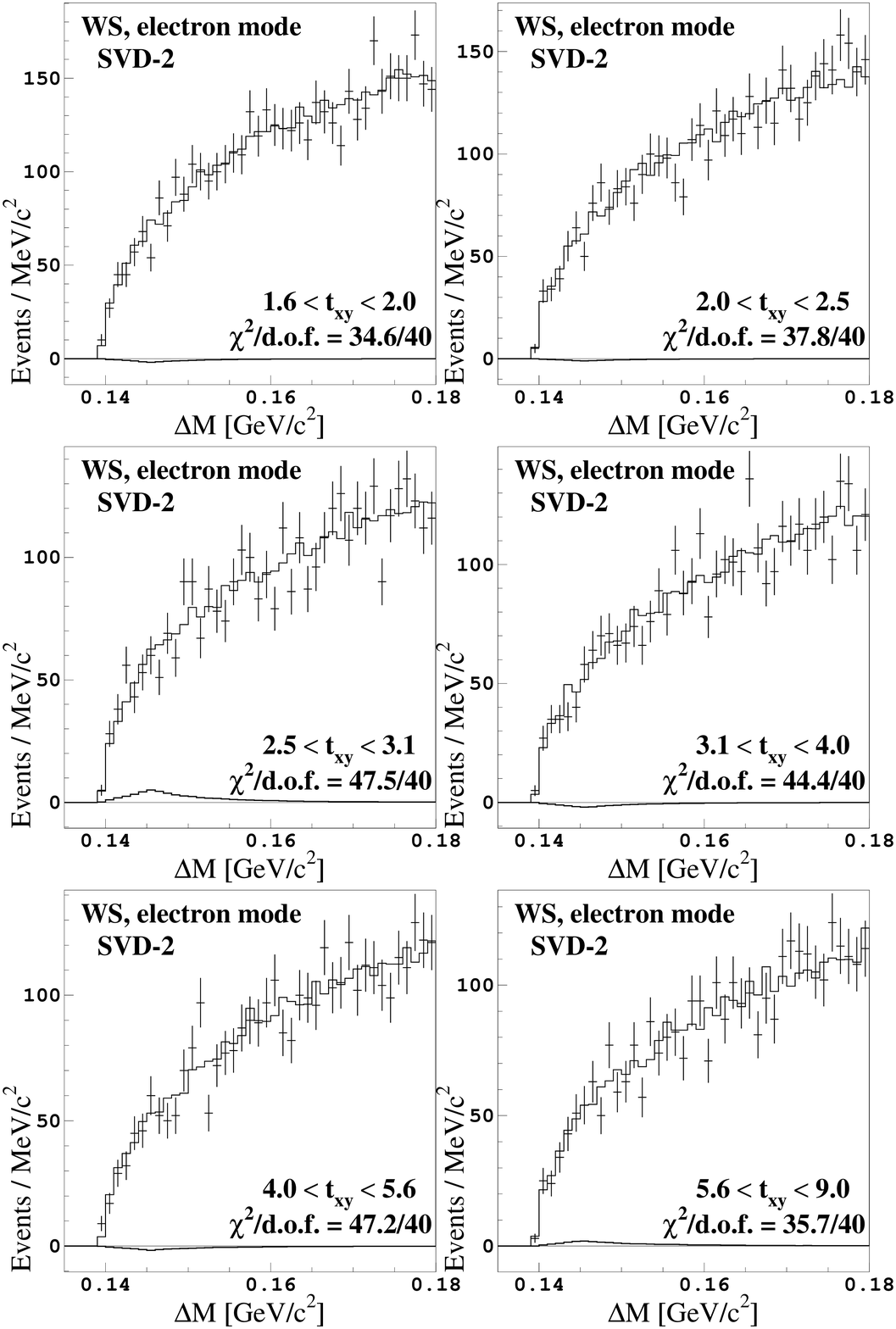}
\end{center}
\vskip -5mm
\caption{The \dm distribution of WS events in the six proper decay time
intervals for the e-1 (left) and e-2 (right) subsamples. The
points with error bars are the data, 
the histogram represents the result of the fit, described in the text,
and the small contribution on the horizontal axis shows the fitted
signal yield.}
\label{fig-dmWS_e1}
\end{figure*}

\begin{figure*}[htbp]
\begin{center}
\includegraphics[width=\ssls]{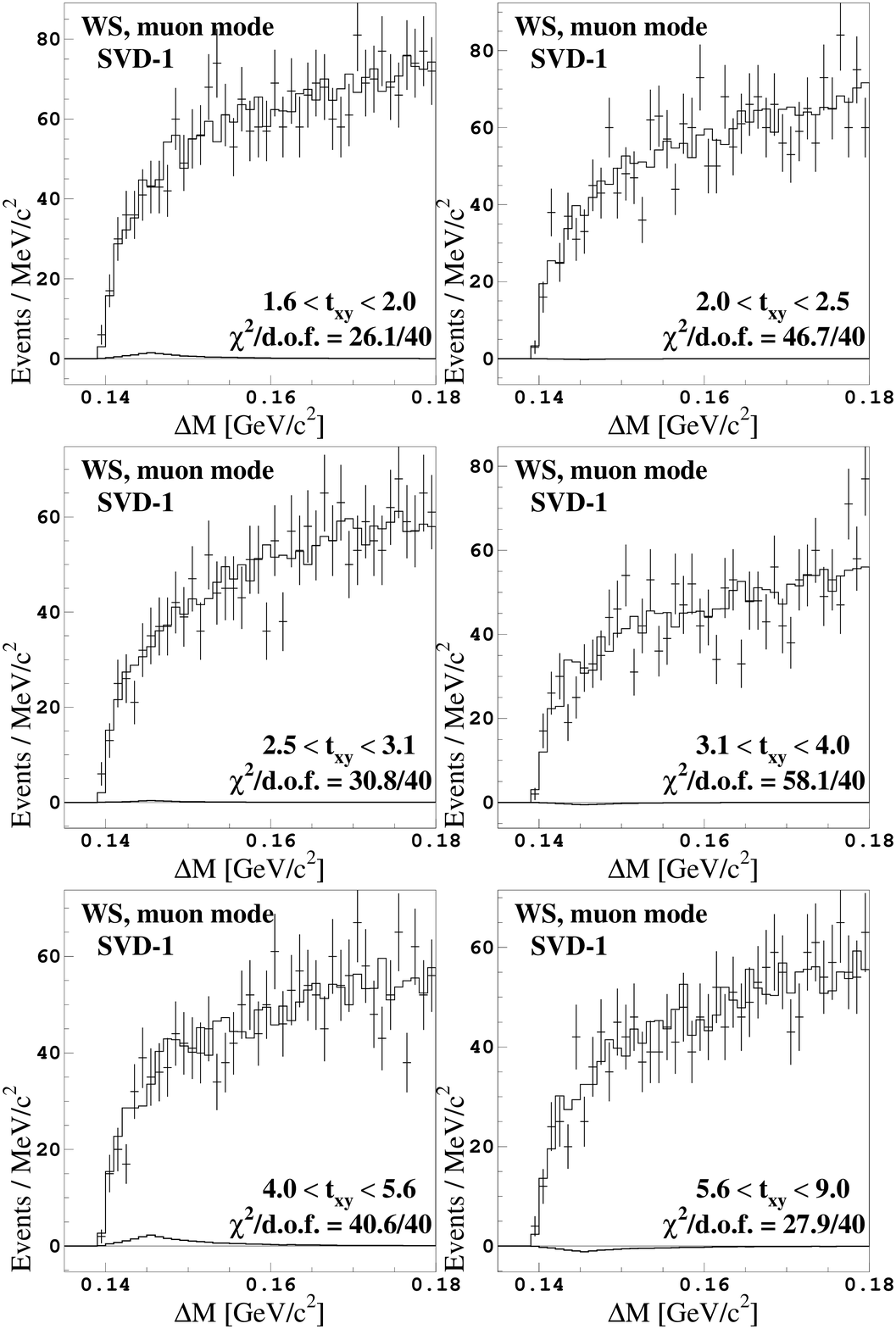}
\qquad
\includegraphics[width=\ssls]{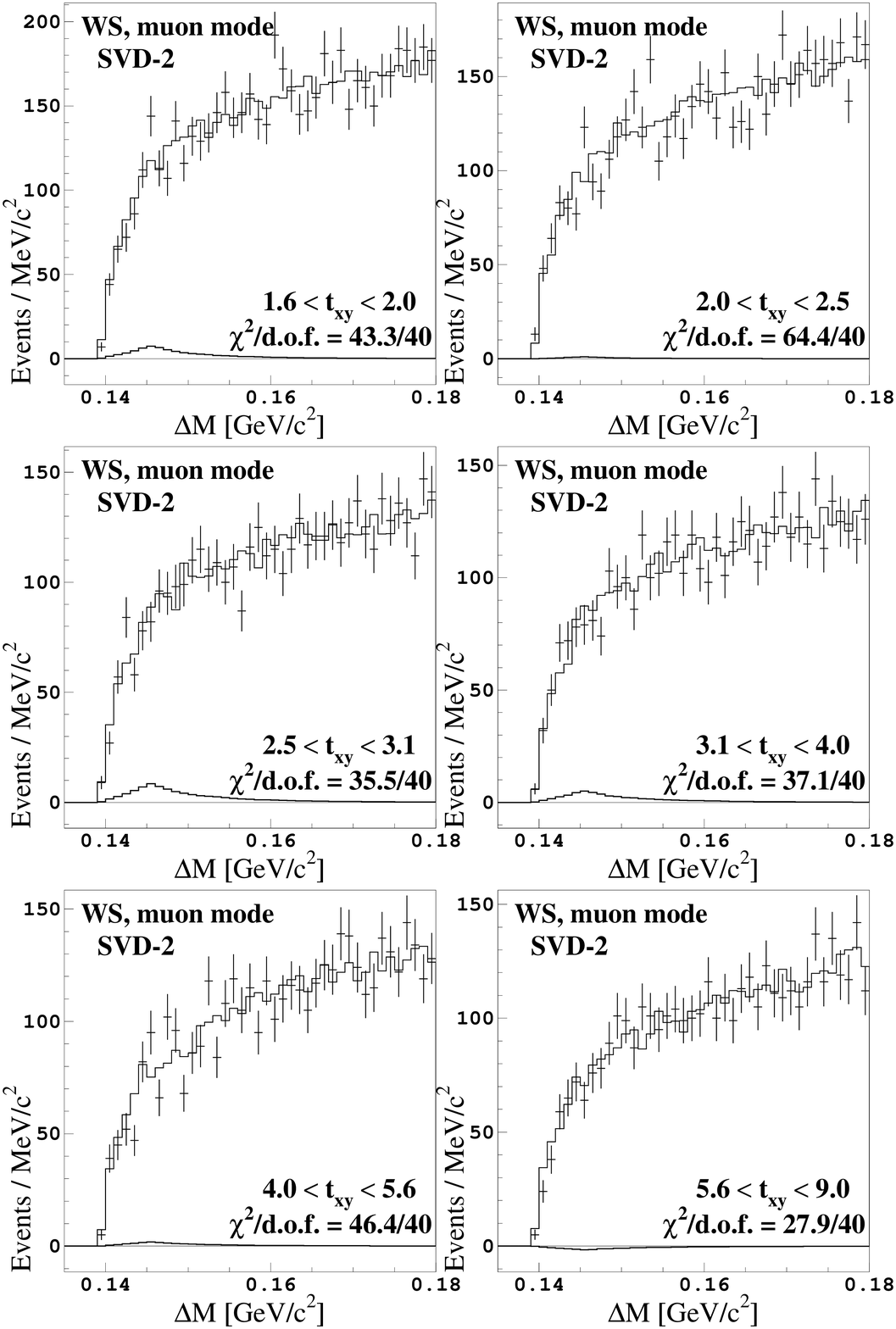}
\end{center}
\caption{Same as Fig.~\ref{fig-dmWS_e1} for the $\mu$-1 and
$\mu$-2 subsamples.}
\label{fig-dmWS_mu1}
\end{figure*}

As the kinematic properties of the RS and WS decays are the same, we use
the \dm shape of the MC simulated RS signal events also for the WS
signal decays. The \dm distribution of the WS background events is obtained as
described in Sec. 
\ref{sec-bkg}.
By fitting the \dm distribution as described in Sec.
\ref{sec-dm-fit}, we extract the number of the mixed signal events in
the four WS subsamples (e-1, e-2, $\mu$-1, $\mu$-2). The \dm
distributions and the $\chi^2$ 
values of the fits for all the subsamples and  different
proper decay time intervals are shown  in Figs.
\ref{fig-dmWS_e1}, \ref{fig-dmWS_mu1}.
The extracted WS signal yields are given in 
Table \ref{tab-mu2}.

For each of the four subsamples we determine the mixing ratio $R_M$ by
three different methods 
which are discussed below.

\begin{figure}[t]
\begin{center}
\includegraphics[width=\ssld]{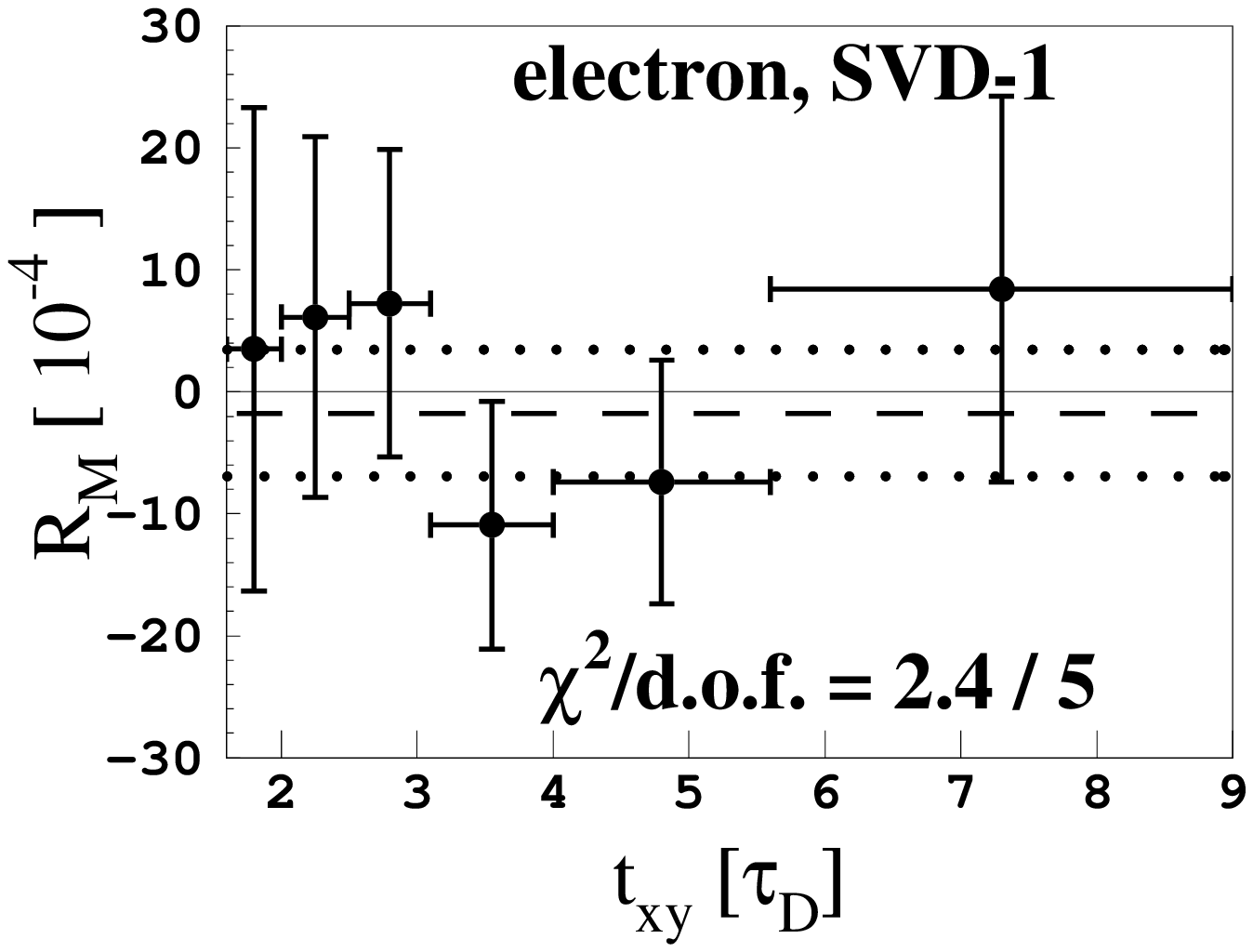}
\includegraphics[width=\ssld]{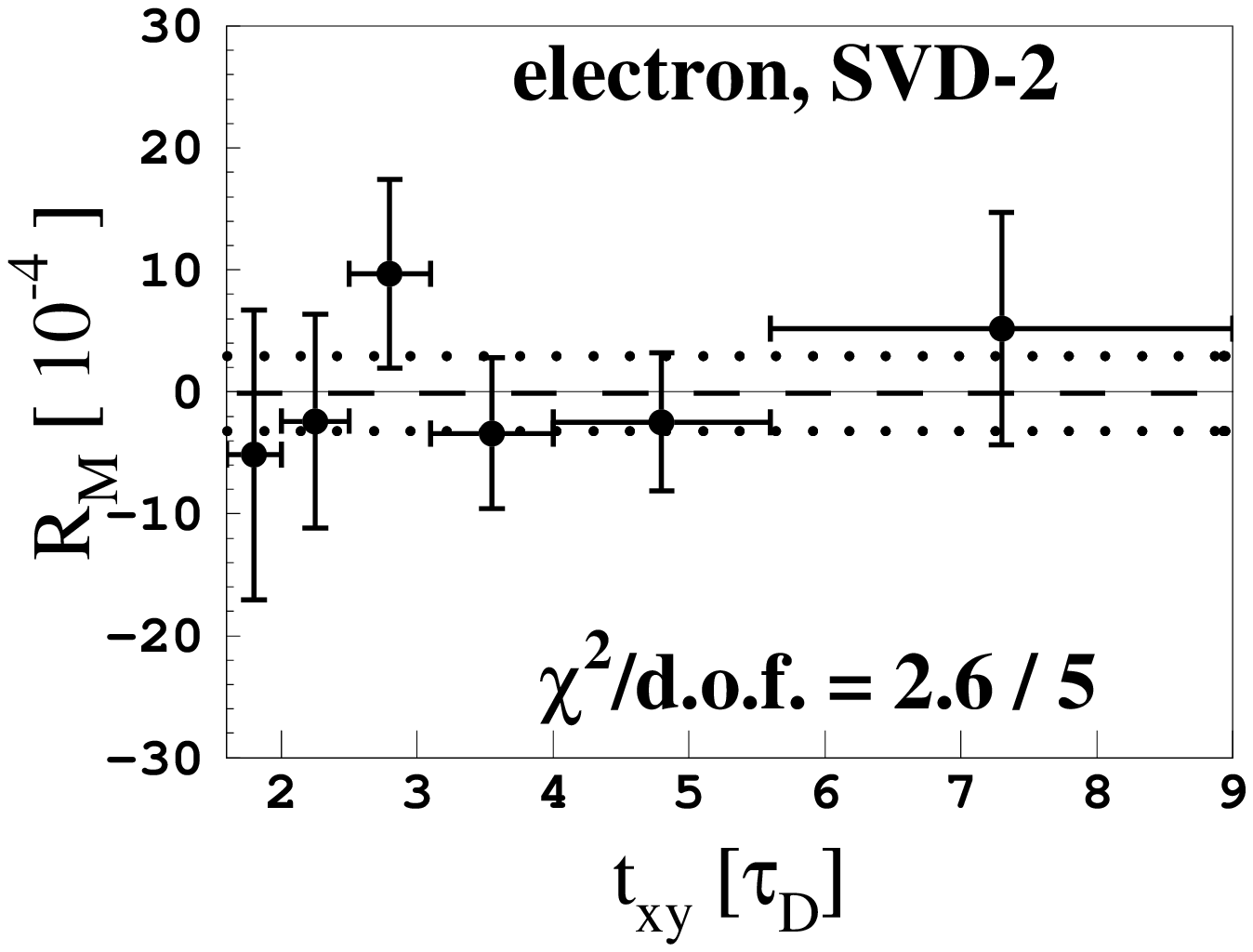}

\includegraphics[width=\ssld]{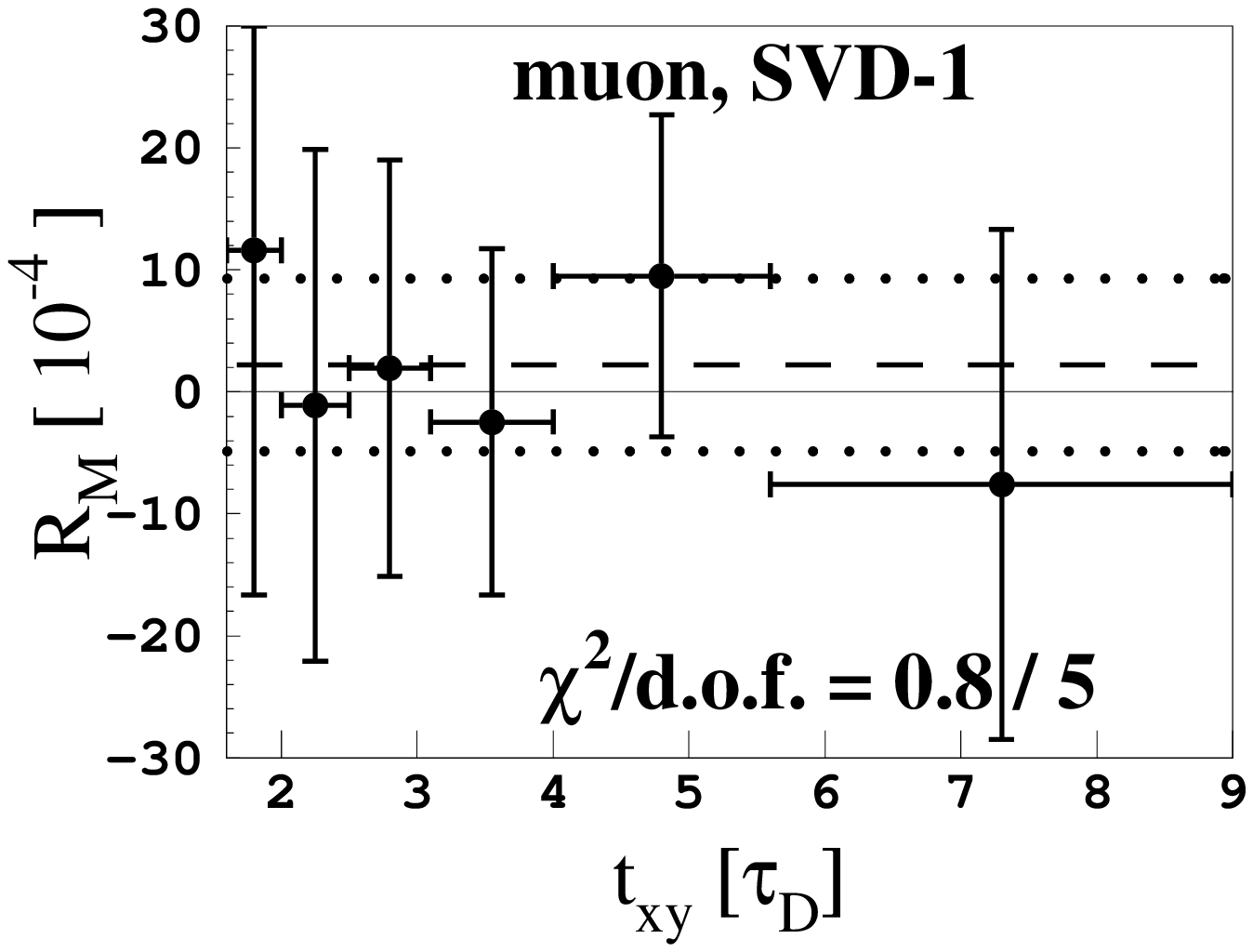}
\includegraphics[width=\ssld]{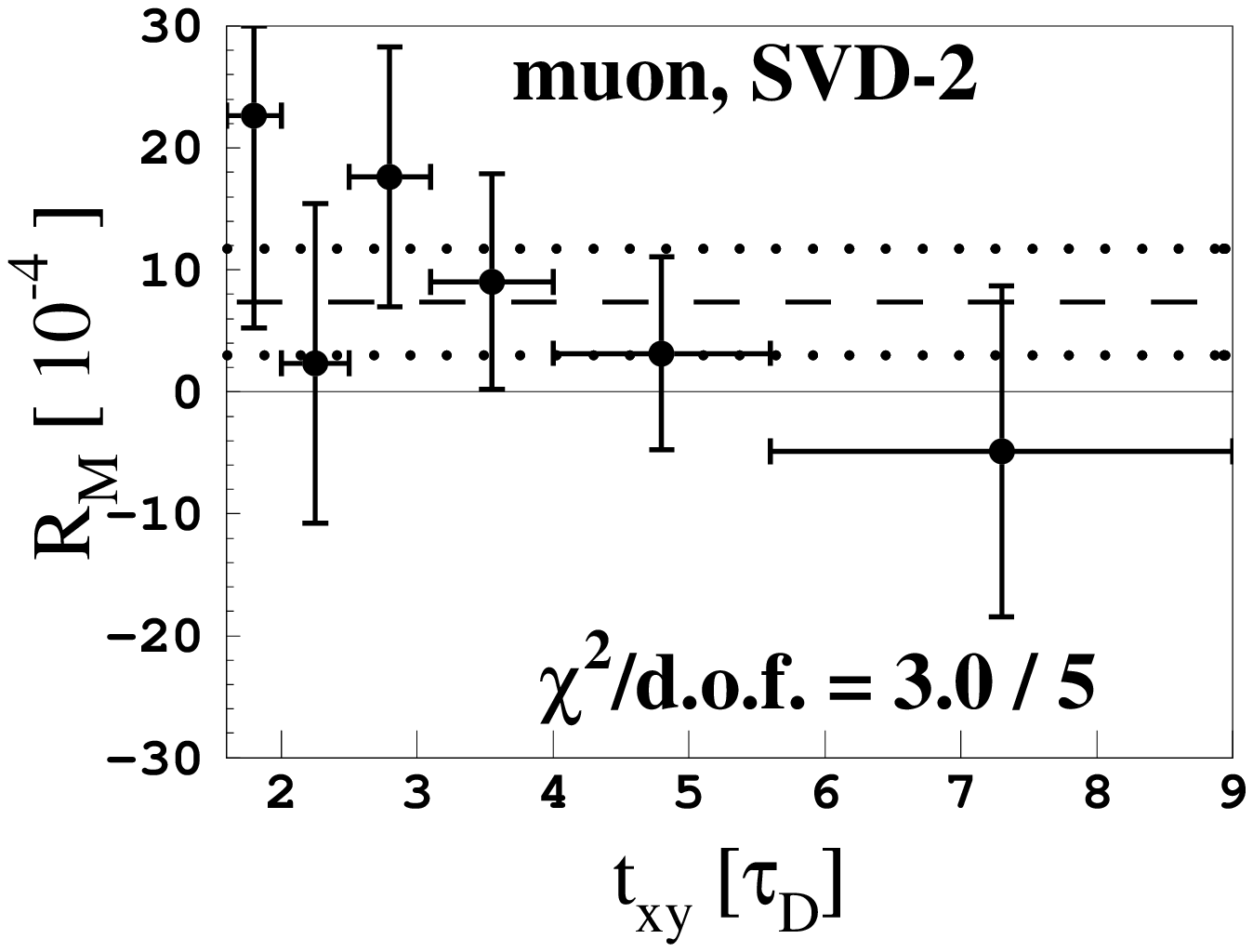}
\end{center}
\caption{ The resulting $R_M^i$ values for the four subsamples and their average
value (dashed line). The dotted lines represent the $\pm 1\sigma$
interval. The solid line corresponds to no mixing.}
\label{fig_RM}
\end{figure}

\begin{itemize}
\item[1)] The fit to $\Delta M$ in the RS and WS samples is performed
  without any 
selection based on the proper decay time measurement. The ratio $R_M$
is calculated as the ratio of the obtained number of WS and RS signal
events, $N_{\rm WS}/N_{\rm RS}$.  The results can 
be found in 
Table \ref{tab-mu2} in rows labeled ``all $t_{xy}$''.

\item[2)] The fit to $\Delta M$ distributions for the RS and WS
sample is performed for events with $1.6 < t_{xy} < 9.0$. The ratio
$R_M$
is calculated as $N^i_{\rm WS}/N^i_{\rm
RS}\times\epsilon^i_{\rm RS}/\epsilon^i_{\rm WS}$.  
 The results are given in 
Table \ref{tab-mu2} in rows labeled ``1.6--9.0''.
The resulting statistical uncertainty of the result is
around 34\% smaller than the one obtained by method 1).

\item[3)] The third result, given in rows labeled ``combined'' of 
Table \ref{tab-mu2}, is a 
$\chi^2$ fit of a constant to the six $R_M^i$ values measured in the six proper
decay time bins. The six $R_M^i$ values and the result of the fit for
each of the four subsamples are shown in Fig.~\ref{fig_RM}.
The statistical uncertainty of this result is 2--3\%
smaller than in method 2), because additional information on proper
decay time is included through the six $\epsilon_{\rm
RS}^i/\epsilon_{\rm WS}^i$ ratios.

\end{itemize}

Using the MC simulation, we verified that method 3) has the best
sensitivity; we therefore 
quote as our final result the value obtained by method
3). To illustrate the effect of including the proper decay time
information, we show also the results of methods 1) and 2).

From  Table \ref{tab-mu2} one can see that the central
values obtained by using the three methods, are slightly different. 
In evaluating the significance of the difference between the result of
method 1) and the result 
of method 2), we have
accounted for the ratio of the proper decay time efficiencies and for
the statistical correlation between the samples. The differences are
within the 
expected statistical fluctuations: they range
between $-0.6$ and $+1.4$ standard deviations. 
By using toy MC simulation, it has also been verified that the
differences in the central values between methods 2) and 3) are within
the range of 
expected statistical fluctuations. 
For the default method 3), results for $R_M$ in all four subsamples
are consistent with the null value. The $\chi^2$ values, shown in
Fig.~\ref{fig_RM}, are in good agreement with the expected $\chi^2$
distribution for 5 d.o.f., which has a maximum at the value of 3.0.


\begin{table}[htbp]
\begin{center}
\caption{The number of fitted signal events in the RS and WS samples,
  the ratio of RS and WS $t_{xy}$ efficiencies, and
the resulting $R_M^i$ value for each proper decay time interval for the
four subsamples. The results of the fit to the six individual
$R_M^i$ values are denoted as ``combined''.}
\begin{tabular}{c  r@{$\pm$}l   r@{$\pm$}l  r@{$\pm$}l  r@{$\pm$}l}
\hline
\hline
$t_{xy}$ &\multicolumn{2}{c}{$N^i_{\rm RS}$}	
	& \multicolumn{2}{c}{$N^i_{\rm WS}$}	
	& \multicolumn{2}{c}{$\displaystyle{{\epsilon_{\rm RS}^i}/ \epsilon_{\rm WS}^i}$}
	& \multicolumn{2}{c}{$R_M^i~[10^{-4}]$}	\\
\hline
\multicolumn{9}{c}{e-1 subsample:}\\
\hline
1.6--2.0	& 12578 & 94&    4.8 & 27.2 & 0.915 & 0.007 &    3.5 & 19.8	 \\
2.0--2.5	& 11273 & 89&   10.9 & 26.3 & 0.634 & 0.004 &    6.1 & 14.8	 \\
2.5--3.1	&  8975 & 84&   14.7 & 25.6 & 0.443 & 0.002 &    7.2 & 12.6	 \\
3.1--4.0	&  7937 & 83& --28.0 & 25.9 & 0.310 & 0.003 & --10.9 & 10.1	 \\
4.0--5.6	&  6394 & 85& --21.2 & 28.6 & 0.223 & 0.003 &  --7.4 & 10.0	 \\ 
5.6--9.0	&  4196 & 89&   15.9 & 29.8 &	0.223 & 0.003 &    8.4 & 15.8	 \\
\hline
combined	&\multicolumn{5}{c}{} &  & {\bf --1.7} & {\bf 5.2}  	 \\
\hline
1.6--9.0	& 51325 & 213& --11.5  & 65.4   & 0.413 & 0.001 &  --0.9 & 5.3	 \\
all $t_{xy}$	& 183496 & 443	& 70.1 & 141 & \multicolumn{2}{c}{1} & 3.8 & 7.7 \\
\hline
\multicolumn{9}{c}{e-2 subsample:}\\
\hline
1.6--2.0	&  32616 & 150 & --19.1 &  44.0  & 0.881 & 0.003 & --5.2 & 11.9	 \\
2.0--2.5	&  28711 & 146 & --11.4 &  41.7  & 0.603 & 0.002 & --2.4 & 8.8	 \\
2.5--3.1	&  22513 & 131 &   52.5 &  41.9  & 0.415 & 0.002 &   9.7 & 7.7	 \\
3.1--4.0	&  18941 & 132 & --22.6 &  41.1  & 0.285 & 0.002 & --3.4 & 6.2	 \\
4.0--5.6	&  14796 & 129 & --18.6 &  42.3  & 0.198 & 0.002 & --2.5 & 5.7	 \\ 
5.6--9.0	&   9072 & 128 &   25.2 &  46.5  & 0.186 & 0.002 &   5.2 & 9.5	 \\
\hline
combined	&\multicolumn{5}{c}{} &  & {\bf --0.1} & {\bf 3.1}  	 \\
\hline
1.6-9.0	& 126539 & 332 &   --10.7 & 102    & 0.389 & 0.001 &  --0.3 & 3.1	 \\
all $t_{xy}$
	& 469947 & 701 &  --369 & 222    &  \multicolumn{2}{c}{1}   &  --7.8 & 4.7	 \\
\hline
\multicolumn{9}{c}{$\mu$-1 subsample:}\\
\hline
1.6--2.0	&  11314 &  111  &   14.2 &  34.7  & 0.921 & 0.005 &   11.6 & 28.2	 \\
2.0--2.5	&  10185 &  109  &  --1.8 &  33.5  & 0.637 & 0.004 &  --1.1 & 21.0	 \\
2.5--3.1	&   7893 &   98  &    3.5 &  30.7  & 0.440 & 0.003 &    1.9 & 17.1	 \\
3.1--4.0	&   6804 &   96  &  --5.5 &  31.8  & 0.303 & 0.002 &  --2.5 & 14.2	 \\
4.0--5.6	&   5350 &   97  &   23.7 &  33.0  & 0.214 & 0.002 &    9.5 & 13.2	 \\ 
5.6--9.0	&   3670 &   90  & --12.8 &  35.4  & 0.217 & 0.003 &  --7.6 & 20.9	 \\
\hline
combined	&\multicolumn{5}{c}{} &  & {\bf 2.2} & {\bf 7.1}  	 \\
\hline
1.6--9.0	& 45181  &  245  &  --11.2  &   79.9   & 0.410 & 0.001 &  --1.0 & 7.2	 \\
all $t_{xy}$
	& 163215 & 485   &  --204 & 180    & \multicolumn{2}{c}{1}	& --12.5 & 11.0	 \\
\hline
\multicolumn{9}{c}{$\mu$-2 subsample:}\\
\hline
1.6--2.0	&  27612 &  180  &   71.4 &  54.8  & 0.876 & 0.015 &  22.7 & 17.4	 \\
2.0--2.5	&  23695 &  170  &    9.3 &  52.3  & 0.595 & 0.010 &   2.3 & 13.1	 \\
2.5--3.1	&  18905 &  154  &   82.3 &  49.8  & 0.405 & 0.006 &  17.6 & 10.7	 \\
3.1--4.0	&  15488 &  150  &   51.1 &  50.1  & 0.273 & 0.004 &   9.0 & 8.8	 \\
4.0--5.6	&  11989 &  144  &   20.4 &  51.1  & 0.186 & 0.007 &   3.2 & 7.9	 \\ 
5.6--9.0	&   7146 &  138  & --20.3 &  56.5  & 0.171 & 0.016 & --4.9 & 13.6	 \\
\hline
combined	&\multicolumn{5}{c}{} &  & {\bf 7.4} & {\bf 4.4}  	 \\
\hline
1.6--9.0	& 104556 &  381 &     192 &    125 & 0.380 & 0.002 &   7.0 & 4.5	 \\
all $t_{xy}$
	& 396151 &  761 &  	170 & 	 284 & \multicolumn{2}{c}{1}	&   4.3	& 7.2	 \\
\hline
\hline
\end{tabular}
\label{tab-mu2}
\end{center}
\end{table}


The combined result for the electron decay mode is obtained by a $\chi^2$ fit
to the 
values for the e-1 and e-2 subsamples, obtained by method
3). The fit yields 
$R_M^e = (-0.6 \pm 2.6)\times 10^{-4}$ 
with a
$\chi^2$ value of 0.1
per 1 degree of freedom.
The combined result for the muon decay mode is obtained in the same way; 
the $\chi^2$ fit yields 
$R_M^\mu = (5.9 \pm 3.7)\times 10^{-4}$ 
with a
$\chi^2$ value of 0.4.

The combined result, taking into account the
statistical uncertainty only, is obtained by a $\chi^2$ fit to
the four values (electron and muon decay mode, SVD-1 and SVD-2); it yields
a value of 
\zac
R_M^{\rm stat. } = (1.6 \pm 2.2)\times 10^{-4},
\label{eq_stat_rez}
\kon
where the quoted uncertainty is statistical only. The $\chi^2$ 
value is 2.5 
for three degrees of freedom. The $R_M$ values for the
four subsamples and the result of the fit are shown in the left
plot of Fig.~\ref{fig_rm_fin}.
To obtain the final result, the partially correlated systematic
uncertainties have to be studied and taken into account.

\subsection{Systematic uncertainties}
\label{sec-systematics}

In the following subsection,
different sources of the systematic uncertainties are 
discussed and the systematic uncertainties are given; they are summarized in
Table~\ref{tab-syst}.

\subsubsection{Finite statistics of the fitting distributions}
\label{sec-finite-stat}

One of the main sources of systematic uncertainty is the limited statistics of
the samples used to obtain the signal and background \dm distributions
used in the \dm fit to data.
To estimate this, we vary the contents of 
all bins of the RS and WS, signal and background \dm
distributions independently in
accordance with each bin's statistical uncertainty. We 
repeat the fit to the RS and WS data, calculate the corresponding $R_M^i$ 
in each proper decay time interval, and obtain a new $R_M$
value.
Repeating the procedure 1000 times, the obtained distribution of
$R_M$ values has a Gaussian shape. The sigma of the Gaussian, fitted to the distribution, is
taken as the systematic uncertainty due to the limited statistics of the
fitting distributions. The uncertainties are listed in Table~\ref{tab-syst},
line 1.
Larger uncertainties in the muon decay mode reflect the fact that, compared to
the electron decay mode, the muon 
background is larger especially in the signal region, and
secondly, the embedded slow pion sample is smaller due to the smaller
$\mathcal N_A$ value used. Since
this uncertainty is statistical in nature, it is considered to be completely
uncorrelated between the four subsamples (e-1, e-2, $\mu$-1, $\mu$-2).

\subsubsection{The amount of WS correlated background }
\label{sec-syst-wscb}

The normalization of the WS-correlated background
is determined by MC simulation, taking into account the central values
of branching fractions \cite{PDG}, of decay modes that contribute to
this background.

From MC simulation studies we find that in the electron decay mode the
largest contributions to the WS correlated background come from the
following decays: 
$D^0 \to K^- e^+ \nu_e $ ($D^0$ mesons mainly from
$D^{*0} \to D^0 \gamma$ decays, 34\% of the correlated background), 
$D^0 \to K^- \pi^+ \pi^0$ (12\%), 
$D^0 \to K^- e^+ \nu_e \pi^0$ (7\%), 
$D^0 \to K^- K^+ $ (7\%).
In total  60\% of the correlated background
comes from these decays. 

In the muon decay mode, the largest contributions to the WS correlated background come from 
$D^0 \to K^- \pi^+ \pi^0$ (19\%), 
$D^0 \to K^- \mu^+ \nu_\mu $ (12\%), 
$D^0 \to K^- \pi^+ \pi^0 \pi^0$ (7\%), 
$D^0 \to K^- \pi^+ \pi^- \pi^+$ (7\%), 
$D^0 \to K^+ \pi^- \overline{K}{}^0$ (4\%), 
$D^0 \to K^+ K^- \overline{K}{}^0$ (4\%).
In total  53\% of the correlated background
comes from these decays.
We calculate the weighted average of the relative uncertainties of the
branching fractions \cite{PDG} for the stated decay modes. For the electron
decay mode, the averaged relative uncertainty is $\pm$3.6\% and for the muon decay mode $\pm$5.9\%. 
To take into account the uncertainties of the branching fractions
used in the MC simulation, we repeat the WS 
fits, changing the amount of the total WS correlated background by the
average uncertainties on the branching fractions. The differences
between the resulting $R_M$ values and the default values are taken
as the systematic uncertainty from this source;
they are listed in Table~\ref{tab-syst},
line 2.

This procedure is conservative for two reasons. First, by
varying the total correlated background instead of varying its individual
components,
the uncertainties on the branching fractions are implicitly
considered to be 100\% correlated, resulting in the maximum possible
systematic uncertainty. Second, the modes comprising the correlated
background contribute significantly also to the uncorrelated
background. Taking this into account would
lead to a smaller change in $R_M$.

This uncertainty is larger in the muon decay mode, 
because the probabilities to misidentify a pion or kaon
as a muon are much larger than the corresponding
probabilities for misidentification as an electron.
Consequently, in the muon decay mode the fraction of the correlated
background is significantly larger (Table \ref{tab-corb}), its
$\Delta M$ shape tends to lower values (Fig.~\ref{fig-WSbkg})
and its averaged uncertainty of the branching fractions is larger.

The systematic uncertainty from this source is the same for the SVD-1 and
SVD-2 subsample. 
Since a significant part of the correlated background is due to
decays common to the electron and the muon decay modes, the systematic
uncertainties for both decay modes are highly correlated. Hence the systematic
uncertainty from this source will be treated as 100\% correlated for all
four subsamples (e-1, e-2, $\mu$-1, $\mu$-2).

\subsubsection{The \dm shape of the WS uncorrelated background }
\label{sec-syst-wsucb}

We also conservatively account for the uncertainty of the \dm
shape of the WS uncorrelated background.
We vary $\mathcal N_A$ within the
limits given by the statistical uncertainties of the $r_{155}$ values
in Table \ref{tab-NA}. 
The $r_{155}$ statistical uncertainty  for the
data is $\pm 0.18$ in the electron decay mode and $\pm 0.16$ in the muon
decay mode. 
For the embedded slow pion sample, the $r_{155}$ value
changes by 0.1 for $\Delta \mathcal N_A = 5$. 
Hence $\mathcal N_A$ is varied by $\pm 9$ in the electron decay mode,
and by $\pm 8$ in the muon decay mode. 
With the new \dm distributions 
we repeat the fit to WS data,
recalculate the $R_M$ values and quote the differences from the
default values as the systematic uncertainties from this source. 
These uncertainties are listed on line 3 of Table~\ref{tab-syst}.

Since $\mathcal N_A$ is
determined for the electron and muon decay mode separately and the
uncertainty on $\mathcal N_A$ is statistical in nature, this
systematic uncertainty is considered to be completely uncorrelated
between both decay modes.
On the other hand, $\mathcal N_A$ is determined for SVD-1 and SVD-2 subsamples
together, hence the uncertainty is treated as completely correlated between
them.

\subsubsection{Proper decay time distribution}
\label{ch-systerr5}

To check the reliability of efficiencies $\epsilon_{\rm RS}^i$ and  
 ratios $\epsilon_{\rm RS}^i/\epsilon_{\rm WS}^i$,
and to
estimate the effect of the imperfect 
fit to the proper decay time distribution, the
values of $\epsilon_{\rm RS}^i$ are compared
to an alternative estimate from the fit to $\Delta M$, 
$\epsilon_{\rm RS}^{i,\Delta M} = N_{\rm RS}^i/N_{\rm RS}^{\rm
tot}$. 
This method accounts for the influence of the
associated signal in the $t_{xy}$ distribution. 
In a majority of the \t~subintervals, $\epsilon_{\rm RS}^i$ and
$\epsilon_{\rm RS}^{i,\Delta M}$
typically agree
within $\pm 2\%$, the largest discrepancies being $-9.8\%$ and $+4.4\%$.
For the integrated $1.6 < t_{xy} < 9.0$ interval, they agree within
0.8\%--1.4\% for the four subsamples.

To estimate the effect of the discrepancies,
the relative difference between $\epsilon_{\rm RS}^i$ and
$\epsilon_{\rm RS}^{i,\Delta M}$ is assigned as
the relative uncertainty on $\epsilon_{\rm RS}^i/\epsilon_{\rm WS}^i$.
Hence we reduce the
six effciency ratios simultaneously by this uncertainty
and repeat the $R_M$ calculation; we then increase the ratios
by this uncertainty, and again recalculate $R_M$.
The difference between the resulting $R_M$ value and the
default fit is quoted as the systematic uncertainty from this
source. It is very small and can be found on line 4 of Table~\ref{tab-syst}.

\subsubsection{The amount of the associated signal}

The systematic error due to the uncertainty in the associated signal
fraction 
is estimated by varying the fraction and repeating the fitting procedure.
Taking into account  the uncertainties on the measured branching fractions
\cite{PDG} of the 
associated signal decay channels, we conservatively vary the amount of
associated signal  by $\pm 40\%$.
We recalculate the $R_M$ values and compare them to the default $R_M$
value; we 
quote the differences as the systematic uncertainty from this source. From
Table~\ref{tab-syst} (line 5) one can see that it is almost negligible.

\begin{figure}[t]
\begin{center}
\includegraphics[width=\ssld]{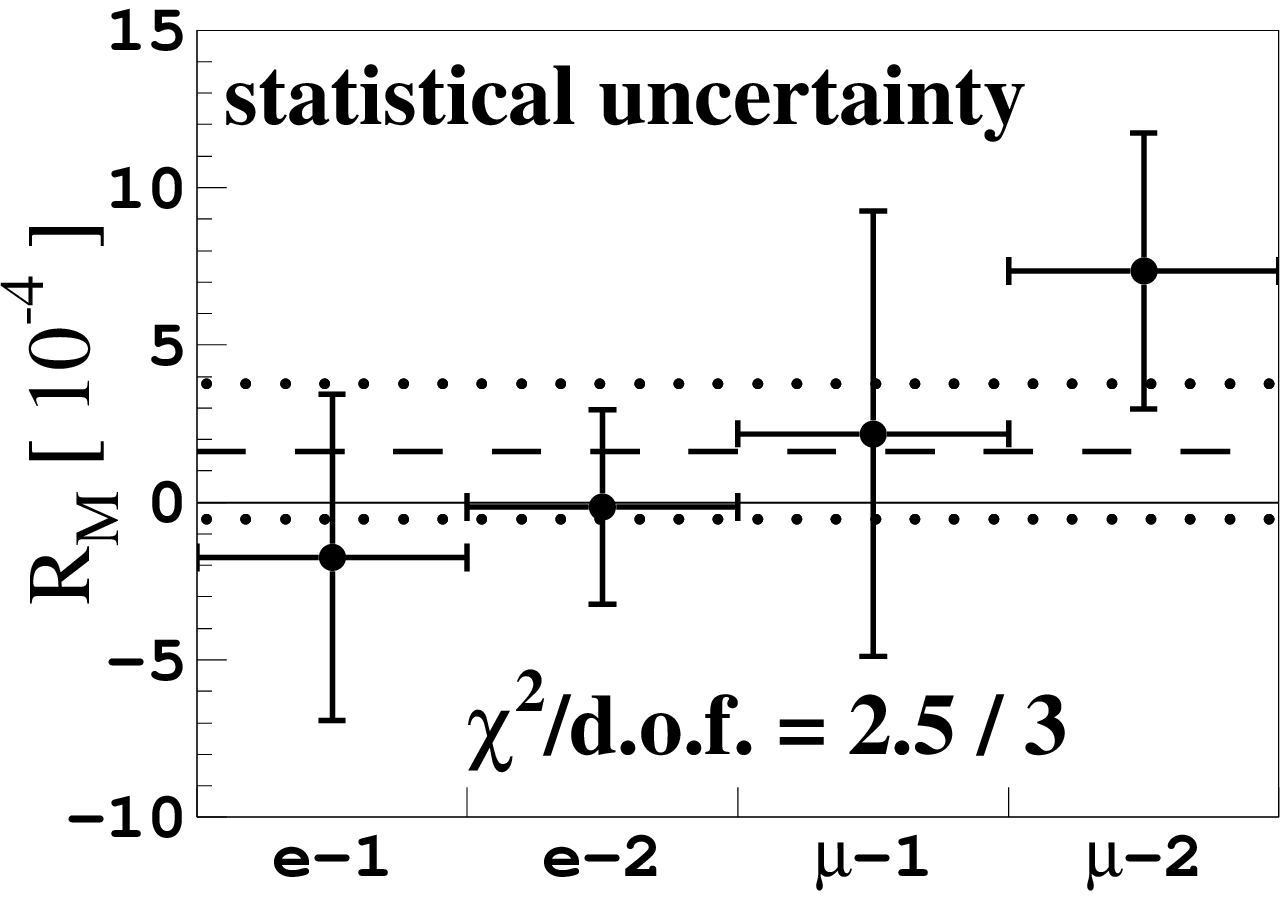}
\includegraphics[width=\ssld]{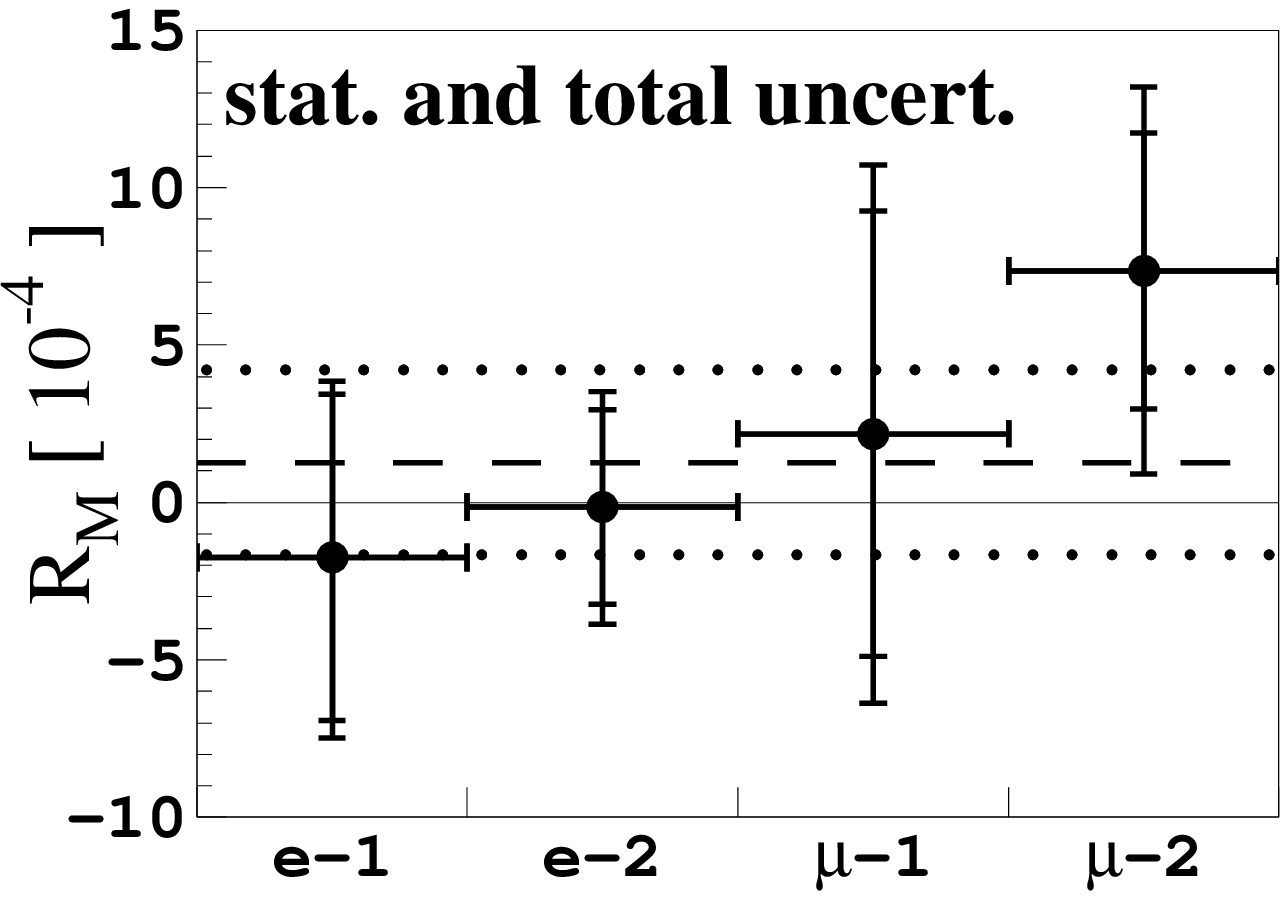}
\end{center}
\caption{ Left: the $R_M$ values of the four subsamples with the statistical
uncertainty only, and the result of the fit to these four values (dashed
line, $\chi^2/{\rm d.o.f.} = 2.5/3$).
Right: the four $R_M$ values
with the systematic uncertainty included and the combined result (dashed line), obtained
as described in Sec. \ref{sec-res-sys}. The dotted lines represent
the $\pm 1 \sigma$ interval. The solid line corresponds to no mixing.}
\label{fig_rm_fin}
\end{figure}

\subsubsection{The amount of the RS correlated background}

From MC simulation studies we find that in the electron decay mode the
largest contributions to the RS correlated background come from the
following decays: 
$D^0 \to K^- \pi^+ \pi^0$ (33\% of the RS correlated background), 
$D^0 \to K^- \pi^+ \pi^0 \pi^0$  (14\%), 
$D^0 \to K^- e^+ \nu_e $ (13\%).
In total  60\% of the RS correlated background 
comes from these three decays.

In the muon decay mode, the largest contributions to the RS correlated background come from 
$D^0 \to K^- \pi^+ \pi^0$ (43\%), 
$D^0 \to K^- \pi^+ \pi^0 \pi^0$ (17\%), 
$D^0 \to K^- \pi^+ \pi^- \pi^+$ (12\%).
In total  72\% of the correlated background
comes from these decays.
We calculate the weighted average of the relative uncertainties of the
branching fractions \cite{PDG} for the stated decay modes. For the electron
decay mode the averaged relative uncertainty is $\pm$4.3\% and for the muon decay mode $\pm$4.4\%. 

We repeat the RS 
fits, changing the amount of the total correlated background by the
average uncertainties on the branching fractions. The differences
between the obtained values of $R_M$ and the default values are taken
as the systematic uncertainty from this source.
They can be found in Table~\ref{tab-syst} (line 6) and are negligible.

\renewcommand{\arraystretch}{1.5}
\begin{table}[htbp]
\begin{center}
\caption{ A summary of the systematic uncertainties on $R_M$, the total
systematic uncertainty in each subsample and the combined (summed in quadrature) statistical and
systematic uncertainty. Values are given
in units of $10^{-4}$.}
\begin{tabular}{r l c c c c}
\hline
\hline
 & source	& e-1	& e-2	& $\mu$-1 	& $\mu$-2 	\\
\hline
1 & fitting histo. statistics	& $\pm$1.54	& $\pm$0.91	& $\pm$2.64	& $\pm$1.81	\\
2 & WS correlated bkg.	& $\pm$0.37	& ${+0.39}\atop{-0.38}$	& ${+2.98}\atop{-2.89}$	& ${+3.05}\atop{-2.97}$ \\
3 & WS uncorrelated bkg.	& ${+1.30}\atop{-1.88}$	& ${+1.70}\atop{-1.85}$	& ${+2.58}\atop{-2.82}$	& ${+1.57}\atop{-3.20}$ \\
4 & imperfect $t_{xy}$		& $\pm$0.05	& $\pm$0.02	& $\pm$0.01
& ${+0.25}\atop{-0.33}$	\\
5 & associated signal & ${+0.01}\atop{-0.00}$	& $\pm$0.00	& ${+0.09}\atop{-0.10}$	&
$\pm$0.02 \\
6 & RS correlated bkg. & $\pm$0.00	& $\pm$0.00	& $\pm$0.01	&
$\pm$0.04 \\
\hline
& total	systematic	& ${+2.05}\atop{-2.46}$	&
${+1.97}\atop{-2.10}$	& ${+4.75}\atop{-4.83}$	& ${+3.89}\atop{-4.74}$	\\
\hline
& statistical + systematic & ${+5.58}\atop{-5.74}$	&
${+3.66}\atop{-3.73}$	& ${+8.53}\atop{-8.57}$ & ${+5.86}\atop{-6.45}$	\\
\hline
\hline
\end{tabular}
\label{tab-syst}
\end{center}
\end{table}
\renewcommand{\arraystretch}{1.1}

\subsubsection{Total systematic uncertainty and the final result}
\label{sec-res-sys}

The final result of the measurement is obtained by averaging the
results for the four subsamples, e-1, e-2, $\mu-1$ and $\mu-2$. As
explained at the beginning of Sec. \ref{ch-result}, the results
obtained by 
method 3) are used 
(quoted in Table \ref{tab-mu2}) as ``combined''.

The contributions to the systematic uncertainty are divided into three
categories:
\begin{itemize}

\item[(a)] The systematic uncertainty that is completely correlated
between all four subsamples. This is the error due to the uncertainty of
the WS correlated background fraction, Sec. \ref{sec-syst-wscb}.

\item[(b)] The systematic uncertainty that is completely correlated
between the SVD-1 and SVD-2 subsamples and is uncorrelated between the
electron and the muon decay mode. Such a contribution comes from the
uncertainty of the  \dm shape of the uncorrelated WS 
background, Sec. \ref{sec-syst-wsucb}.

\item[(c)] The systematic uncertainties that are uncorrelated between
the four subsamples, or are very small. The main contribution comes from
the uncertainty due to the finite statistics  of the fitting
distributions, Sec. \ref{sec-finite-stat}. The uncertainties from
all the remaining sources are also added.

\end{itemize}

\noindent To obtain the final result and its uncertainty, taking into
account the systematic uncertainties,  we adopt the following
procedure:

\begin{itemize}
\item[(1)] For each of the four subsamples, we add to the statistical uncertainty in
quadrature all the uncertainties from category (c).

\item[(2)] We perform the $\chi^2$ fit to the SVD-1 and SVD-2 $R_M$
values in the electron and muon decay mode to obtain the averaged 
value for the electron and muon decay mode, 
$(-0.56 \pm 2.76) \times 10^{-4}$ and 
$( 5.89 \pm 4.02) \times 10^{-4}$, respectively.
The quoted uncertainties include the statistical uncertainty and the uncertainties (c).

\item[(3)] To add the uncertainty (b) for the electron decay mode, we
first simultaneously increase and then simultaneously decrease the
results for the e-1 and e-2 
subsamples by the 
uncertainty (b) and repeat step (2). 
The difference from the
default result of step (2) is added in quadrature to the uncertainty
obtained in step (2). The result for the electron decay mode, including
the statistical uncertainty and systematic uncertainties (b) and (c), is 
$(-0.56^{+3.19}_{-3.33} ) \times 10^{-4}$. 
We perform the same procedure also for the muon
decay mode; the result is $(5.89^{+4.43}_{-5.07} ) \times 10^{-4}$.

\item[(4)] We perform a $\chi^2$ fit to the results for the electron
and muon decay mode, obtained in step (3); the result is 
$(1.27 \pm 2.70)\times 10^{-4}$. The obtained mean value is the
final result, but the uncertainty needs to be increased by the uncertainty (a).

\item[(5)] To account for the uncertainty (a) we first simultaneously
increase and then  simultaneously decrease the initial four $R_M$ values by the
uncertainties (a) and repeat the steps (1)--(4). The difference from
the default result, ${}^{+1.13}_{-1.11}\times 10^{-4}$, is added in
quadrature to the uncertainty 
previously obtained from step (4) to obtain the final uncertainty of the result.

\end{itemize}

The total uncertainty of the final result is 
$\pm 2.93 \times 10^{-4}$. 
We calculate the contribution of the systematic
uncertainty  as
the difference between the total uncertainty and the statistical uncertainty (Eq.~(\ref{eq_stat_rez})), 
$2.93^2 - 2.16^2 = 1.98^2$. 
The final result is then
\zac
R_M = (1.3 \pm 2.2 \pm 2.0)\times 10^{-4},
\label{eq-res}
\kon
where the first uncertainty is statistical and the second systematic. 
As this value is close to the boundary of the physical region
($R_M \geq 0$) we use the Feldman-Cousins approach
\cite{FeldmanCousins} to calculate upper limits:
\zac
R_M < 6.1 \times 10^{-4}~{\rm  at~the~90\% ~confidence~level,}\\ 
\label{eq-ul1}
R_M < 7.0 \times 10^{-4}~{\rm  at~the~95\% ~confidence~level.} 
\label{eq-ul2}
\kon
With systematic uncertainties included,
the final results for the electron and muon decay modes are:
\zac
\label{eq-res-e}
R_M^e = (-0.6 \pm 2.7 ^{+1.8}_{-2.1})\times 10^{-4},\\
\label{eq-res-mu}
R_M^\mu = (5.9 \pm 3.7 ^{+3.9}_{-4.5})\times 10^{-4}.
\kon
The $R_M$ values for the
four subsamples, including the systematic uncertainty, and the combined result
are shown in the right plot in Fig.~\ref{fig_rm_fin}.

The increase in the sensitivity of the current result, compared to the
one published in \cite{Urban},
is caused partially by the larger statistical power of the sample, but
also by the improvements in the measurement method.
The statistical uncertainty of the present result in the electron
sample (Eq.~(\ref{eq-res-e})) is about 22\% smaller than one would
expect by appropriately rescaling the uncertainty of the result \cite{Urban}
by the increase of the data set used.
The improvement is mainly due to improved selection criteria,
improved neutrino reconstruction and improvements in using the \d
proper decay time measurement.
The systematic error of the result in the electron sample is, however,
larger than the one published in \cite{Urban} as it is estimated more
conservatively.

\section{SUMMARY}

Using a data sample with
an integrated luminosity of 492.2~fb$^{-1}$,
collected
by the Belle detector, 
we have searched for \mix using semileptonic decays of the neutral
charmed meson,
$D^0 \to K^{(*)+} e^- \bar\nu_e $
and
$D^0 \to K^{(*)+} \mu^- \bar\nu_\mu.$
We select $D^0$ mesons produced via the decay
$D^{*+} \to \pi^+_s D^0$,
and tag the flavor of the $D$ meson at production by the charge of
the accompanying slow pion. The measured mixing rate $R_M$ is
consistent with no mixing in both electron and muon decay modes.
The combined result accounts for the partially correlated systematic
error and yields
$R_M = (1.3 \pm 2.2 \pm 2.0)\times 10^{-4}$.
Since it is consistent with zero we set upper limits on the mixing
rate of $R_M < 6.1 \times 10^{-4}$  at the 90\% confidence level.  

This result supersedes that published in Ref.~\cite{Urban}
and represents the most stringent experimental limit on
$R_M$ obtained to date from semileptonic $D^0$ decays.
Its accuracy is significantly better than that of the world 
average of previous measurements in semileptonic decays, 
$R_M = (1.7 \pm 3.9) \times 10^{-4}$ \cite{hfag}.
Although the sensitivity is not sufficient to
observe a positive mixing signal, 
it is worth noting that in semileptonic
decays  no model uncertainties can influence the result.
The reported value of $R_M$  is
in agreement with the world average values of 
$x = (0.87^{+0.37}_{-0.34})\%$
and
$y = (0.66^{+0.21}_{-0.20})\%$
\cite{hfag}
and 
it will help in further constraining the $D^0$ mixing parameters
in combination with the results of the measurements in other decay
channels.

\begin{center}
{\bf Appendix}
\end{center}

In the proper decay time fit the following functions are used:

The Lorentz function, centered at 0:
\zac
L(t;b) = \frac{b}{\pi}\cdot\frac{1}{1 + (bt)^2}
\label{eq-L}
\kon

The asymmetric Lorentz function, centered at 0:

\zac
L_a(t;b_l,b_r) = \left\{ \begin{array}{rr}
	  \displaystyle{\frac{b_l b_r}{(b_l+b_r)\pi}\cdot\frac{1}{1 +
(b_l t)^2}}\quad ; \quad t  <  0 \\
	\\
	  \displaystyle{\frac{b_l b_r}{(b_l+b_r)\pi}\cdot\frac{1}{1 + (b_r t)^2}}\quad ; \quad t \ge 0
	\end{array}\right.
\label{eq-La}
\kon

~\\

\begin{acknowledgments}

We thank the KEKB group for the excellent operation of the
accelerator, the KEK cryogenics group for the efficient
operation of the solenoid, and the KEK computer group and
the National Institute of Informatics for valuable computing
and Super-SINET network support. We acknowledge support from
the Ministry of Education, Culture, Sports, Science, and
Technology of Japan and the Japan Society for the Promotion
of Science; the Australian Research Council and the
Australian Department of Education, Science and Training;
the National Natural Science Foundation of China under
contract No.~10575109 and 10775142; the Department of
Science and Technology of India; 
the BK21 program of the Ministry of Education of Korea, 
the CHEP SRC program and Basic Research program 
(grant No.~R01-2005-000-10089-0) of the Korea Science and
Engineering Foundation, and the Pure Basic Research Group 
program of the Korea Research Foundation; 
the Polish State Committee for Scientific Research; 
the Ministry of Education and Science of the Russian
Federation and the Russian Federal Agency for Atomic Energy;
the Slovenian Research Agency;  the Swiss
National Science Foundation; the National Science Council
and the Ministry of Education of Taiwan; and the U.S.\
Department of Energy.
\end{acknowledgments}


\begin{thebibliography}{99}


\bibitem{belleKK} M.~Stari\v{c} {\it et al.} (Belle Collaboration),
Phys. Rev. Lett. {\bf 98} 211803 (2007).

\bibitem{babarKpi} B.~Aubert {\it et al.}, (BaBar Collaboration),
Phys. Rev. Lett. {\bf 98}, 211802 (2007).

\bibitem{LP07} See the talks by
W. Lockman   (http://chep.knu.ac.kr/ lp07/htm/S4/S04\_13a.pdf) and 
K. Tollefson (http:// chep.knu.ac.kr/lp07/htm/S4/S04\_14.pdf) 
at Lepton Photon 2007;
W. M. Sun (for CLEO), arXiv:0712.0498v1.

\bibitem{hfag} Heavy Flavor Averaging Group (Charm Decays
subgroup), http://www.slac.stanford.edu/xorg/hfag/ charm/index.html

\bibitem{small_mix} For a review see 
I.I.~Bigi and N.~Uraltsev, Nucl. Phys. B {\bf 592}, 92 (2001);
S.~Bianco, F.L.~Fabbri, D.~Benson and I.~Bigi,
Riv. Nuovo Cim. {\bf 26N7}, 1 (2003);
A. Falk {\it et al.}, Phys. Rev D {\bf 69}, 114021 (2004).

\bibitem{Xing} Z.-Z.~Xing, Phys. Rev. D {\bf 55}, 196 (1997).

\bibitem{Urban} U.~Bitenc {\it et al.} (Belle Collaboration), 
Phys. Rev. D {\bf 72}, 071101 (2005).

\bibitem{Babar} B.~Aubert {\it et al.} (BaBar Collaboration), Phys. Rev. D
{\bf 70}, 091102 (2004); B.~Aubert {\it et al.} (BaBar Collaboration),  Phys. Rev. D
{\bf 76}, 014018 (2007) .

\bibitem{Cleo}C. Cawlfield {\it et al.} (CLEO Collaboration),
Phys. Rev. D {\bf 71}, 077101 (2005).

\bibitem{E791} E. M. Aitala {\it et al.} (E791 Collaboration),
Phys. Rev. Lett. {\bf 77}, 2384 (1996).

\bibitem{liming-kpi} L. Zhang {\it et al.} (Belle Collaboration),
 Phys. Rev. Lett. {\bf 96}, 151801 (2006).

\bibitem{babar-k2pi}  B.~Aubert {\it et al.} (BaBar Collaboration),
Phys. Rev. Lett. {\bf 97}, 221803 (2006).

\bibitem{babar-k3pi} B. Aubert {\it et al.} (BaBar Collaboration),
 hep-ex/0607090.


\bibitem{pospesevalnik} S.~Kurokawa and E. Kikutani, Nucl. Instr. Meth. A
{\bf 499}, 1 (2003), and other papers included in this volume.

\bibitem{detektor} A.~Abashian {\it et al.} (Belle Collaboration),
Nucl. Instr. Meth. A {\bf 479}, 117 (2002).

\bibitem{SVD-2} Z.~Natkaniec {\it et al.} (Belle SVD-2 group),
Nucl. Instr. Meth. A {\bf 560}, 1 (2006).

\bibitem{MC} Events are simulated with the EvtGen generator, D.-J.
Lange, Nucl. Instr. Methods Phys. Res., Sect. A {\bf 462},
152 (2001); the detector response is simulated with
GEANT, R. Brun et al., GEANT 3.21, CERN Report
No. DD/EE/84-1, 1984.

\bibitem{naboj} The charge-conjugate modes are implied throughout the
paper unless otherwise stated.

\bibitem{hadron_B} K.~Abe {\it et al.} (Belle Collaboration),
Phys. Rev. D {\bf 66} 032007 (2002).

\bibitem{fox-wolfram} G.~C.~Fox, S.~Wolfram, 
Phys. Rev. Lett. {\bf 41}, 1581 (1978).

\bibitem{e_id}  K.~Hanagaki {\it et al.}, Nucl.\ Instr.\ Meth. A \
{\bf  485}, 490 (2002). %

\bibitem{muid} A. Abashian {\it et al.},
 Nucl. Instr. Meth. A {\bf 491}, 69 (2002).

\bibitem{PDG} W.-M.~Yao  {\it et al.} (Particle Data Group),
 J. Phys. G {\bf 33}, 1 (2006).

\bibitem{momenta} In the following, $P$ denotes the 
particle's 4-momentum while 
$\vec{p}$ and $p$ denote the 3-dimensional momentum and its magnitude,
respectively.

\bibitem{nu_reco_mass_assignment} The actual mass assignment 
of tracks included in $P^*_{\rm rest}$ has only a 
marginal effect on the neutrino reconstruction and thus on the $\Delta M$ 
resolution. If all charged particles are assigned a mass of the pion the 
$\Delta M$ resolution remains almost the same.

\bibitem{FeldmanCousins} G.J.~Feldman and R.D.~Cousins, 
Phys. Rev. D {\bf 57}, 3873 (1998).


\end{thebibliography}
\end{document}